\RequirePackage{fix-cm}
\documentclass[natbib,smallextended]{svjour3}       
\smartqed  
\usepackage{graphicx}
\usepackage{amssymb}
\usepackage{longtable}
\usepackage{lscape}
\usepackage{color}
\usepackage{changepage}
\begin{document}
\title{Magnetic White Dwarfs}
\author{Lilia Ferrario \and Domitilla deMartino \and Boris T. G\"ansicke}
\authorrunning{Ferrario, deMartino, G\"ansicke} 

\institute{L. Ferrario \at
              Mathematical Sciences Institute, The Australian National University, ACT2601, Australia  \\     
              \email{Lilia.Ferrario@anu.edu.au} \\
              \and
            D. de Martino \at INAF - Osservatorio Astronomico di
            Capodimonte, Via Moiariello 16,  I-80131, Naples, Italy \\
            \email{demartino@oacn.inaf.it}
            \and  B.T. G\"ansicke \at Department of Physics,
            University of Warwick, Coventry CV4 7AL, UK\\
            \email{Boris.Gaensicke@warwick.ac.uk}
}

\date{Received: date / Accepted: date}

\maketitle

\begin{abstract}
  In this paper we review the current status of research on the
  observational and theoretical characteristics of isolated and binary
  magnetic white dwarfs (MWDs).

  Magnetic fields of isolated MWDs are observed to lie in the range
  $10^3-10^9$\,G.  While the upper limit cutoff near $10^9$\,G appears
  to be real, the lower limit is more difficult to investigate. The
  incidence of magnetism below a few $10^3$\,G still needs to be
  established by sensitive spectropolarimetric surveys conducted on
  8\,m class telescopes.

  Highly magnetic WDs tend to exhibit a complex and non-dipolar field
  structure with some objects showing the presence of higher order
  multipoles. There is no evidence that fields of highly magnetic
  WDs decay over time, which is consistent with the estimated
  Ohmic decay times scales of $\sim 10^{11}$\,yrs.  The slow rotation
  periods ($\sim 100$\,yrs) inferred for a large number of isolated
  MWDs in comparison to those of non-magnetic WDs (a few days) suggest
  that strong magnetic fields augment the braking of the stellar
  core. 

  MWDs, as a class, {also appear to be more massive
    ($0.784\pm 0.047$\,M$_\odot$) than their weakly or non-magnetic
    counterparts ($0.663\pm0.136$\,M$_\odot$).}

  MWDs are also found in binary systems where they accrete matter from
  a low-mass donor star. These binaries, called magnetic Cataclysmic
  Variables (MCVs) and comprise about 20-25\% of all known CVs.  Zeeman
  and cyclotron spectroscopy of MCVs have revealed the presence of
  fields in the range $\sim 7-230$\,MG. Complex field geometries have been inferred
  in the high field MCVs (the polars) whilst magnetic field strength
  and structure in the lower field group (intermediate polars, IPs)
  are much harder to establish.

  {The MCVs exhibit an orbital period distribution
    which is similar to that of non magnetic CVs. Polars dominate the
    distribution at orbital periods $\lesssim$4\,h and IPs at longer
    periods. It has been argued that IPs above the $2-3$\,hr CV period
    gap with magnetic moments $\gtrsim 5\times 10^{33}$\,G\,cm$^3$ may
    eventually evolve into polars. It is vital to enlarge the still
    incomplete sample of MCVs to understand not only their accretion
    processes but also their evolution.}

  The origin of fields in MWDs is still being debated. While the
  fossil field hypothesis remains an attractive possibility, field generation within
  the common envelope of a binary system has been gaining momentum, since it
  would explain the absence of MWDs paired with non-degenerate
  companions and also the lack of relatively wide pre-MCVs.

\end{abstract}


\section{Introduction} 
\label{s:introduction}

{The Sloan Digital Sky Survey \citep[SDSS,][]{York00}
  has increased the number of known MWDs with fields $B$ in the range
  $2-1\,000$\,MG from fewer than 70, as listed in
  \citet{wickramasinghe00} to over 600 in 2015} \citep[see][ and this
work]{Gaensicke2002,Schmidt2003,Vanlandingham2005,Kulebi2009,
  Kepler2013,Kepler2015}. However, their space density estimates are
still debated. Volume-limited samples suggest that $\sim10-20$\% of
WDs are magnetic \citep{Kawka2007, giammicheleetal12-1, sionetal14-1},
whereas magnitude-limited samples indicate that only $\sim 2-5$\% are
magnetic \citep{Liebert2003,Kepler2013,Kepler2015}. This discrepancy may be partly
resolved by correcting for the difference in search volume for the
MWDs, since, on average, they are more massive than
their non-magnetic counterparts \citep[see Sect.\,\ref{mass_mwds}
and][]{Liebert2003}.

A spectropolarimetric survey of a small sample of cool
($\lesssim $14,000\,K) WDs conducted by \citet{Landstreet2012} found that
the probability of detecting a kG field in DA\footnote{DA type WDs
  have a hydrogen rich atmosphere.} WDs is $\sim 10$\% per decade of field
strength but also stress the inability of current precision measures
to reveal whether there is a lower cutoff to the field strengths in
WDs or there is a field below which all WDs are magnetic.

Furthermore, there seems to be a paucity of young MWDs in the
intermediate field range \citep[$0.1-1$\,MG, see][]{Kawka2012}. The
reason for this dearth of objects is not clear since they should be
easily detected in most spectropolarimetric surveys such as that
conducted by \citet{Aznar2004}.

The ESO SNIa Progenitor Survey (SPY) also pointed to a peculiarly low
percentage of MWDs in the field range $0.1-1$\,MG
\citep{Koester2001}. Therefore, the magnetic field distribution of WDs
could be bimodal exhibiting a high field ($1-1,000$\,MG) population
and a low field ($<0.1$\,MG) one.  Intensive searches for
magnetic fields in the direct progenitors of MWDs have so far drawn a
blank, both among planetary nebulae
\citep{asensioramosetal14-1,leoneetal14-1}, and hot subdwarfs
\citep{mathysetal12-1,savanovetal13-1}.

MWDs can also be found in CVs, which are close binary systems where a
WD accretes matter from a late-type main sequence companion through
Roche-lobe overflow.  Orbital periods are less than a day and orbital
separations of the order of the solar radius making these binaries
relatively compact.  Because of their abundance\footnote{1166 CVs are
  reported in the 7.20 (dec.2013) version of the \citet{ritter_kolb}
  catalogue} and proximity, CVs are key objects to understand close
binary evolution. In particular, those hosting a strongly magnetic WD
($B \gtrsim$ 1\,MG) allow us to study accretion and emission processes
in a strong magnetic field environment as well as improving our
understanding of the influence of magnetic fields in close binary
evolution. 

{The number of known MCVs has also increased
  dramatically over the years – from about 60 as listed in
  \citet{wickramasinghe00} to about 170. Magnetic field measures are
  however available only for about half of them most of which in the
  range $\sim 7-230$\,MG.}

Here we review the current status of research on MCVs and MWDs and
outline key aspects and open problems to be investigated in the
future. Previous reviews on the topic can be found in
\citet{Angel1978,Angel1981,cropper90a, chanmugam92-1, Patterson94,
  wickramasinghe00}.

\section{Historical background}

\subsection{Isolated Magnetic white dwarfs}

Grw$+70^\circ 8247$, discovered by \citet{Kuiper1934}, was the first
WD to be classified as magnetic when \citet{Kemp1970} demonstrated that
its light was strongly circularly polarised. Although the spectrum of
Grw$+70^\circ 8247$ appeared to be nearly featureless, close
inspection by \citet{Minkowski1938} and \citet{Greenstein1957}
revealed the presence of unusual shallow broad absorption bands near
3,650\AA, 4,135\AA and 4,466\AA, which became known as
``Minkowski bands''.  The spectral features of Grw$+70^\circ 8247$
remained unidentified till the mid-80s, when the first computations of
the hydrogen transitions in strong magnetic fields became available
(see Sect.\,\ref{ss:fieldmeasuresWD}). These calculations allowed the
spectral features of Grw$+70^{\circ} 8247$ to be identified as Zeeman
shifted hydrogen lines in a magnetic field of $100-320$\,MG
\citep{Angel1985,Greenstein1985,Wickramasinghe1988}.  In particular, the famous
Minkowski band near 4,135\AA\, was shown to be a Zeeman
component of H$\beta$ shifted some 700\AA\, from its zero field
position.

Following the discovery of polarisation in Grw$+70^\circ 8247$,
\citet{Angel1971a} detected circular polarisation in another object,
G$195-19$. Follow-up observations conducted by \citet{Angel1971b}
revealed that the circular polarisation in G$195-19$ varies at the
spin period of the star under the assumption that the magnetic axis is
inclined with respect to the spin axis of the WD \citep[the
`oblique rotator' of][]{Stibbs1950}. In both Grw$+70^\circ 8247$ and
G$195-19$, some linear polarisation was also detected, although at a
much lower level than circular polarisation. Since the observations
showed that the polarisation data in G$195-19$ did not follow a
sinusoidal behaviour, \citet{Landi1976} raised the possibility that
magnetic spots were responsible for the observed asymmetries.

\citet{Landstreet1971} soon discovered circular polarisation in
a third WD, G$99-37$ and a few years later, polarisation was also
detected in GD$229$ \citep{Swedlund1974}. The spectral features of
GD$229$ remained a mystery for more than 20 years until the
calculations for He I transitions in a strong magnetic field became
available (see Sect.\, \ref{ss:fieldmeasuresWD}). The
absorption structures were thus interpreted as stationary line
transitions of helium in a magnetic field of $300-700$\,MG
\citep{Jordan1998,Wickramasinghe2002}.

The sample of known MWDs has rapidly increased since Kemp's first
discovery in 1970. Over the past 40 years, thanks to surveys such as
the Hamburg/ESO Quasar Survey \citep[HQS][]{Wisotzki1991}, the
Edinburgh-Cape survey \citep{Kilkenny1991}, and, as already noted, the
SDSS, their number has grown to more than 600 (if we also count
objects with uncertain or no field determination) thus providing a large
enough sample to allow meaningful statistical studies of their
characteristics (see Table\,\ref{tab:mwds}).

\subsection{Magnetic Cataclysmic Variables}

Known as a Novalike since 1924, AM\,Her was the first CV discovered to
emit soft X-rays by the \emph{UHURU} and \emph{SAS-3} satellites in
1976 \citep{Hearn76}.  Follow-up optical observations revealed
variable linear (up to 7\%) and circular (up to 9\%)
polarisation at the 3.09\,h binary orbital period
\citep{Tapia77}. Systems with similar characteristics were named
AM\,Her-type variables.  The name ``polar" was introduced later for
AM\,Her and other objects identified as X-ray sources that also
showed polarised optical light \citep[see][]{Warner95}.  

The binary system DQ\,Her was discovered in the mid-50s to display a
71\,s periodic variability \citep{Walker56} and only about two decades
later was also found to be weakly polarised \citep{Swedlund74}. In
1978, 33\,s optical pulsations were detected in AE\,Aqr, but not in
polarised light \citep{Patterson79}.  In the following years, fast
optical periodic variations at periods much shorter than the orbital
one ($P_{\rm rot} << P_{\rm orb}$) were found in other CVs. These systems were
first called DQ\,Her-type variables and then renamed ``intermediate
polars" (IPs) \citep{Patterson83,Patterson94,Warner95}.  This led to
the division of MCVs into two groups, polars and IPs.

The properties of polarised radiation were first studied by
\citet{Chanmugam_Dulk81} and \citet{Meggit_Wickramasinghe82}.  The
magnetic moment of the accreting WD in polars ($\mu \gtrsim 5\times
10^{33}$\, G\,cm$^3$) is sufficient to lock the stars into synchronous
rotation with the orbital period ($P_{\rm orb}\sim 70-480$ mins). On
the other hand, the magnetic moment of the WD in IPs is not high
enough to phase-lock the stars into synchronous rotation with the
orbit, resulting in WD spin periods $P_{\rm s}$ that are shorter than their
orbital periods.

Because of their strong soft X-ray emission, the number of polars
increased significantly thanks to soft X-ray surveys, and in
particular that conducted in the nineties by the \emph{ROSAT}
satellite \citep{beuermann99}. To date $\sim$\,110 systems of this
class are known hosting WDs with surface field strengths
$B \sim$\,7-230\,MG (see Sect.\ref{ss:fieldmeasures} and
Table\,\ref{tab:mcvs} { for a complete list of known
  systems as of December 2014).}  The IPs constituted a minor group
of harder X-ray sources and remained elusive objects until the recent
hard X-ray surveys conducted by the \emph{INTEGRAL}/IBIS and
\emph{Swift}/BAT satellites \citep{bird10,Baumgartner13}. The flux
limits ($\sim 10^{-11}$\,erg\,cm$^{-2}$\,s$^{-1}$) of these surveys
can detect sources up to 1\,kpc away for X-ray luminosities
$L_x \sim 10^{33}$\,erg\,s$^{-1}$. IPs, being the brightest and
hardest X-ray sources among CVs, account for $\sim 20$\% of the
Galactic hard X-ray sources discovered \citep{barlow06}. Thus, the
number of identified IPs has now increased to $\sim 55$ systems
{\citep[see Table \ref{tab:ips} and reference
  therein and also][]{Bernardini12,Bernardini13}, with $\sim$\,60
  candidates still awaiting confirmation through X-ray follow-ups with
  sensitive facilities such as XMM-Newton and
  NuSTAR. \footnote{Further details on IP type CVs can be found at
    http://asd.gsfc.nasa.gov/Koji.Mukai/iphome/iphome.html}.}

\begin{figure}[t]
\begin{center}
\includegraphics[width=\columnwidth]{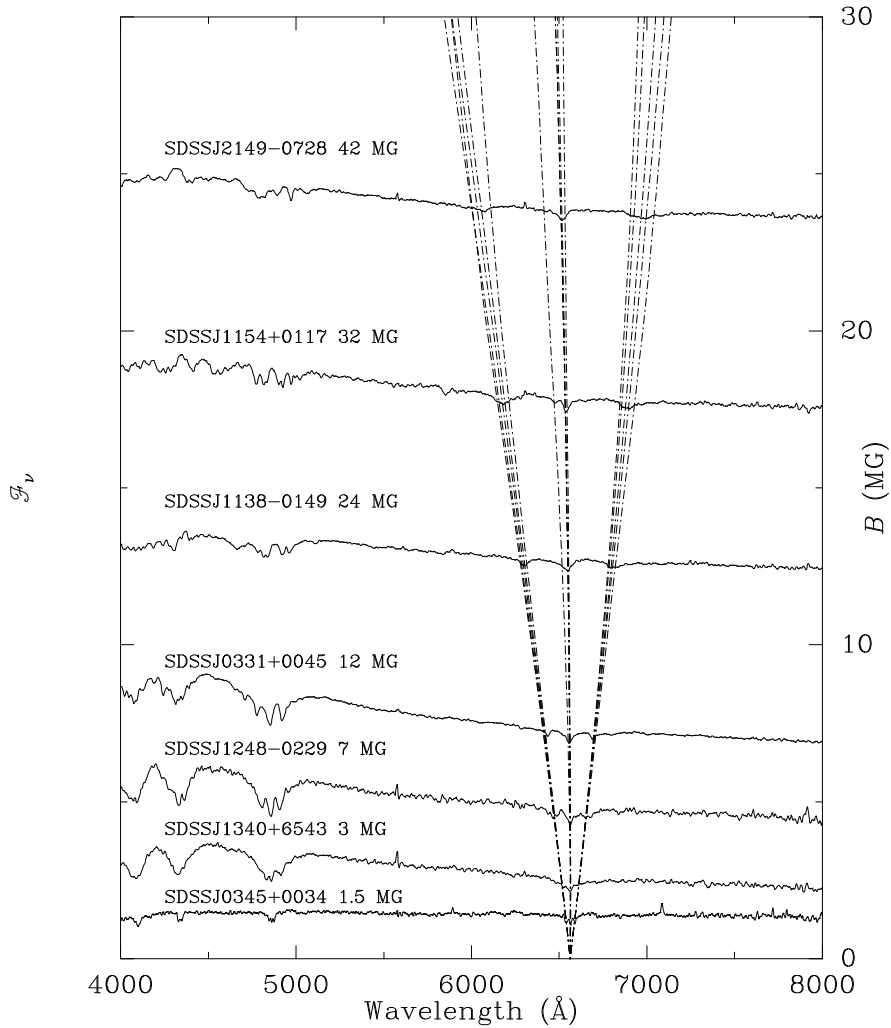} 
\caption{The Zeeman effect on H$_{\alpha}$ in the linear and quadratic
  regimes for fields of $1.5-42$\,MG). The quadratic effect becomes
  gradually more important in the higher members of the Balmer series
  and as the field strength increases \citep{Schmidt2003}}
\label{Fig1:zeem_triplets_low}
\end{center}
\end{figure}

\begin{figure}[t]
\begin{center}
\includegraphics[width=\columnwidth]{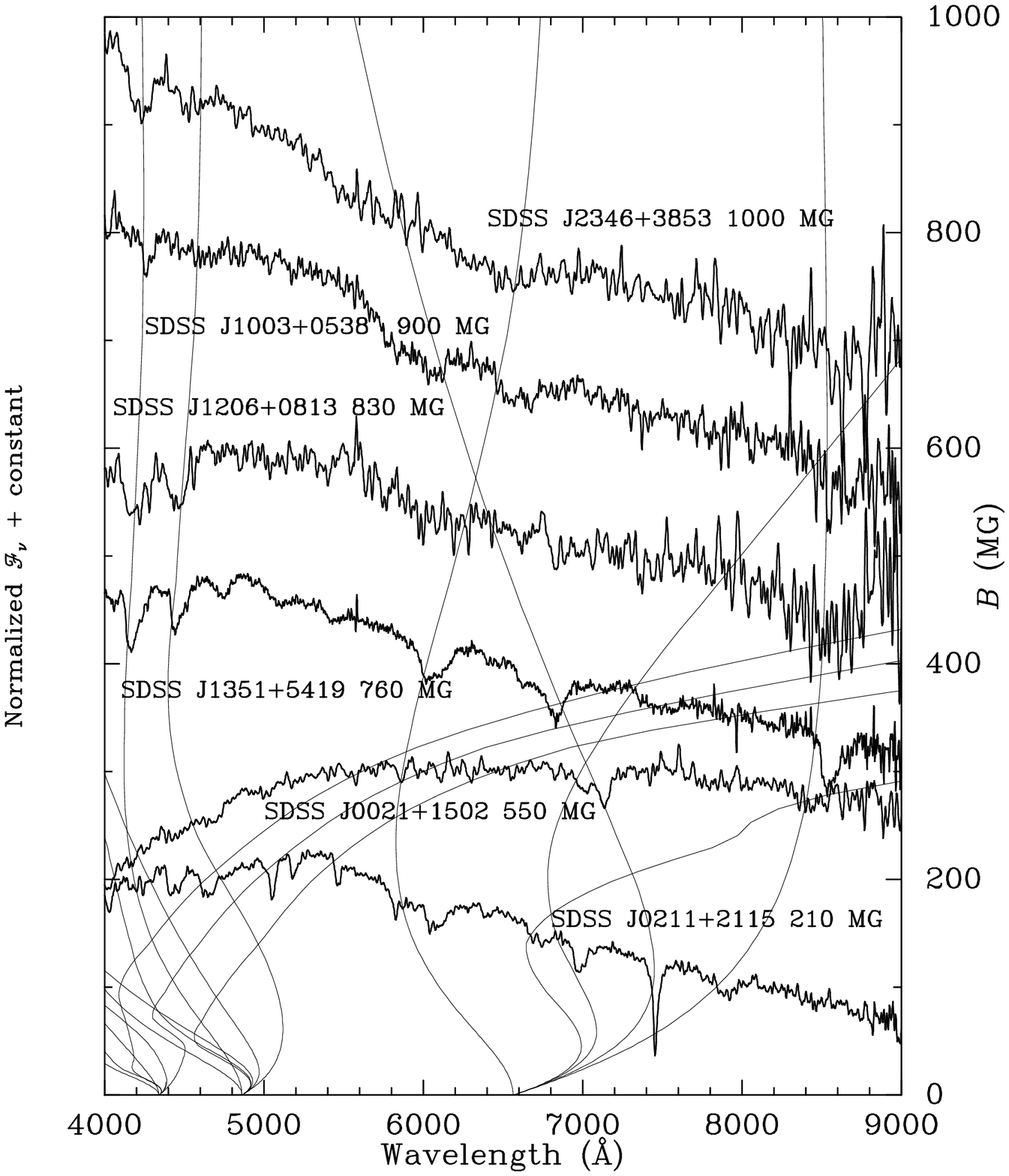}
\caption{Stationary Zeeman components of $H_{\alpha}$ and $H_{\beta}$
  \citep[from][]{Vanlandingham2005} in the spectra of strongly
    magnetic MWDs}
\label{Fig2:zeem_triplets_high}
\end{center}
\end{figure}

\section{Magnetic field measures} 
\label{ss:fieldmeasures}

In this section we shall review the techniques that are routinely
adopted to determine the magnetic field strength of isolated MWDs and
of accreting MWDs in binary systems.

\subsection{Field determination in isolated magnetic white dwarfs}
\label{ss:fieldmeasuresWD}

Similarly to their non-magnetic counterparts, most MWDs are
hydrogen-rich (DA). They are classified as DAp or DAH, indicating the
method used in their discovery. Thus, ``p'' stands for
polarisation measurements and ``H'' for Zeeman splitting
\citep[see][for more information on the WD spectral classification
system]{Sion1983}.  An understanding of their properties relies
heavily on the theory of the hydrogen atom at strong magnetic fields.

Depending on field strength, different Zeeman regimes become relevant
to the understanding of MWD spectra.  If $(n,l,m_l)$ are the quantum
numbers corresponding to zero field, then the removal of the $m_l$
degeneracy will give rise to the linear Zeeman regime ($\sim 1-5$\,MG
for Balmer lines).  Here the energy levels are shifted by
$\frac{1}{2}m_lh\omega_C$, where
$\omega_C = \displaystyle{\frac{eB}{m_e c}}$ is the cyclotron
frequency of a free electron, $m_e$ and $e$ are the mass and charge of
the electron respectively and $c$ is the speed of light.  A line is
split into three components: an unshifted central $\pi$ component
($\Delta m_l=0$), a redshifted $\sigma_+$ ($\Delta m_l=+1$) component
and a blueshifted $\sigma_-$ ($\Delta m_l=-1$) component. Some Zeeman
triplets observed in MWDs are shown in
Fig.\,\ref{Fig1:zeem_triplets_low} \citep{Schmidt2003}. Circular
spectropolarimetry across lines can be effectively used to detect low
fields ($B\lesssim 1$\,MG) when Zeeman splits are unresolvable in the
flux spectra. When viewed along the magnetic field, the $ \sigma_-$
and the $\sigma_+$ components are circularly polarised with opposite
signs.

As the field strength increases and/or $n$ increases, the quadratic
effect becomes gradually more important and the $l$ degeneracy is also
removed (inter-$l$ mixing regime). The energy shifts depend on the
excitation of the electron and the $\pi$ and $\sigma$ components are
shifted by different amounts from their zero field positions. The
quadratic shift is comparable to the linear shift for,
e.g. H$_\delta$, at $B\sim 4$\,MG (see
Fig.\,\ref{Fig1:zeem_triplets_low}). The first Zeeman calculations in
this intermediate field strength regime were conducted in 1974 by
\citeauthor{Kemic1974} for fields up to $\sim20$\,MG.

As the field progressively increases, the Coulomb and magnetic field
forces become comparable in strength and neighbouring $n$ manifolds
overlap (inter-$n$ mixing regime). In the ``strong field
mixing regime'' the magnetic field dominates
\citep[see][]{wickramasinghe00}. The first published data of
wavelengths and oscillator strengths of hydrogen transitions in the
infrared to ultraviolet bands in the presence of very strong magnetic
fields (up to $10^6$\,MG) were published in the mid 80s by
\citet{Rosner84,Forster84,Henry84,Henry85,Wunner85}, and more recently
by \citet{schimeczek+wunner14-1}.

An important characteristic which is clearly visible in the field
against wavelength curves diagrams is that the $\sigma^+$ components become
nearly `stationary'. That is, appreciable changes in $B$ only yield
small changes in wavelength.  This is very useful in establishing the
field of MWDs, since the features corresponding to these turning
points will have the smallest amount of magnetic broadening and will
have the largest influence on the observed field averaged spectrum.
We display in Fig.\,\ref{Fig2:zeem_triplets_high} some spectra of strongly magnetic WDs showing
the presence of stationary components \citep{Vanlandingham2005}.

Another interesting effect in the presence of strong
magnetic fields ($\gtrsim100$\,MG) and local electric fields in highly
ionised plasmas, is an increase of the oscillator strength of the
``forbidden'' $1{s_0}\rightarrow2{s_0}$ component at the expense of
the $\pi~(1{s_0}\rightarrow2{p_0})$ component. This was first detected
in \textit{HST} observations of RE\,J0317--853
\citep{Burleigh1999}, and later seen in the MCV AR\,UMa
\citep{Gaensicke01,Schmidt96}.

\begin{figure} [h]
\begin{center}
\includegraphics[width=\columnwidth]{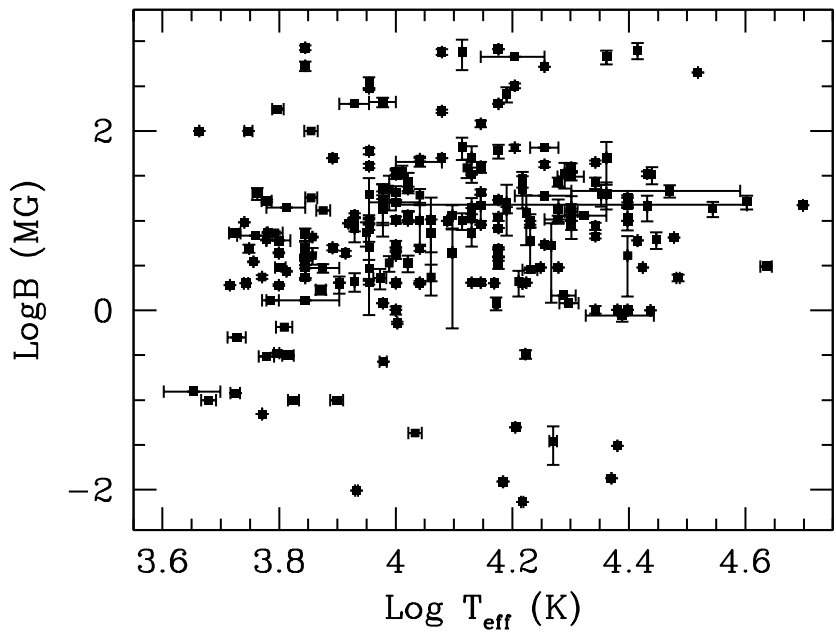}
\caption{Magnetic field strength against effective temperature in MWDs
  showing no indication for field evolution with cooling age (this work)}
\label{allB_T}
\end{center}
\end{figure}

\subsubsection{Magnetic field evolution in isolated magnetic white dwarfs}

There is no evidence for field evolution along the cooling curve, that
is, the mean field strength and the distribution about this
mean appear to be independent of effective temperature (see Fig.\,
\ref{allB_T}, this work). The Pearson Product
Moment Correlation test gives a correlation coefficient $r=0.03$,
which indicates that these variables are not related.

The free Ohmic decay time can be estimated from 
\[
t_{\rm{ohm}}\sim \frac{4\pi\sigma L^2}{c^2}
\]
where $L$ is the length scale over which the magnetic field varies and
$\sigma$ is the electrical conductivity.  If we set $L\sim\,R$ (where
$R$ is the stellar radius) and $\sigma$ equal to the value expected in
the degenerate cores of WDs then we have $t_{\rm{ohm}}\sim 2-6\times
10^{11}$\,yr almost independently of mass \citep{Cumming2002}.  The
lack of evidence for correlation between magnetic field strength and
effective temperature is consistent with these long decay time scales.

\begin{figure} [h]
\begin{center}
\includegraphics[width=\columnwidth]{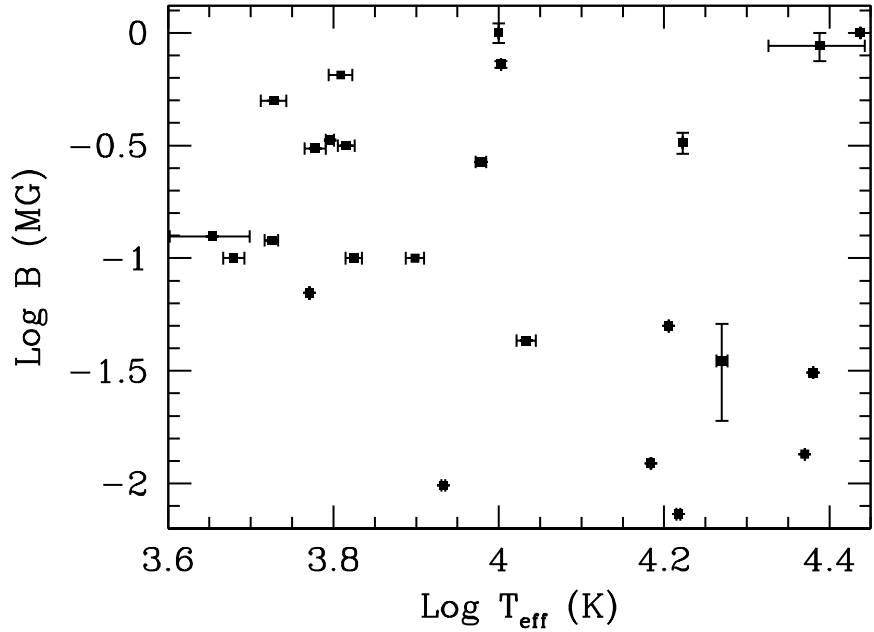}
\caption{Magnetic field strength against effective temperature in MWDs
  with $B\lesssim1$\,MG showing some correlation between field and
  temperature (this work)}
\label{weakB_T}
\end{center}
\end{figure}

{We note that \citet{Kawka2012} find that the
  distribution of field strengths below 1\,MG versus cooling ages may
  show some selections effects (see their Fig. 10). That is, objects
  with fields $\lesssim50$\,kG appear to be younger (that is hotter)
  than those with field $\gtrsim50$\,kG. We have calculated a
  correlation coefficient $r=-0.192$ for objects with
  $B\lesssim1$\,MG, indicating that this effect does exist also in our
  sample of weakly magnetic WDs listed in Table\,\ref{tab:mwds} (see
  Fig.\,\ref{weakB_T}). This trend could be caused by the fact that
  cool WDs ($\lesssim7000$\,K) do not have narrow and deep lines in
  their spectra that are good magnetic field tracers unless heavy
  element lines are present \citep{Kawka2012}.  Another possibility is
  that the estimation of parameters such as effective temperature and
  gravity using models for non-magnetic WDs \emph{may} introduce
  biases even in the presence of very low fields. However, the latter
  possibility is more difficult to ascertain at the present time. A
  possible explanation for why this effect is not apparent if we take
  all MWDs could be because their temperatures are estimated using a
  wide range of methods so that biases cancel each other out. }

\begin{figure} [h]
\begin{center}
\includegraphics[width=\columnwidth]{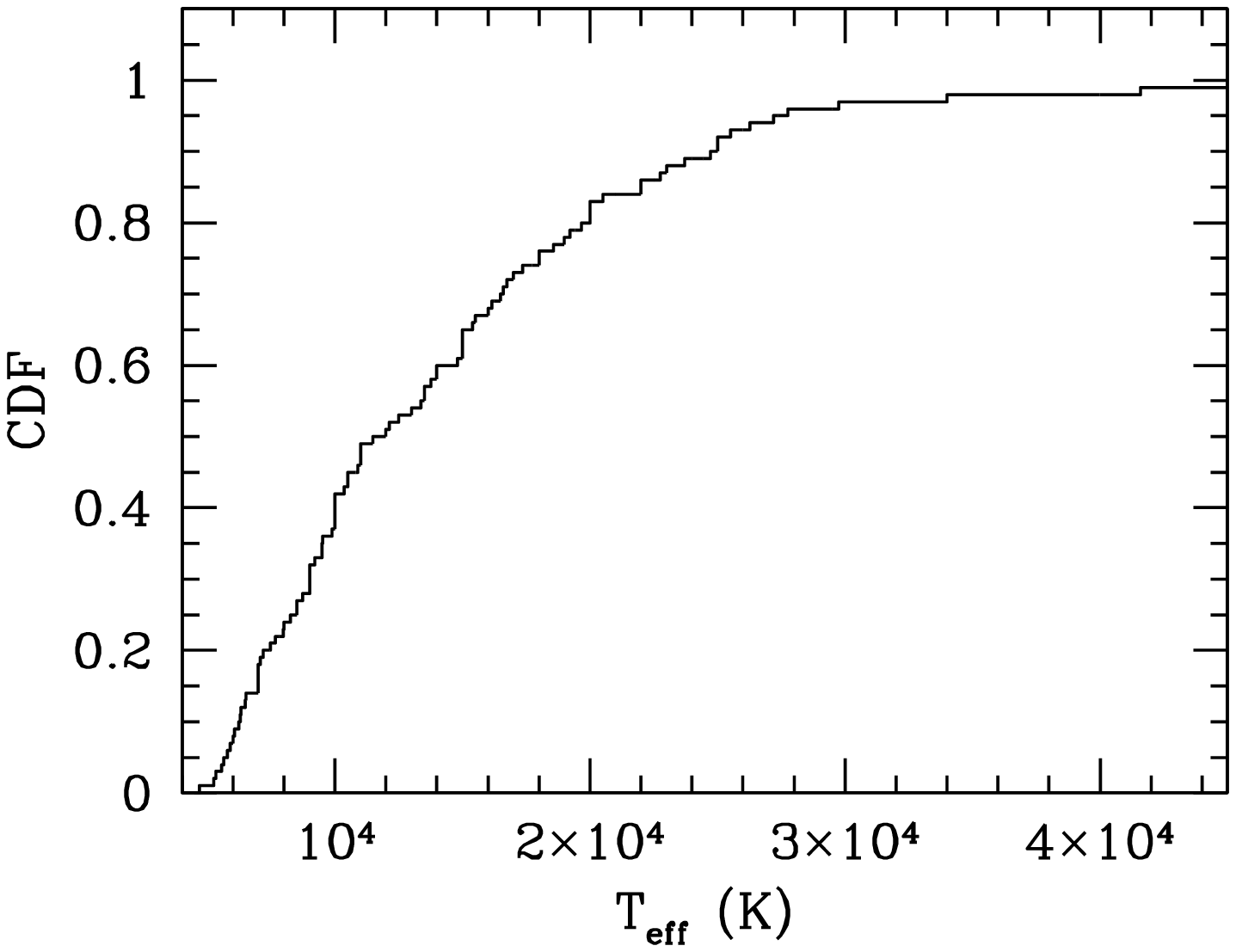}
\caption{Cumulative distribution function of MWD effective
  temperatures. The observed distribution is smooth implying that the
  birthrate of MWDs has not changed over time (this work)}
\label{CDF_MWDs}
\end{center}
\end{figure}

{Finally, we can state that the claim originally made
  by \citet{LiebertSion1979} and supported by
  \citet{FabrikaValyavin1999} that there is a higher incidence of
  magnetism among cool WDs than among hot WDs does not appear to be
  corroborated by the present enlarged sample of MWDs. We show in
  Fig.\,\ref{CDF_MWDs} the cumulative distribution function of the
  effective temperatures of the observed sample MWDs (see
  Table\,\ref{tab:mwds}).  We note that this function is smooth over
  the entire range of effective temperatures
  $T_{\rm eff}=4,000-45,000$\,K thus indicating that the birthrate of
  MWDs has not significantly changed over the age of the Galactic
  disk.}

\subsection{Field determination of white dwarfs in binary systems}
\label{ss:fieldmeasuresMCV}

Direct measurements of the WD magnetic field strength in the high field
magnetic CVs, the polars, can be obtained either (i) through Zeeman
splitting of the photospheric hydrogen absorptions lines when these
systems enter low accretion states (as described in
Sect.\,\ref{ss:fieldmeasuresWD}) or (ii) through the modelling of cyclotron
emission features that characterise the optical to IR spectra during
intermediate and high accretion states \citep[see][]{wickramasinghe00}
{or (iii) via the study of Zeeman features arising from the
  halo of matter surrounding the accretion shock.}

\begin{figure}[t]
\includegraphics[width=\columnwidth]{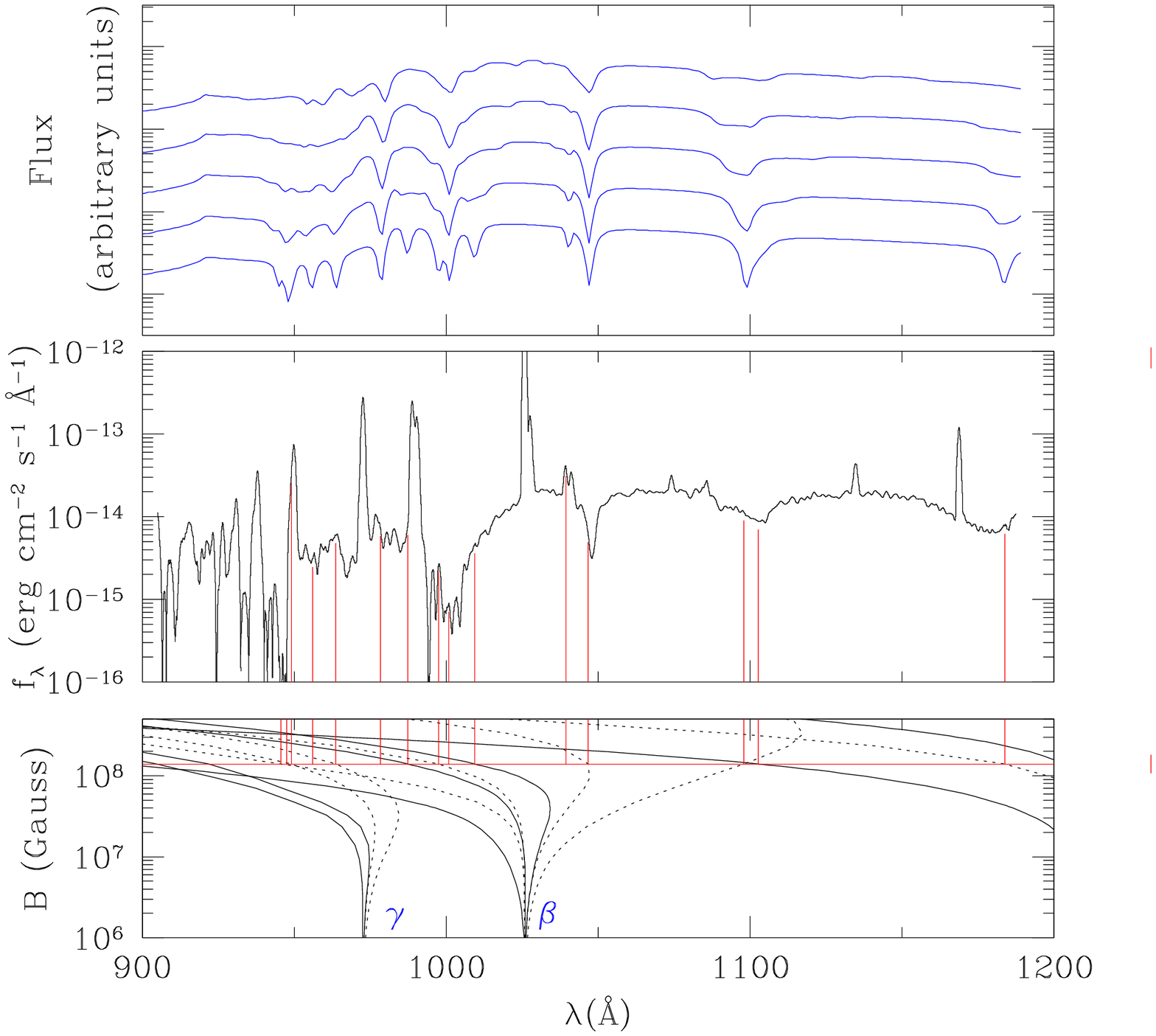}
\caption{FUSE spectrum of the high field MCV AR\,UMa (middle panel)
  compared with hydrogen Lyman transitions for 1MG\,$\le
  B\le$\,500\,MG (bottom panel) and with photospheric model spectra
  (top panel). The model correspond (from top to bottom) to dipole
  field strengths and fractional offsets of 200\,MG, 0.0; 215\,MG,
  0.10; 235\,MG, 0.15; 260\,MG, 0.20; and 280\,MG, 0.25. Negative
  offsets imply that we view the weaker field hemisphere, where the
  field distribution is more uniform. Dashed lines represent normally
  forbidden components that are enabled by the strong electric fields
  present in highly magnetic WDs
  \citep{Hoard04}} 
\label{Fig4:aruma_fuse}
\end{figure}

When accreting at low rates, polars reveal the WD photosphere and thus
can allow the detection of the Zeeman $\sigma_{+}$, $\sigma_{-}$ and
$\pi$ components of Balmer line absorptions. Thus some polars have
their field determined through Zeeman spectroscopy of photospheric
lines in the optical wavelength range
\citep{Wickramasinghe_Martin85,Ferrario92,Schwope95a}.  The highest
magnetised polars, AR\,UMa \citep[230\,MG,][]{Schmidt96} and AP\,CrB
\citep[144\,MG,][]{Gaensicke04}, were instead detected through Zeeman
split absorption features in the UV range. We show in
Fig.\,\ref{Fig4:aruma_fuse} an ultraviolet spectrum of AR\,UMa
covering the range 917--1,182\,\AA\, when the system was in its
typical low-accretion state.  The spectral absorption features are
caused by Ly$\alpha$--Ly$\gamma$ Zeeman transitions. The modelling
indicates a dipolar field strength of about 235\,MG offset along its
axis by a 0.15 of the stellar radius \citep{Hoard04}.

\begin{figure}[t]
\includegraphics[width=\columnwidth]{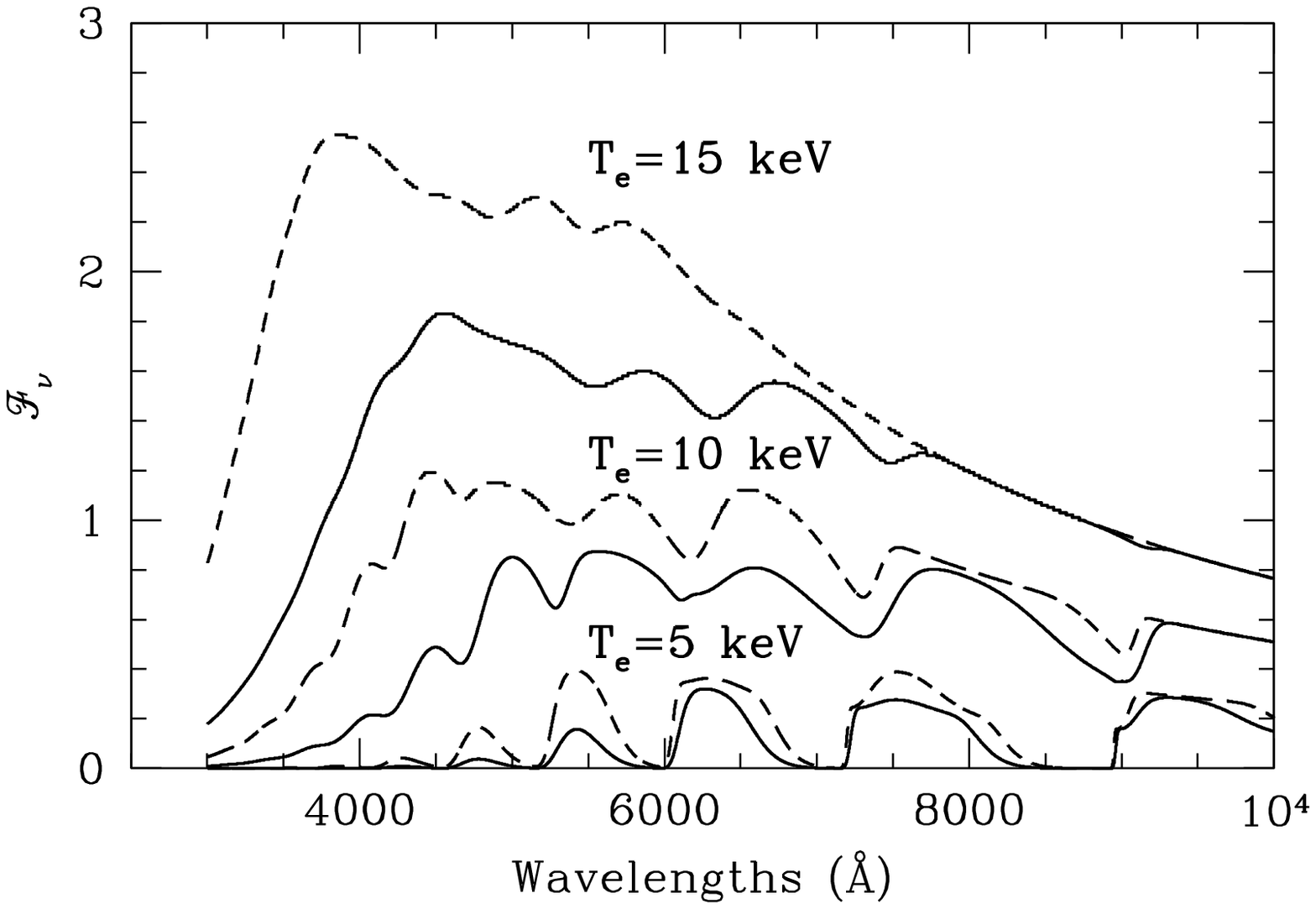}
\caption{Theoretical cyclotron spectra for a field $B=30$~MG as a
  function of electron temperature $T_e$ for a viewing angle $\theta
  =90^\circ$. The solid and dashed curves are for optical depth
  parameters $\Lambda=2\times 10^5$ and $\Lambda=10^6$ respectively
  \citep[from][]{wickramasinghe00}} 
\label{Fig:harmonics}
\end{figure}

The optical to IR spectra of polars may also reveal the typical
undulation of cyclotron humps.   At low temperatures, the
position of the $n^{\rm th}$ harmonic for a magnetic field $B$ and viewing
angle near $90^\circ$ can be estimated using
\[
\lambda_{\rm n}={10\,710\over n}\left({10^8 {\rm G}\over {B}
  }\right)\,{\rm \AA}.
\]
Cyclotron lines in polars are only seen when the shocks are viewed at
large angles $\theta$ to the field direction.  The intensity is at its
maximum at the fundamental and rapidly decreases as $n$ increases, with
the rate of decline depending on temperature.

We show in Fig.\,\ref{Fig:harmonics} a number of theoretical cyclotron
emission spectra. {These have been obtained using the
  \citet{WickMeggitt1985} constant $\Lambda$ (or ``point source'')
  models which assume uniform conditions in the shock but allow for
  optical depth effects. The parameters of these models have been
  chosen to yield the characteristic cyclotron undulation in the
  optical band}.

At long wavelengths these spectra display an optically thick
Rayleigh-Jeans tail while at shorter wavelengths they are
characterised by a power law spectrum modulated by cyclotron
lines. The flat-topped profiles of the harmonic peaks at low harmonic
numbers implies optically thick emission. As the harmonic number
increases the opacity drops and the shock becomes optically thin so
that the harmonic structure becomes clearly visible. The switch from
optically thin to thick emission is a strong function of the optical
depth parameter $\Lambda$
\[
\Lambda= 2.01\times10^6 \left({s\over 10^6{\rm
      cm}}\right)\left({N_{\rm e}\over 10^{16}{\rm cm}^{-3}}\right)\left({3\times 10^7 {\rm G}\over B}\right) 
\]
where $s$ is a characteristic path length through the post-shock
region and $N_e$ is the electron density number. The parameter
$\Lambda$ is approximately equal to the optical depth at the cyclotron
fundamental at a viewing angle $\theta=90^\circ$ to the field
direction. Cyclotron emission in MCVs has been observed at infrared
wavelengths
\citep[eg][]{Bailey91,Ferrario93,Ferrario96,Campbell2008a,Campbell2008b},
optical
\citep[eg][]{Visvanathan1979,Wickramasinghe89,Ferrario94,Schwope90a,Schwope95c},
and in a few systems with the highest field strengths also at
ultraviolet wavelengths \citep[eg][]{rosenetal01-1,Gaensicke01,
  ferrarioetal03-1}.

{The values of $\Lambda$ inferred from the modelling of
  cyclotron emission ($\sim10^5-10^6$) are much lower than those
  expected for a bremsstrahlung-dominated shock
  \citep[$\sim 10^8-10^9$, e.g.,][]{lamb_masters79,Chanmugam_Dulk81},
  thus indicating that the cyclotron radiation mainly comes from
  strongly cyclotron cooled shock regions characterised by low
  specific accretion rates.}

{The lack of harmonic features seen in the intensity
  spectra of most polars also supports the hypothesis that
  the emission regions of these systems are extended and
  structured. This flat energy distribution has been attributed to
  magnetic field spread and to the optical depth parameter varying
  across a wide region characterised by different specific
  mass flow rates. More realistic models that take into consideration
  the effects of extension and temperature and density distribution
  across the emission region were constructed by, e.g.,
  \citet{WickFerrario88, FerrarioWick1990,Rousseau1996}.  }

Through the careful fitting of cyclotron harmonic features it is
possible to determine the magnetic field strength and physical
parameters of the post-shock region \citep[see][and references
therein]{wickramasinghe00}. Many polars have measured field strengths
through time-resolved cyclotron
spectroscopy. {Information on the accretion geometry
  can also be gained through the study of harmonics that shift with
  orbital phase due to different parts of the
  accretion region becoming visible as the WD rotates
  \citep[e.g.][]{SchwopeBeuermann1990,Burwitz96}. The cyclotron study
  of the phase resolved spectra of eclipsing systems such as the
  bright polar HU\,Aquarii by \citet{Schwope2003} has proven to be
  particularly important to establish the accretion geometry of these
  objects.}

\begin{figure}[t]
\includegraphics[width=\columnwidth]{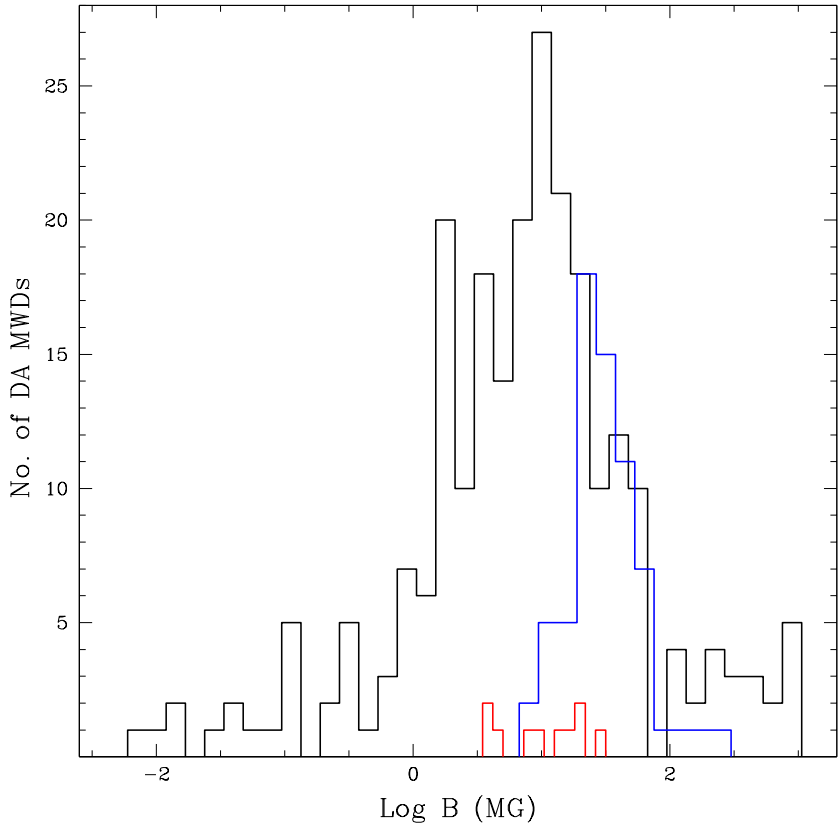}
\caption{Distributions of magnetic field strength in polars (Blue
  line) and IPs (red line) compared to that of single magnetic WDs
  (black line). This figure has been prepared using the data 
  in Tables\,\ref{tab:mwds}, \ref{tab:mcvs}, and \ref{tab:ips} (this
  work)}
\label{fig:b_field}
\end{figure}

\begin{figure}[t]
\includegraphics[width=\columnwidth]{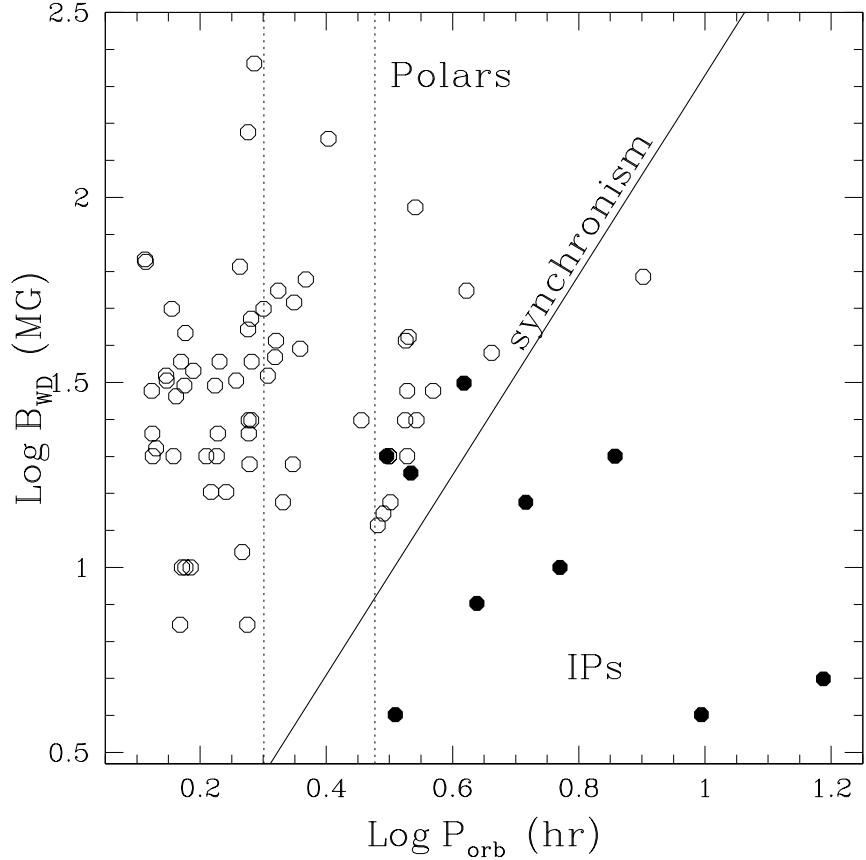}
\caption{The magnetic field strength and orbital period diagram for
  polars (empty circles) and IPs (filled circles). The line at which
  synchronism is expected for a mean mass accretion rate as a function
  of $P_{\rm orb}$ is reported as a solid line \citep[adapted
  from][]{beuermann99}. The dotted vertical lines mark the orbital
  period gap } 
\label{fig:porb_b}
\end{figure}

In many cases two independent sets of cyclotron lines arising from
regions with different magnetic field strengths have been found, with
the main accreting pole possessing a weaker field
\citep{Ferrario93,Schwope95a,Schwope96,Ferrario96,Campbell2008a}.
These studies indicate that the magnetic field distribution is not
that of a centred dipole and offsets as high as 10-20\% are often
inferred \citep{wickramasinghe00}.  Evidence of non-centred dipole
field distribution also comes from Zeeman components which are seen
against strong cyclotron emission and are formed in the free fall
material surrounding the WD pole, often named ``halo"
\citep{Achilleos1992a, Schwope95a}. The study of all these different
components have shown that the field strength obtained from the
modelling of photospheric Zeeman lines, $B_{\rm Zeem, phot}$, is
different from the field strength obtained from the study of halo
Zeeman features, $B_{\rm Zeem, halo}$, with the latter comparable to
the field strength $B_{\rm cyc}$ obtained through the modelling of
cyclotron humps. This is because the field strength measured from
photospheric Zeeman split lines is averaged over the entire visible
hemisphere of the WD while the fields derived from cyclotron modelling
or from the study of halo Zeeman features arise from regions close to
the visible accreting pole.

From time--resolved polarimetry \citep[e.g.][]{Potter04} and
spectropolarimetry \citep[e.g.][]{Beuermann07} detailed information on
the complexity (quadrupole or even multipoles) of magnetic field
topology in these systems can be obtained (see also
Sect.\,\ref{s:topology}).  However, in systems with ages $\gtrsim
1$\,Gyr a substantial decay of multipole components could be expected
and thus short period MCVs may not have complex fields
\citep{Beuermann07}.  Magnetic field strengths have been measured
{ or estimated for $\sim$\,86 WDs in binaries (see
  Tables \ref{tab:mcvs} and \ref{tab:ips} for a complete list of known
  systems as of December 2014).} Using the main pole magnetic field
strength, Fig.\,\ref{fig:b_field} depicts the magnetic field
distribution of polars compared to that of single MWDs listed in Table
\ref{tab:mwds} with the latter having fields in the range $0.1-
1,000$\,MG.  The polars clearly populate a more restricted range of
field strengths, $7-230$\,MG, with a mean value of 38\,MG.

{Fields strengths above $230$\,MG, which are detected in
  single magnetic WDs, are not found in polars and there is no clear
  explanation for this yet.  High magnetic field polars could be
  difficult to identify due to selection effects because these systems
  would be highly intermittent soft X-ray sources such as
  AR\,UMa. \citet{Hameury89} explained the paucity of very high field
  polars in terms of their very short lifetimes due to efficient loss
  of angular momentum via magnetic braking mechanism.} However, it
appears more likely that the strong fields in polars may
\emph{decrease} the efficiency of magnetic braking, which would result
in a slower evolution of their orbital periods and in lower
  accretion luminosities \citep{li+wickramasinghe98-1,
  webbink+wickramasinghe02-1, Araujo05}. On the other hand, if the
magnetic field is generated during the CE phase, then the highest
fields could only be produced when the two stars merge to give rise to
an isolated MWD (see Sect.\,\ref{s:ORIGIN} and the chapter on the
origin of magnetic fields in this book).

The lowest surface averaged magnetic field strength measured in a
polar is 7\,MG in V2301\,Oph which was modelled by \citet{Ferrario95}
with a dipolar field of $12$\,MG offset by 20\% from the centre of the
WD. The lack of lower field synchronous systems could be explained if
the asynchronous IPs represent the low field tail of the magnetic
field strength distribution in MCVs.  However this is difficult to
prove, because the absence of low accretion states in IPs prevents the
WD photosphere to become visible, thus precluding the detection of
photospheric Zeeman split lines. Most of these systems do not show
polarised optical/IR emission or cyclotron features, which also
prevents the determination of the WD magnetic field.  Whether the lack
of polarisation in the optical/IR is caused by weaker magnetic fields
or is due to efficient depolarisation mechanisms is difficult to
ascertain.  So far only ten IPs are known to be circularly polarised
at a level $\lesssim 1-3\%$ \citep{penning86, piirolaetal93,
  buckleyetal95, potter97, katajainen07, piirolaetal08, butters09,
  potter12}. In these IPs, the field strengths are estimated to be in
the range $\sim 5-20$\,MG, with V405\,Aur \citep{piirolaetal08}
possessing the highest ($\sim$30\,MG) field {(see Table
  \ref{tab:ips}). }  Seven are also found to show a soft X-ray
blackbody component.  The fields of these IPs, shown in
Fig.\,\ref{fig:b_field}, are at the low-field end of the distribution
and partially overlap with the low field polars.  Their orbital
periods are all above the CV period gap (see Fig.\,\ref{fig:porb_b})
and, given the large uncertainties in the estimates, they are below or
close to the line at which synchronism is expected to occur \citep[see
Fig.\,\ref{fig:porb_b}][]{beuermann99}.  These systems could in fact
be the progenitors of the low-field polars, as suggested by
\citet{norton04}.  As the number of polarised IPs has increased by a
factor of three in the last few years, further deep polarimetric and
spectropolarimetric surveys of all known IPs are crucial to establish
their field strengths and accretion properties.

\begin{figure}[t]
\begin{center}
\includegraphics[width=\columnwidth]{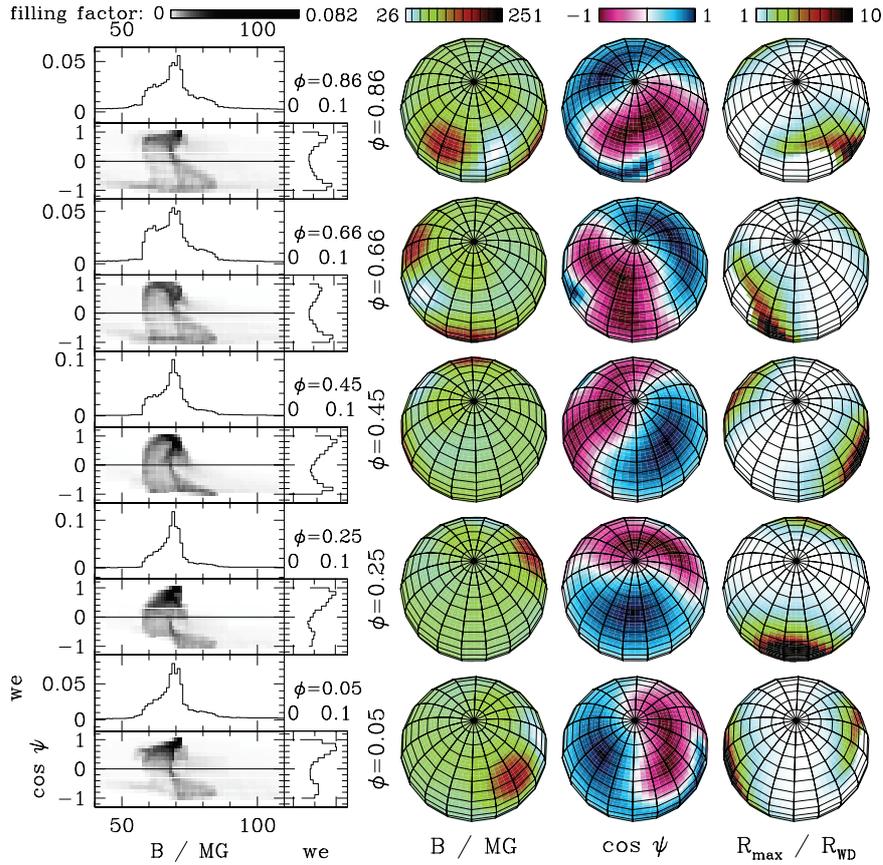} 
\caption{Field topology of the MWD PG\,1015+014, derived from a
  tomographic analysis of time-resolved spectropolarimetry
  \citep{euchneretal05-1}. Fields in the range 50--80\,MG are detected
  (left panel) with a highly non-dipolar configuration (middle panels:
  surface field and angle between the line-of-sight and the magnetic
  field direction). The  maximum radial distance reached by field
  lines in units of the WD radius is shown in the right-most
  panel}
\label{f:pg1015_tomo}
\end{center}
\end{figure}

\subsection{Field topology\label{s:topology}}

When discussing the magnetic field strength in isolated MWDs, some
attention has to be given to the definition of $B$, as the field
strength measured from observations is usually an average value over
the visible hemisphere of the WD. In the absence of any other
information, it is common practice to assume a dipolar field
configuration, in which case the field varies by a factor two between
the magnetic pole and the equator. When averaging over the visible
hemisphere, the relative weighting of regions with different fields
strengths depends on the angle between the observer's line of sight
and the magnetic field axis \citep[see. Fig.\,2
of][]{Achilleos1992b}.

However, detailed spectroscopy and spectropolarimetry have demonstrated
in a number of cases that the field topologies can be very
complex \citep[e.g.][]{putney+jordan95-1}. Several sophisticated
tomographic reconstruction methods have been developed to map the
field topology \citep{donatietal94-1, euchneretal02-1}. The
application of these methods to both single MWDs
\citep{euchneretal05-1, euchneretal06-1} and MWDs in polars
\citep{Beuermann07} reveals a startling complexity of the field
topologies (Fig.\,\ref{f:pg1015_tomo}).

\subsection{Beyond hydrogen}

The bulk of all WDs have hydrogen-dominated atmospheres, both
in magnitude-limited \citep{kleinmanetal13-1} and volume-limited
samples \citep{giammicheleetal12-1}, and the same is true for MWDs.

\begin{figure}[t]
\begin{center}
\includegraphics[width=\columnwidth]{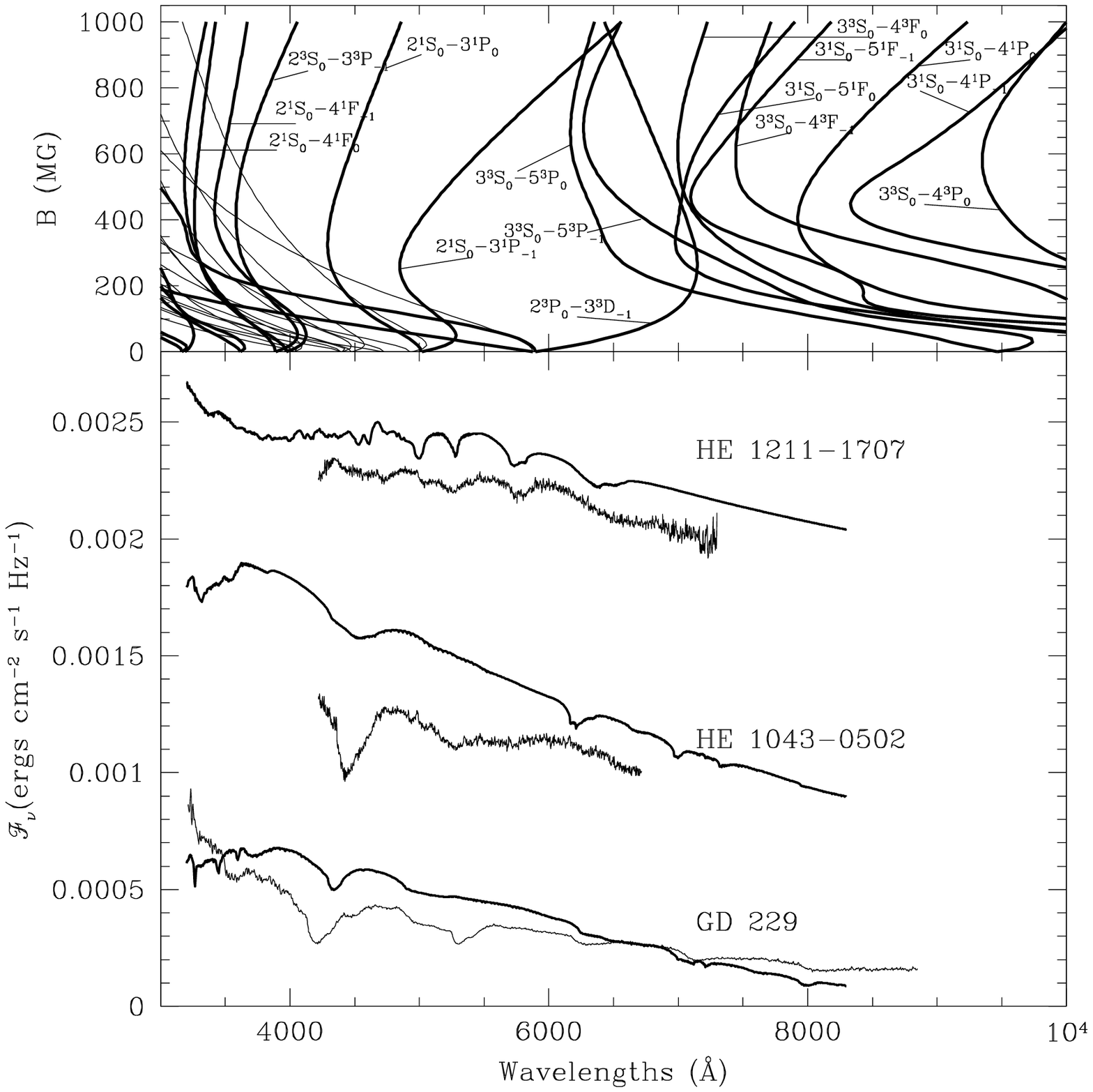}
\caption{A comparison of centred dipole models for helium rich WDs
  with observations of GD229, HE\,1043-0502, and HE\,1211-1707
  \citep{Wickramasinghe2002}. The models have polar fields $B_d=
  520$~MG (GD 229), 820~MG (HE~1043-0502) and 50~MG
  (HE$1211-1707$,). The observation and theory mismatch in GD229 could
  be due to resonances in the HeI bound-free opacities for which there
  is at present no adequate theory}
\label{Fig3:helium_fits}
\end{center}
\end{figure}

At temperatures above $\sim10,000$\,K, MWDs can also exhibit
Zeeman split HeI lines in their spectra (DBp WDs).  The two electron
problem is much more difficult to treat and calculations for $n\le 5$
singlet and triplet states for $m=0,\pm 1,\pm 2,\pm 3$ only became
available in the late 90s \citep{Jordan1998, jonesetal99-1, Becken99,
  Becken00a, Becken00b, Becken01, AlHujaj03}.  We show in
Fig.\,\ref{Fig3:helium_fits} a comparison between centred dipole
models and observations for three helium-rich MWDs.

\begin{figure}[t]
\includegraphics[width=\columnwidth]{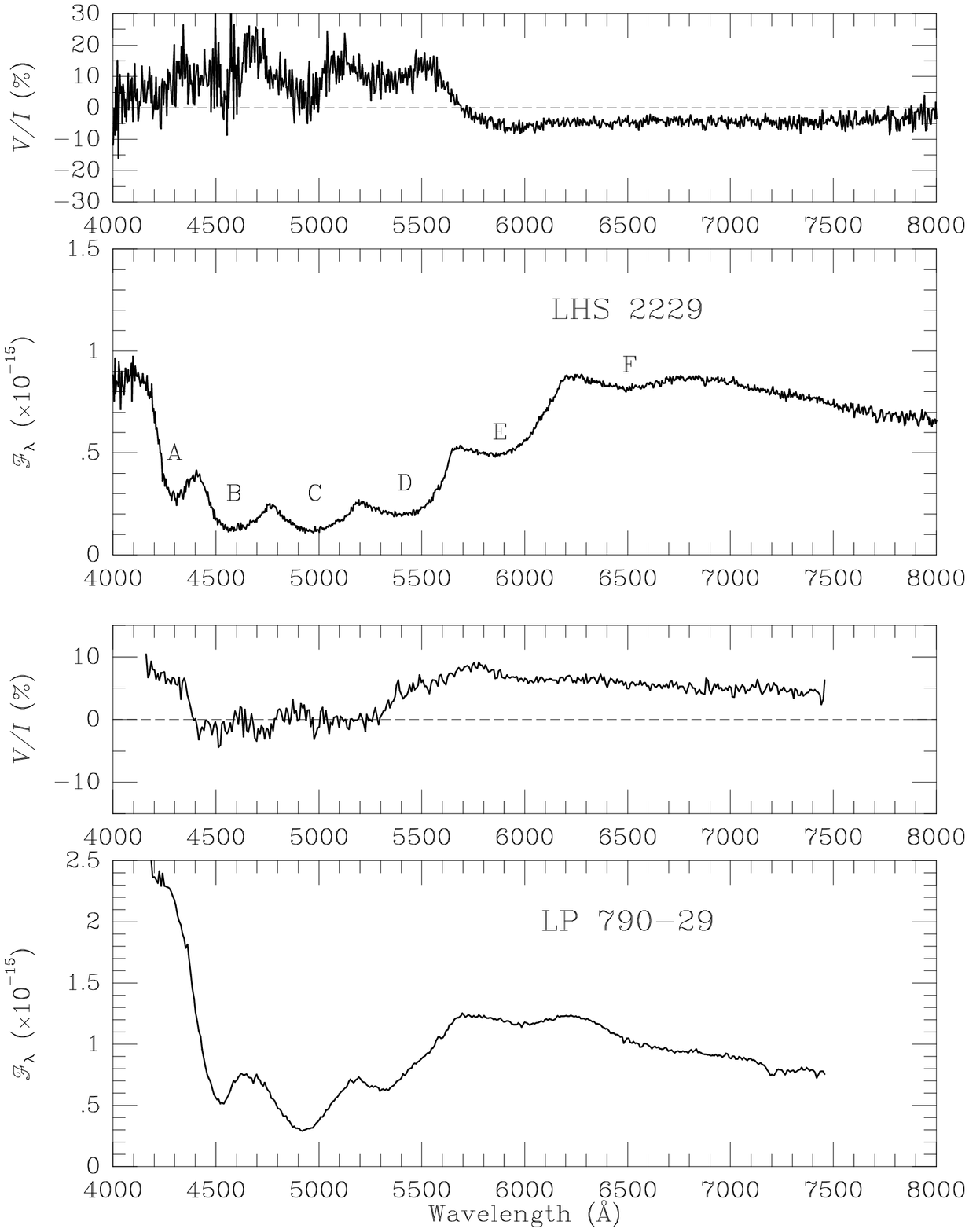}
\caption{Flux and polarization spectra of two confirmed magnetic DQp
  stars, with crude $B$-field estimates of $\sim100$\,MG \citep{schmidtetal99-1}}
\label{f:dqp}
\end{figure}

\begin{figure}[t]
\includegraphics[width=\columnwidth]{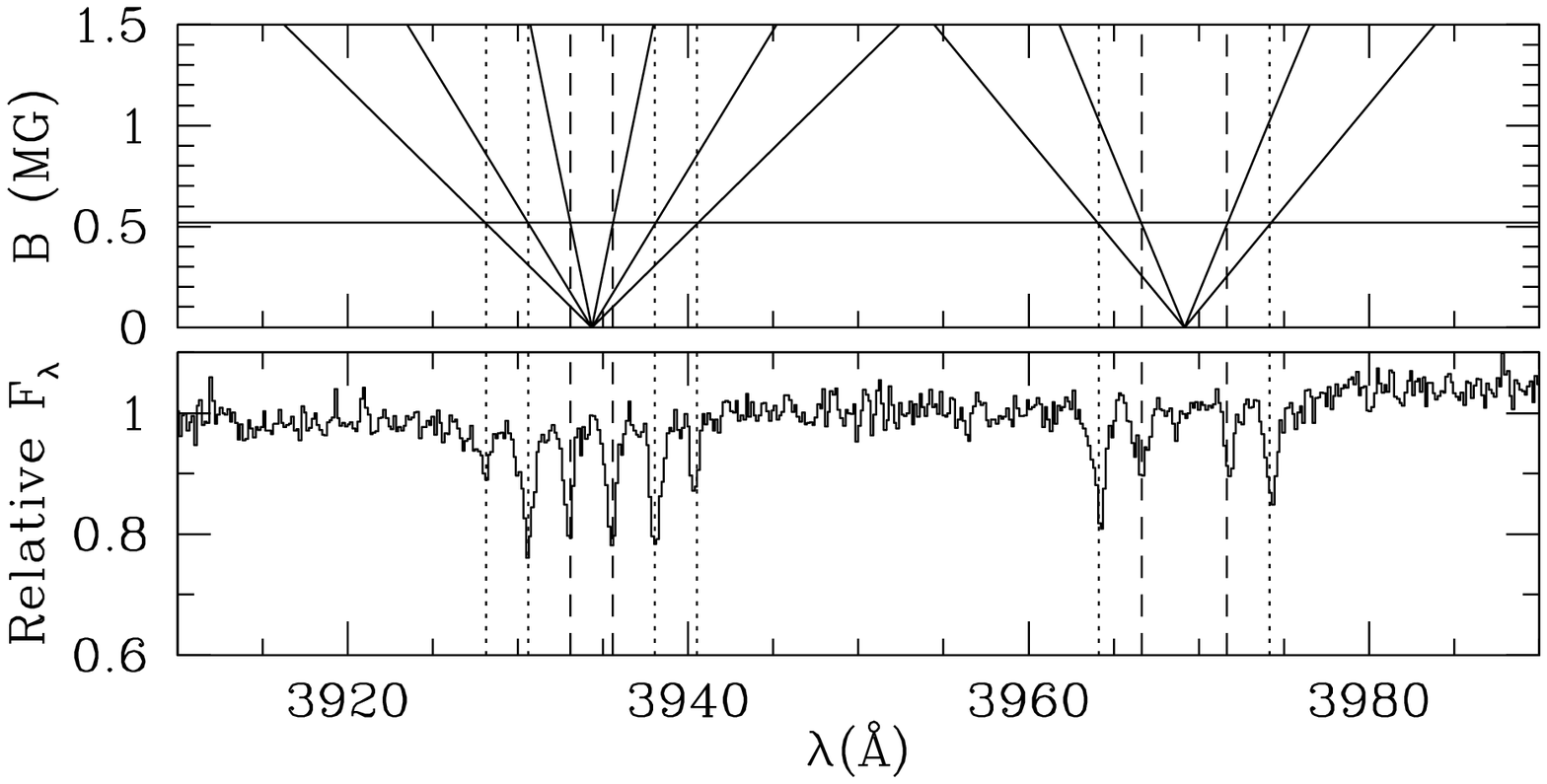}
\caption{The Ca H/K absorption lines in the cool ($\simeq 5,000$\,K)
  hydrogen-dominated atmosphere Zeeman-split in a field of $B=0.5$\,MG
  \citep{Kawka2011}}
\label{f:mdaz}
\end{figure}
Cool helium-dominated atmospheres develop deep convection zones that
may dredge-up core material into the photosphere, resulting in atomic
or molecular carbon features. The maximum carbon contamination in
these DQ WDs is expected around $\simeq12\,000$\,K. A number of very
cool DQs show broad absorption troughs reminiscent of C$_2$ Swan
bands, without however matching the laboratory wavelengths of the Swan
bands. Spectropolarimetry has revealed magnetism in some of these DQp
stars \citep[see Fig.\,\ref{f:dqp};][]{liebertetal78-2,
  schmidtetal99-1}, with suggested (but rather uncertain) field
strengths of $\sim100$\,MG. However, not all DQp WDs show
polarisation, and the nature of the peculiar absorption bands is not
fully settled -- suggestions are absorption by C$_2$H
\citep{schmidtetal95-2} or pressure-shifted Swan bands
\citep{lieberetal83-1,Hall2008, kowalski10-1}.  A number of cool WDs
exhibit polarisation in the absorption band of CH \citep{Angel1974a,
  Vornanen2010}. \citet{berdyuginaetal07-1} applied new calculations
of molecular magnetic dichroism to observations of G\,99-37, a cool
helium-rich MWD showing strongly polarised molecular bands, and
estimated a field of $7.5\pm0.5$\,MG.

Recently, \cite{dufouretal07-1} discovered a new class of WDs with
carbon-dominated atmospheres, that are substantially hotter than the
``classic'' cool DQ stars. Short-periodic photometric variability
detected in several of the hot DQs \citep{montgomeryetal08-1,
 Dufour2008, dunlapetal10-1} was initially interpreted as
non-radial pulsations, but the discovery of a 2.11\,d modulation in
SDSS\,J000555.90-−100213.5 \citep{Lawrie2013} casts some doubt on
this hypothesis. The additional discovery of magnetic fields among the
hot DQs \citep{DufourHotDQ2008,Dufour2008, williamsetal13-1}
suggests that they may represent a peculiar and rare path in WD
evolution. Currently, 5 out of 14 known hot DQs are magnetic
\citep{Vornanen2013}.

Because of their high surface gravity, WDs should not have any
photospheric elements apart from hydrogen, helium, and dredge-up of
carbon or, much more rarely, oxygen \citep{gaensickeetal10-1}. Yet,
$\sim25$\% of all WDs show traces of metals -- most commonly Ca and
Si, but also Mg, Fe, Na and other elements \citep{zuckermanetal03-1,
  koesteretal14-1}. This photospheric pollution requires recent or
ongoing accretion and the widely accepted hypothesis for the origin of
the material is planetary debris \citep{jura03-1}. This hypothesis is
corroborated by the detection of close-in circumstellar dust and gas
\citep{gaensickeetal06-3, farihietal09-1}. A very small fraction of
these debris-polluted WDs show magnetic fields \citep[][see
Fig.\,\ref{f:mdaz}]{Reid2001,Kawka2011, Farihi2011}. It is interesting
to note that all known metal-polluted MWDs are very cool
$(T_\mathrm{eff}<7,000$\,K, \citealt{Kawka2014}), including the sample
of strongly metal-polluted and highly magnetic DZ WDs recently
discovered by \citet{Hollands2015}.

\section{Mass distribution of isolated magnetic white dwarfs}\label{mass_mwds}

\begin{figure}[t] 
\includegraphics[width=\columnwidth]{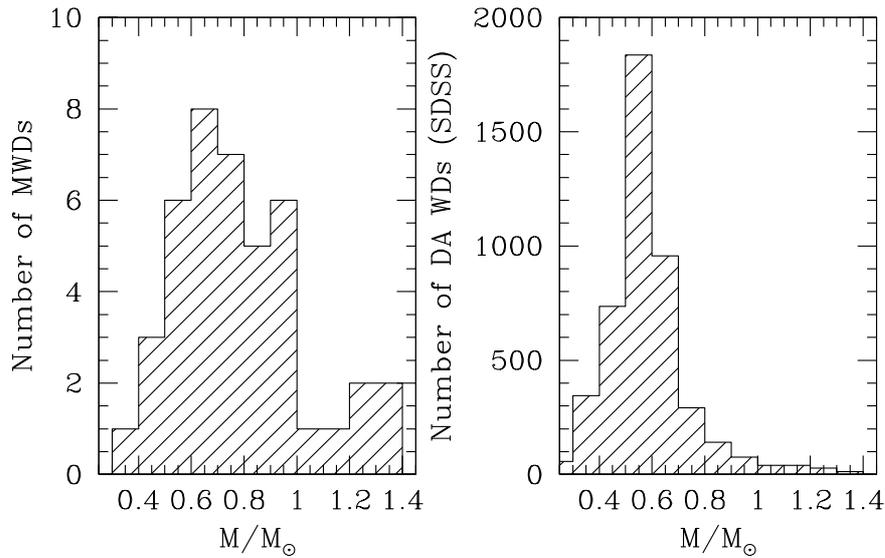} 
\caption{Left panel: The mass distribution of MWDs
  \citep{Ferrario2010}. Right panel: The mass distribution of
  non-magnetic DA WDs from SDSS \citep{Kepler2007}.}
\label{mass_distribution} 
\end{figure}

\begin{figure}[t]
\begin{center}
\includegraphics[angle=-90,width=\columnwidth]{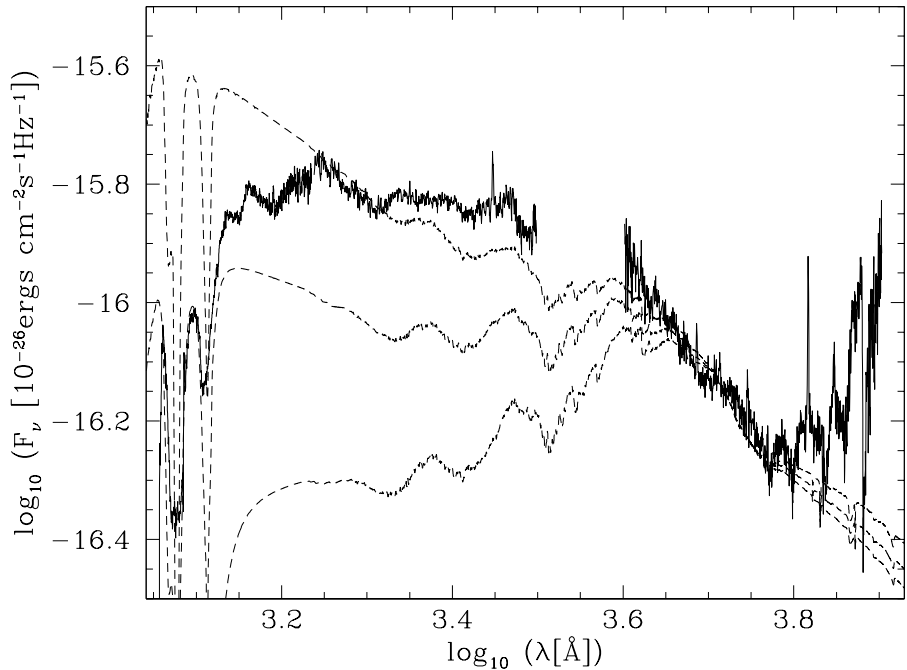} 
\caption{The combined ultraviolet/optical spectral energy distribution
  of the high-field (230\,MG) MCV AR\,UMa (solid line), along with
  model spectra for $T_\mathrm{eff}=15,000, 20,000, 25,000$\,K
  (dashed lines, top to bottom). Magnetic effects cannot be fully
  described by current atmosphere models, and consequently the
  effective temperature of high-field MWDs remains poorly constrained
  even in the presence of excellent data \citep{Gaensicke01}}
\label{f:aruma_sed}
\end{center}
\end{figure}

Mass estimates of isolated MWDs with fields $\gtrsim$ 1\,MG are
available only for a small number of objects. The mass determination
is not straightforward since there is no Stark broadening theory for
high field MWDs. Thus, even state-of-the art models fail to fully
account for the magnetic effects in MWD atmospheres. As a consequence,
the effective temperatures derived often remain inherently uncertain
(Fig.\,\ref{f:aruma_sed}), and the same is true for the implied MWD
masses \citep[e.g.][]{kuelebietal10-1}.  The standard procedure of
fitting the Balmer lines for $T_{\rm eff}$ and $\log g$ can only be
applied to MWDs with fields below a few MG and even in these cases the
results have to be treated with some caution
\citep[e.g.][]{Ferrario1998,dupuisetal03-1}.  For higher field
strengths, mass estimates are derived from the combination of
effective temperatures, parallaxes, and a mass-radius relation. While
only a small number of MWDs have accurate parallaxes, an estimate of
the distance can be determined for MWDs that have non-degenerate WD
companions \citep[e.g.]{Girven2010,Dobbie2012,Dobbie2013}, or for MWDs
in open clusters \citep{kuelebietal13-1}. The situation will
dramatically improve in the next few years, when parallaxes for
practically all known MWDs will become available from the ESA
satellite Gaia.

Taking into account the caveats mentioned above, the mean mass of high
field isolated MWDs ($B\gtrsim1$\,MG) is $0.784
\pm0.047$\,M$_\odot$. High field MWDs also exhibit a strong tail that
extends to the Chandrasekhar limit. The most recent estimate for the
mean mass of non-magnetic DA WDs is $0.663\pm0.136$\,M$_\odot$
\citep{tremblayetal13-1}.  That the mean mass of MWDs is higher than
that of their non-magnetic counterparts was first noted by
\citet{Liebert1988}. The mass distribution of all magnetic and
non-magnetic WDs is shown in Fig.\,\ref{mass_distribution}.

\section{Spin periods of isolated magnetic white dwarfs}

The majority of non-magnetic WDs are slow rotators, with even
high-resolution spectroscopy usually only providing lower limits on
$v\sin i$ \citep{karletal05-1, bergeretal05-2}. Asteroseismology shows
that the spin periods are typically a few days
\citep{fontaine+brassard08-1, greissetal14-1} and that the angular
momentum of the stellar core is lost before the WD stage
\citep{charpinetetal09-1}.

In MWDs, the magnetic effects in their atmospheres can give rise to
noticeable spectroscopic \citep{Liebert1977, latteretal87-1},
photometric \citet{Brinkworth2004, Brinkworth2005}, and
polarimetric variability \citep{SchmidtNorsworty1991,
  piirola+reiz92-1}.

The observed rotation periods of MWDs span a wide range, from 725\,s
\citep[RE\,J0317--853][]{Barstow1995, Ferrario1997} to lower limits of
decades, if not centuries \citep{berdyugin+piirola99-1,
  beuermann+reinsch02-1, JordanFried2002} (see Table \ref{tab:mwds}). 

\begin{figure}
\begin{center}
\includegraphics[width=\columnwidth]{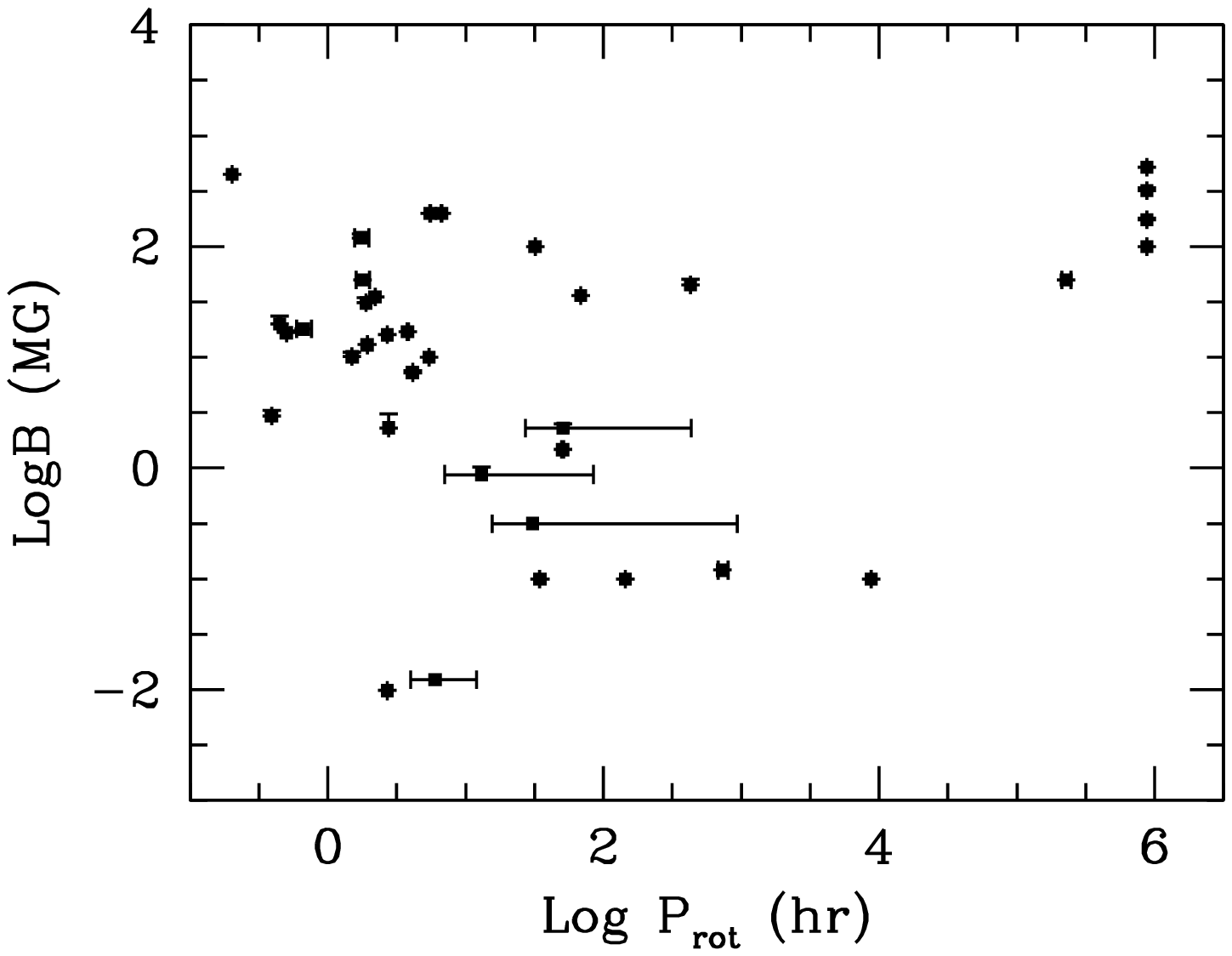}
\caption{Rotation periods of isolated MWDs against their magnetic field
  strength (this work)}
\label{rotation_distribution}
\end{center}
\end{figure}

We show in Fig.\,\ref{rotation_distribution} (this work) the MWDs with
known magnetic fields and rotation (spin) periods, the latter
determined from polarimetry and photometry
\citep[e.g.][]{Brinkworth07,Brinkworth2013}. Inspection of the
magnetic field versus spin period distribution {may
  suggest the existence of two groups. One of them characterised by
  ``short'' periods of hours to weeks, and the other by rotation
  periods of decades to centuries as estimated for objects that have
  been monitored over many decades \citep[e.g., Grw$+70^\circ 8247$,
  discovered over 80 years ago;][]{Kuiper1934,Brinkworth2013}.  It
  seems that very long period MWDs tend to possess high fields while
  short period ones do not show any preferred field strength. Clearly
  further observations to measure rotation periods of MWDs are needed
  to ascertain the existence of two groups of MWDs with different
  rotational properties.}

The rotation rate of MWDs holds potentially some crucial clues on
their nature and origin. For instance, slow rotators could be the
descendants of the magnetic main sequence Ap/Bp stars (and their
fields would thus be of fossil origin), whereas fast rotators could be
the products of binary interaction, though \citet{kuelebietal13-2}
have argued that in the case of a merger magnetospheric interactions
of the MWD with the debris disk may slow down the rotation rather
quickly. Unfortunately, the statistics of rotation periods have only
slowly improved, however \citet{Brinkworth2013} have demonstrated that the
search for photometric variability is relatively cheap in terms of
observational requirements, and could extensively be used for future
work. An interesting hypothesis is that rapidly rotating MWDs may be
detectable as sources of gravitational wave radiation
\citep{heyl00-1}.

For the sake of completeness, we recall that the spin period of the
MWD in MCVs is dictated by the interaction between the magnetic field
of the MWD and that of the secondary and/or by the torque of the
accretion flow. That is, for strong fields the rotation of the MWD is
locked to the orbital period (polars), and for weaker fields the MWD
is rotating faster than the orbital period (IPs).

\section{Origin of magnetic fields in white dwarfs and magnetic cataclysmic variables}
\label{s:ORIGIN}

WDs are often found paired with Main-Sequence (MS) companion stars
\citep[generally M dwarfs, but see][]{Ferrario12}.  A glaring anomaly
is that there are no examples of fully detached MWD-MS pairs, as first
noted by \citet{Liebert05} through the study of 501 objects with
composite WD+MS spectra from the Sloan Digital Sky Survey (SDSS)
DR3. Even the most recent work of \citet{Rebassa13}, which has yielded
3,419 SDSS DR8 WD-MS binary candidates, does not contain objects
consisting of a MWD with a non-degenerate companion. Further searches
conducted through visual inspection of all SDSS spectra of WDs with a
red excess have confirmed the hypothesis that magnetic field and
binarity (with M or K dwarfs) are independent at a $9\,\sigma$ level
\citep{Liebert2015}.  Such a pairing is also absent from catalogues of
high field MWDs \citep[][and this work]{Kawka2007, Kepler2013}.

Thus, although all magnitude-limited surveys of WDs have lead to the
discovery of at least $\sim 2$\% of strongly magnetised WDs ($B\gtrsim
1$\,MG), these objects are never found paired with a non-degenerate
companion star. Yet, about 20-25\% of CVs host a MWD, thus raising
some serious questions regarding the progenitors, and thus the origin,
of MCVs \citep[see][]{Liebert09}.

In the late 90s, surveys such as the HQS and the SDSS have revealed the existence of a small number of cool
MWDs which accrete matter from the wind of their low-mass MS
companions \citep[e.g.][]{Reimers99, Reimers00, Schwope2002,Schmidt05,
  Schmidt07, Vogel07, Schwope09, Vogel11, Parsons13}. The accretion
rate, which is about $10^{-13}-10^{-14}$M$_\odot$ yr$^{-1}$, is a few
orders of magnitude larger than that observed in detached non-magnetic
Post Common Envelope Binaries \citep[PCEBs, e.g.][and references
therein]{Parsons13}.  {Comprehensive studies of these
  systems have unveiled that the secondary is an active late-type main
  sequence star underfilling its Roche-lobe \citep[see][and references
  therein]{Schwope09}. Accretion onto the cool MWD primary, constant
  over years \citep{Schwarz01}, is consistent with what is expected
  from the wind emanating from the active companion
  \citep{Schwope2002} and captured by the MWD. The spectra of these
  systems exhibit strong cyclotron harmonics humps superimposed on the
  WD+M dwarf stellar continuum. Their very peculiar colour is the
  reason why the first such systems were uncovered in surveys whose
  science goal was to identify active galactic nuclei. The suggestion
  is that these systems could be the progenitors of the high field
  MCVs.  Thus their initial class name ``Low-Accretion Rate Polars''
  \citep[``LARPS'',][]{Schwope2002} is a misnomer, so they were
  renamed ``Pre-Polars'' \citep[``PREPS'',][]{Schwope09}. }
  
{There are now ten systems that have been classified as
  PREPS (see Table \ref{tab:mcvs}). The WD magnetic fields determined
  for these systems cluster in the 60-70\,MG range with only one
  system above \citep[108\,MG][]{Schwope09} and two below
  \citep[42\,MG and 8\,MG][]{Schmidt07,Parsons13}. This field
  clustering may be due to selection effects. In any case, the field
  strengths are certainly consistent with the hypothesis that these
  wind accreting magnetic systems are the progenitors of the high
  field MCVs. The low magnetic field of 8\,MG found in
  SDSS\,J030308.35+0054441.1 could instead suggest that this system
  may evolve into an IP \citep{Parsons13}.}

{There are however a few polars that undergo
  prolonged low-accretion states that cannot be reconciled with wind
  accretion and therefore these can be rightfully named LARPS (see
  e.g. \citealt{Schmidt2005, Breedt12}, and Table \ref{tab:mcvs}).}

The PREPS hypothesis is consistent with the scenario first proposed by
\citet{Tout08} and further developed by \citet{Wick14} for the origin
of fields in MCVs. They have raised the possibility that the strong
differential rotation expected in the CE phase may lead to the
generation, by the dynamo mechanism, of a magnetic field that becomes
frozen into the degenerate core of the pre-WD in MCVs \citep[see
also][about field generation in MWDs]{Nordhaus2011, Garcia2012, Kissin2015}. The
dynamo mechanism responsible for magnetic field generation during the
CE phase proposed by \citet{Wick14} is presented in the chapter of
this book on the origin of magnetic fields in stars.

{The binary population synthesis calculations of
  \citet{Briggs2015} are compatible with the hypothesis that MWDs
  originate from stars merging during common envelope evolution and
  are also consistent with the observation that MWDs are on average
  more massive than their non-magnetic counterparts (see
  Sect.\,\ref{mass_mwds}).}

{However, other formation channels for
  MWDs may be at work and, for instance, the fossil field hypothesis for the origin
  of high fields MWDs, as proposed by \citet{Wick05}, cannot be
  dismissed. However, the fact that there is no known MWD paired with
  a non-degenerate companion of M to K spectral type is a serious
  challenge to the fossil field hypothesis. Of course it is possible
  that some MWDs could be hidden in the glare of luminous companions
  \citep[e.g.][]{Ferrario2012}. Should large enough numbers of these
  'Sirius-type systems' hosting a MWD be discovered, this finding
  could point to different (or additional) formation channels for
  MWDs. However here we need to stress that Sirius-type systems could
  not be the progenitors of the MCVs.}

{According to the stellar merging hypothesis, the
  absence of relatively wide pre-MCVs can be explained if magnetic
  systems are born either in an already semi-detached state or with
  the two stars close enough for the MWD to capture the wind of its
  companion as it happens in PREPS. Any other alternative would
  trigger the question ``where are the progenitors of the
  PREPS?'' \citep{Schwope09}}. In the merging star picture, the highest fields are
expected to occur when the two stars merge to produce an isolated
MWD. The MCVs would then arise when the two stellar cores come close
enough to each other to generate a strong field in the WD but fail to
merge. Wider core separations after the end of the CE phase would
result in non magnetic pre-CVs or in systems where the two stars will
never interact. These systems, single stars and binaries that never
undergo a CE evolution would account for the populations of low-field
WDs in CVs, wide binaries and single isolated WDs. The few known wide
binaries consisting of a MWD paired with a non-magnetic WD would
result from triple star evolution where two stars merged to produce
the MWD and the third one evolved as a single star to become a
non-magnetic WD \citep[e.g. LB 11146, RXJ\,0317-853, EUVE\,J1439+750,
CBS\,229;][]{Glenn1994, Barstow1995, Ferrario1997, Vennes1999,
  Dobbie2013}.

One would expect that some PREPS should have hotter WDs. However, as
\citet{Tout08} have pointed out, WDs cool down to an effective
temperature of $15,000$\,K in only $\sim 10^7$ years and at a period
of 2\,h, the orbital decay time scale due to gravitational wave
radiation is about $3\times10^9$ years which is sufficient for a WD to
cool down considerably and reach the observed effective temperatures
of the PREPS before Roche lobe accretion begins.
{Therefore, although some $20-25$\% of CVs are
  magnetic, their birthrate may be considerably lower than that of
  non-magnetic CVs and their presence in large numbers could simply be
  a reflection of their longer lifespans because of the reduction in magnetic
  braking (see Sect\,\ref{ss:evolution}).}

Extensive searches for magnetic fields in central stars of planetary
nebulae and in hot subdwarfs have until now yielded negative results
\citep{asensioramosetal14-1,leoneetal14-1,mathysetal12-1,savanovetal13-1}.
Should this finding be verified by further spectropolarimetric
observations, it would again be consistent with the view that the
origin of fields in MWDs is intimately related to their binary nature.

Another possibility that cannot be a priori excluded may involve the
screening of the WD magnetic field by an unmagnetised layer of
material at the end of the CE phase.  Should this be the case, magnetic
systems would not distinguish themselves as such in the early post CE
phase until their field re-emerges when the stars are in near contact
(as in PREPS) or in contact (as in MCVs). Because of the observed low
effective temperatures of the WDs in PREPS this phase must last
$\gtrsim 10^9$ years. Estimates of the diffusion and advection rates
of a dipolar magnetic field would thus suggest the presence of $\sim
0.2$M$_\odot$ of hydrogen-rich material screening the field in order
to explain the observations \citep{Cumming2002}. However, we need to
stress that the calculations of \citet{Cumming2002} are for accreting
WDs under the assumption that they retain the accreted mass. The
situation in the context of CE evolution is different, in so far as
the screening would be caused by a layer of matter which is retained
following an incomplete ejection of the envelope.

\section{Evolutionary aspects of magnetic white dwarf binaries}
\label{ss:evolution}

\begin{figure}[t]
\includegraphics[width=\columnwidth]{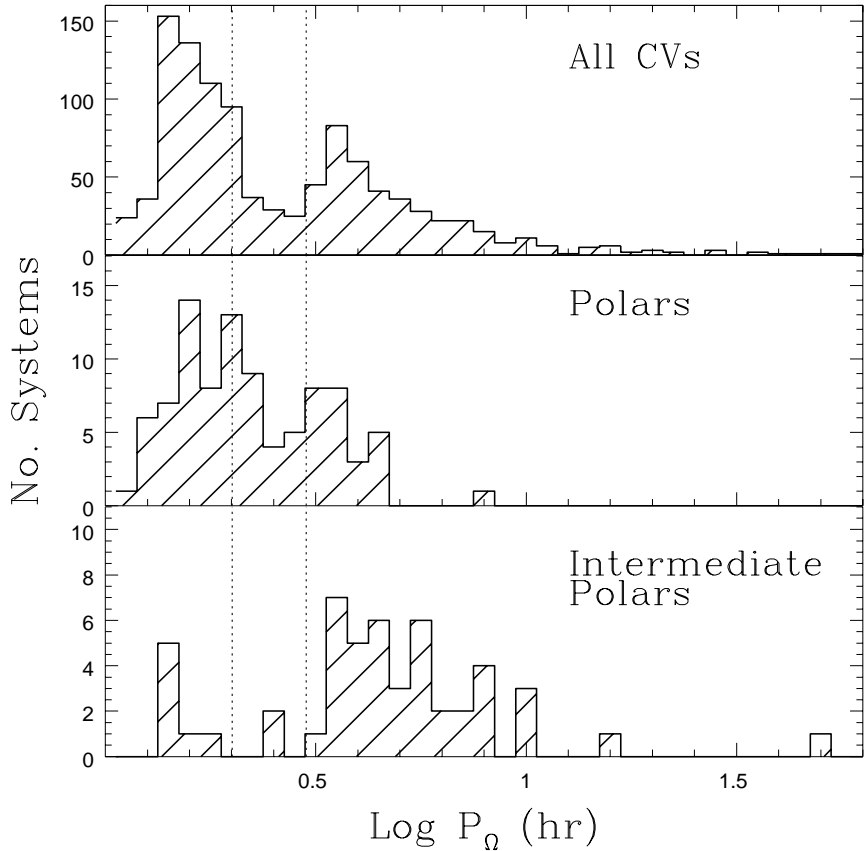}
\caption{The orbital period distribution of CVs (top) and of the
magnetic types Polars (middle) and IPs (bottom). The latest version
(v7.20) of the \citet{ritter_kolb} CV catalogue is used. A few
identifications were corrected. The vertical lines mark the 2-3\,h
orbital period gap } 
\label{fig:orbdistr}
\end{figure}

Close binary evolution theory predicts compact binary systems with
low-mass donors to evolve towards short orbital periods through
angular momentum losses \citep{king88}.  At long orbital periods
($\gtrsim$ 3\,h) magnetic braking \citep{Verbunt81,Rappaport83} acts
as main mechanism that ceases at $\sim$3\,h when secondaries become
fully convective and shrink within their Roche lobe. Mass transfer
stops until contact is re-established near $\sim$2\,h. This gives rise
to the so-called 2-3\,h CV orbital period gap
\citep{Rappaport83,Spruit1983,Davis08}.  Below the period gap
gravitational radiation drives the systems towards the orbital period
minimum, $\sim$70-80\,min, as first suggested by \citet{Paczynski81}
\citep[but see also][]{Townsley_Bildsten03,Townsley09} until the mass
of the donor star becomes too small to sustain nuclear burning and the
star becomes degenerate. The present day CV population is expected to
densely populate the orbital period distribution close to the period
minimum (the ``orbital minimum spike''). These systems would then be
expected to `bounce back' and evolve toward longer periods
\citep{Kolb_Baraffe99}.  The discrepancy between observations and
theory has been a major issue for a long time and only recently
mitigated by deep optical surveys, such as the SDSS which has unveiled
the long-sought low accretion rate and short orbital period CVs
\citep[][see their Fig.\,\ref{fig:orbdistr}, top
panel]{Gaensicke09}. The CV space density also suffers from
discrepancies between theoretical predictions
\citep{dekool92,politano96} and observations
\citep{Schreiber03,Pretorius07}, although there are recent claims by
\citet{Reis13} that some of these disagreements may now be
resolved. The study of \citet{Reis13} of Swift X-ray spectra of an
optically selected sample of nearby CVs has revealed the existence of
a population of objects whose X-ray luminosities are an order of
magnitude fainter than found in earlier studies indicating that the
space density of CVs may be larger than previously forecast and thus
in better agreement with population synthesis calculations.

Among the MCVs, the polars dominate the period distribution at short
($\lesssim$4\,h) orbital periods while the IPs dominate the
distribution at longer periods (see Fig.\,\ref{fig:orbdistr}, mid and
bottom panels). It is not clear yet whether the IPs, as a class, have
generally lower fields than polars or the field strengths of both
sub-classes are similar but IPs still need to synchronise
\citep{King_Lasota91,norton04}.  Unlike polars, the majority of IPs
has not been detected in polarisation searches except for ten systems
that may eventually evolve into polars (see
Sect.\ref{ss:fieldmeasures}).  \citet{norton04, norton08} argued that
IPs above the gap with WD surface magnetic moments
$\mu_{\rm WD} \gtrsim 5\times 10^{33}$\,G\,cm$^3$ and
$P_{\rm orb} \gtrsim$3\,h will eventually evolve into polars. IPs
below the gap, however, are not expected to evolve into
polars. Recently new short period IPs have been discovered both in
X-ray \citep{deMartino05, Bernardini12, Woudt12, Bernardini13,
  Thorstensen13} and optical bands \citep{rodriguez-giletal04-1,
  araujo-betancoretal05-1, Southworth07}. From these findings
\citet{Pretorius_Mukai14} have suggested the existence of a low-luminosity
(hence low-accretion rate) population of IPs that still need to be
unveiled. To enlarge the sample of short period IPs is clearly crucial
to understand MCVs and their evolution.

\begin{figure}[t]
\includegraphics[width=\columnwidth]{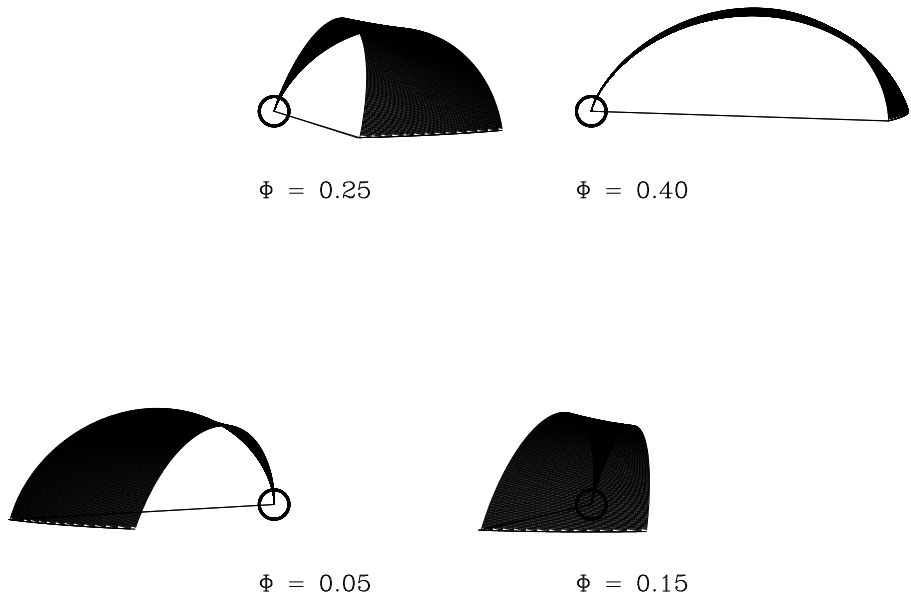}
\caption{Accretion funnel in a polar obtained using an orbital
  inclination, $i=85^\circ$, and a dipole inclination,
  $\theta=15^\circ$. The diagrams for phases $\phi>0.5$ are mirror
  images of those shown here \citep{Ferrarioetal93}}
\label{fig:funnel}
\end{figure}
Although the evolution of MCVs was expected to be similar to that of
non-magnetic CVs, \citet{Wickramasinghe_Wu94} predicted a scenario
where the strong magnetic field in polars reduces the efficiency of
the magnetic braking mechanism \citep[see also][]{wickramasinghe00,
  webbink+wickramasinghe02-1}.  As a consequence, the mass transfer
rates would be lower requiring longer evolutionary timescales than
those of non-magnetic systems.  {The first strong observational
evidence in support of this scenario was presented by \citet{Sion1999} on the
basis of simple considerations of compressional heating and structure
changes in response to accretion in non-magnetic CVs versus the 
known effective temperatures of the WDs in MCVs. This finding was later
confirmed by \citet{Araujo05} for a larger sample of MCVs with
exposed MWDs. They showed that polars possess systematically cooler
WDs than non-magnetic CVs.} Since the WD effective temperature is a
good proxy of the secular mass transfer rate \citep{Townsley09} it
turns out that CVs with highly magnetised WDs accrete at lower rates
and thus evolve on longer timescales.  This could explain the
higher incidence of magnetism observed among CVs ($\sim 20-25$\%) as
compared to that of isolated MWDs \citep{Araujo05}.

The MCVs, being stronger X-ray emitters than non-magnetic CVs, are
claimed to be important constituent of the galactic X-ray source
population at low luminosities. The deep \emph{Chandra} survey of the
galactic centre has revealed a large number of low-luminosity hard
X-ray sources attributed to MCVs of the IP type
\citep{muno04,hong2012a}. The hard X-ray surveys of the Galactic Ridge
conducted by the \emph{INTEGRAL} \citep{revnivtsev09,revnivtsev11},
\emph{Rossi-XTE} \citep{sazonov06} and \emph{Suzaku} \citep{Yuasa12}
satellites have also resolved a large fraction of the diffuse emission
into discrete low luminosity hard X-ray sources largely attributed to
coronally active stars and MCVs of the IP type.  The true contribution
of these systems to the X-ray luminosity function of the Galactic
X-ray source population at low luminosities is still under
investigation
\citep{Muno2004,Revnivtsev2006,Yuasa12,Reis13,Pretorius13,Pretorius_Mukai14,Warwick2014}.

\begin{figure}[t]
\includegraphics[width=\columnwidth]{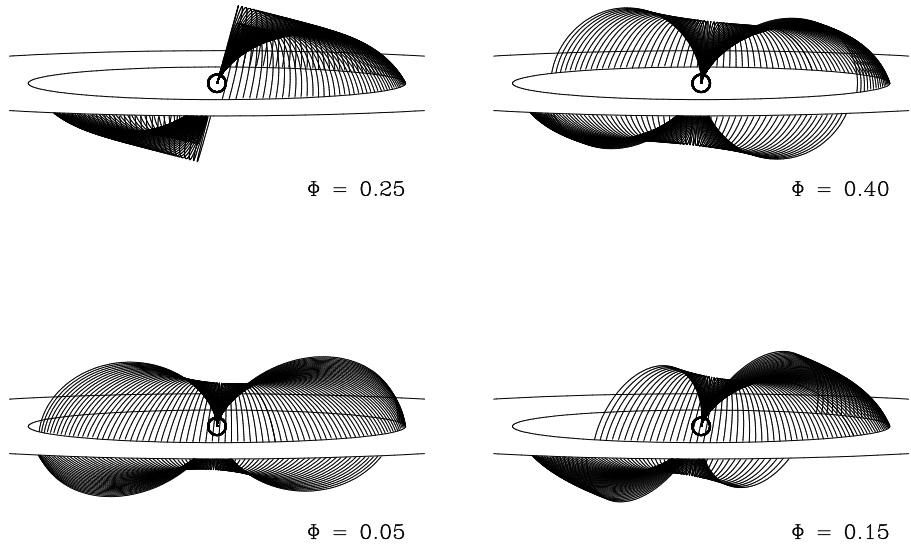}
\caption{Visible portions of the two accretion curtains in IPs
  obtained using a viewing angle $i=70^\circ$ and a magnetic
  colatitude of $\theta=15^\circ$. The diagrams for $\phi>0.5$ are
  mirror images of those shown here \citep{Ferrarioetal93}}
\label{fig:curtains}
\end{figure}
\section{Magnetic accretion}
\label{s:magaccr}

The magnetic field of the WD primaries influences the accretion
geometry and emission properties of MCVs. It determines the dynamics
of accretion when the dynamical timescale is of the order of the
timescale of magnetic interaction.  This occurs at the magnetospheric
boundary where the magnetic pressure balances the ram pressure of the
infalling material: $\rho\,v^2 \simeq B^2 / 8\,\pi$
\citep{Frank_King_Raine85}.  The magnetospheric radius, $R_{\rm mag}$
depends on the magnetic field strength of the WD and the system
parameters \citep[see the review by][]{Wickramasinghe14}. Among
polars, the WD magnetic moment is strong enough ($\mu_{\rm WD} \gtrsim
10^{34}$\,G\,cm$^3$) to prevent the formation of an accretion disc,
because $R_{\rm mag}$ is of the order of the orbital separation. Thus
the accretion stream flowing into the primary Roche lobe along a
ballistic trajectory is channelled by the field lines towards the
magnetic poles of the WD \citep{FerrarioTuohyWick89,Schwope96} forming
an accretion ``funnel''.  The radiation in the optical and IR bands is
characterised by strongly circularly and linearly polarised cyclotron
emission from the stand-off shocks located at the base of the
accretion funnels. Because of phase locking, radial velocity and light
variations are seen only at the orbital period.

{There are currently four confirmed polars that show a
  small degree (several percent) of asynchronism between the spin and
  orbital periods and two candidate systems \citep[][and see also
  Table \ref{tab:mcvs}]{Campbell1999}.
  Doppler tomography of the near-synchronous polar BY\,Cam
  \citep{Schwarz2005} has uncovered that most of the matter flowing
  towards the magnetic WD is accreted via a funnel that extends by
  $\sim180^\circ$ in azimuth. This implies that the stream can travel
  around the WD, making the accretion pattern in this system resemble
  that of IPs (see below). Thus, asynchronous polars are particularly
  important for our understanding of magnetic accretion since they
  display characteristics common to both Polars and IPs.}

The reason for this asynchronism is not clear, but it has been
proposed that it could be caused by recent nova eruptions \citep[e.g. Nova
V1500\,Cyg, ][]{Schmidt_Stockman91,Schmidt95}. Furthermore among newly
identified MCVs there are a handful of systems with weakly
desynchronised MWDs (see Table \ref{tab:ips}), such as Paloma
\citep{Schwarz07}, IGR\,J19552+0044 \citep{Bernardini13}, and V598 Peg
\citep{Southworth07} indicating that the distinction between polars
as synchronous rotators and IPs is not as sharp as formerly believed.

\begin{figure}[t]
\includegraphics[width=\columnwidth]{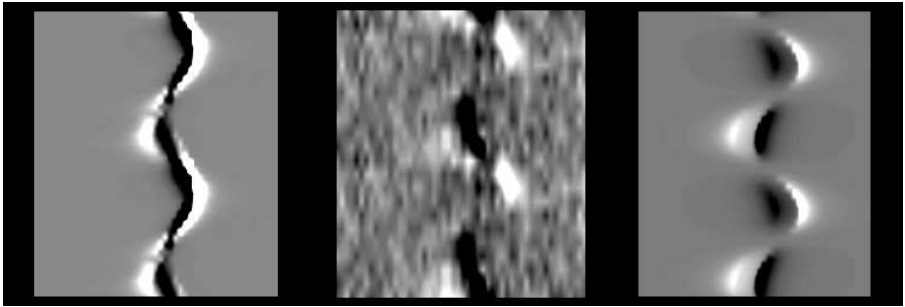}
\caption{Middle panel: observed phase-dependence of the circularly
  polarised flux for the region surrounding H$_\beta$. Wavelength runs
  from 4750~\AA\ to 4950~\AA, and orbital phase advances upward, with
  $\phi=0.25$ at the bottom and $\phi=0.25$ at the top of the figure,
  covering two complete orbital cycles. The polarised flux is white
  for negative circular polarisation and black for positive circular
  polarisation. Left panel: ``standard'' funnel model with
  $\theta_d=30^\circ$, $\phi_d=90^\circ$ and $i=60^\circ$. Right
  panel:idealised plane parallel slab model for field aligned flow in
  the transition region from ballistic stream to funnel flow \citep{Ferrario02}}
\label{ARUMA_spectro}
\end{figure}
In most IPs, accretion discs or rings usually form \citep[but see][for
a  review on accretion in IPs]{Hellier2014} and $R_{\rm
  mag} \lesssim R_{\rm cor}$ where $R_{\rm cor} = (G\,M_{\rm
  WD}/\Omega^2_{\rm rot})^{1/3}$ is the co-rotation radius at which
the Keplerian angular velocity equals the spin angular velocity of the
primary and $R_{\rm mag} = \phi\, 2.7\times 10^{10}\,\mu_{33}^{4/7}\,
\dot M_{16}^{-2/7}\, \,M_{\rm WD,\odot}^{-1/7}$\,cm, where $\phi \sim
1$ is a parameter that takes into account the departure from the
spherically symmetric case, $\mu_{33}$ is the WD magnetic moment in
units of $10^{33}$\,G\,cm$^3$, $M_{16}$ is the mass accretion rate in
units of $10^{16}$\,g\,s$^{-1}$ and $M_{{\rm WD},\odot}$ is the WD
mass in solar units \citep{norton89, hellier95}.  The details of the
threading region depend on the magnetic diffusivity of the disc and
the toroidal field produced by the shear between the Keplerian disc
and the co-rotating disc \citep[see][]{Wickramasinghe14}. Close to the
WD, the flow in IPs consists of two magnetically confined accretion
``curtains'' fed by the disk or ring \citep{Ferrarioetal93, hellier95}.

Since the WD in IPs is not phase-locked into synchronous rotation with
the orbit, the emission variations are observed to occur at the spin
period of the WD, $P_{\rm s}$, at the beat period $P_{\rm
  beat}=(P_{\rm s}^{-1}-P_{\rm orb}^{-1})^{-1}$ and often at both. The
multi-component model of \citet{FerrarioWick99} to calculate the
optical and X-ray power spectra of disced and discless MCVs has
revealed that as the magnetic field strength of the WD increases, the
cyclotron emission from the shocks becomes comparable to the optical
radiation from the magnetically confined flow and the dominant power
shifts from the beat frequency to the WD spin frequency. {Thus those
MWDs in IPs with sufficiently large magnetic moments, such as V\,2400
Oph \citep{Buckley1997}, do not accrete via a disc while in many other IPs,
such as FO Aqr, the stream from the companion star can flow over the
disc \citep{norton92, Beardmore1998, Hellier2014}.}

One of the most significant characteristic that distinguishes disc-fed
accretion from stream-fed accretion is the amplitude of the radial
velocity variations.  Disc-fed accretion is characterised by low
radial velocity amplitudes ($\sim 50 - 100$\,km\,s$^{-1}$), while
stream-fed accretion by high radial velocity amplitudes ($\sim
500-1,000$\,km\,s$^{-1}$).

{Here we need to stress that although curtains are
  generally associated with asynchronous systems, under certain
  accretion conditions even polars may exhibit extended coupling
  regions and a curtain-type structure for the magnetically confined
  infalling matter. Similarly, some IPs may sometime show funnel-like
  flows typical of polars. In this context we note that
  \citet{Schwope2004} have shown how tomographic techniques can be
  effectively used to infer the accretion geometry of MCVs.}

We show schematic diagrams of magnetically channelled accretion flows
in typical polars and IPs in Fig.\,\ref{fig:funnel} and
Fig.\,\ref{fig:curtains} respectively.  As shown in these figures, the
observed modulation arises from projection area effects and viewing
aspect. Depending on the orbital phase $\phi$, parts of the accretion curtains
are either self-occulted or are hidden by the disc or the body of the
WD \citep{Ferrarioetal93, FerrarioWick93}.

The discovery of Zeeman-split emission lines in the circular
polarisation spectra of AR\,UMa \citep{Schmidt99} and V884\,Her
\citep{Schmidt01} have enabled the investigation of the magnetic,
thermal and dynamical structure of accretion funnels. The modelling of
\citet{Ferrario02} of these two very high field polars using the
funnel structures developed by \citet{Ferrario99} has revealed that
the polarisation spectrum is very sensitive to velocity and field
gradients, as shown in Fig.\,\ref{ARUMA_spectro}. They find that the
bulk of the observed emission arises from two components, (i) the
material in the magnetically channelled funnel flow and (ii) the
threading region at the base of the funnel where the stream changes
from ballistic to co-rotational. The latter makes the main
contribution in the high field system AR\,UMa which is viewed at large
inclination angles while the former dominates at the low orbital
inclinations of V884\,Her \citep{Ferrario02}. {Thus,
  the study of polarised line emission, which is only possible for
  systems which possess very strong fields, allows one to unravel the
  contributions from the stream, funnel, and transition region.}

\begin{figure}[t]
\includegraphics[angle=-90, width=\columnwidth]{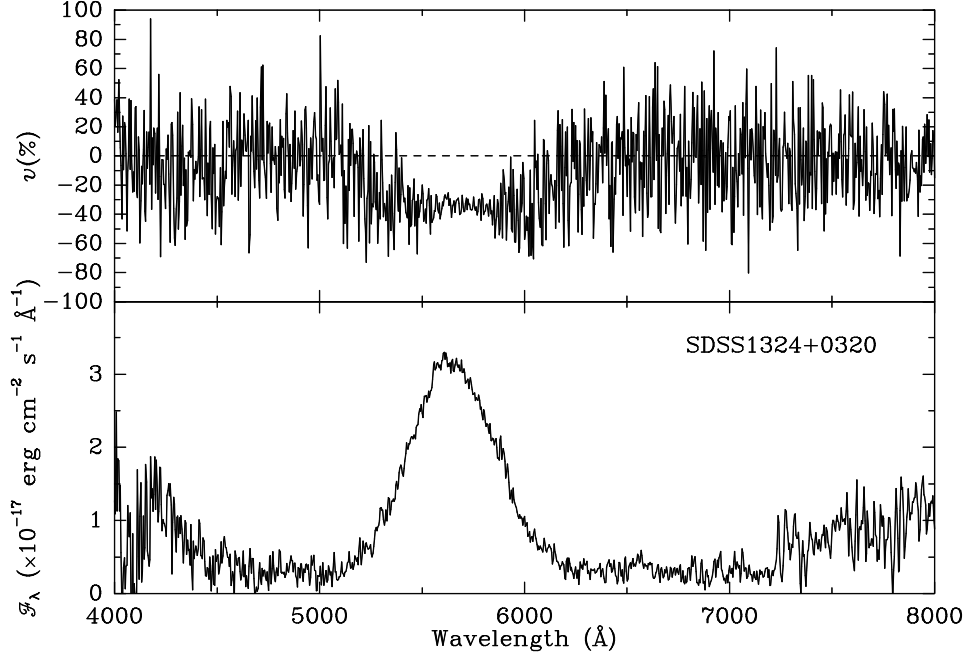}
\caption{Bottom panel: Phase-averaged flux spectrum of the PREP PZ\,Vir
  (=SDSS~J1324+0320). Top panel: phase-averaged circular polarisation
  spectrum \citep{Szkody03}}
\label{fig:SDSS1324_obs}
\end{figure}
Close to the WD surface {the densest parts} of the
supersonic infalling matter produce a strong shock where the gas is
heated up to temperatures of $\sim$ 10-40\,keV.  The shock temperature
is $T_{\rm s} = 3.7\times 10^8\, M_{\rm WD,\odot}/R_9\,$\,K, where
$R_9$ is the WD radius in units of $10^9$\,cm.  The post-shock flow
becomes subsonic and cools via thermal bremsstrahlung (hard X-rays)
and cyclotron radiation (optical/IR) \citep{aizu73, King_Lasota79,
  lamb_masters79}.  The relative proportion of cyclotron to
bremsstrahlung radiations depends on the WD field strength and
specific mass accretion rate. Cyclotron dominates at high field
strengths and/or low local mass accretion rates, thus,  if $L_{\rm
  cyc}$ and $L_{\rm brem}$ are the cyclotron and bremsstrahlung
luminosities respectively, then $L_{\rm cyc} > L_{\rm hard}$. The
temperatures of ions and electrons are different in this case
(two-temperature fluid).  Bremsstrahlung instead dominates at low
field and/or high local mass accretion rates and the ion and electron
temperatures are about the same (one-temperature fluid).  The WD
surface also intercepts these primary emissions, which are partially
reflected and thermalised. The poles of the WD are then heated at
temperatures $\sim20-40$\,eV and emit a blackbody-like spectrum.
However, the prediction that the black body luminosity, $L_{\rm BB}$,
should be $\sim L_{\rm brem}+L_{\rm cyc}$ was not confirmed in AM\,Her
and other polars which showed a prominent soft X-ray component.  This
was referred to as the ``soft X-ray puzzle'' \citep[see
also][]{beuermann99}.  \citet{kuijpers_pringle82} suggested that at
high local mass accretion rates ($\dot m \sim 100$\,
g\,cm$^{-2}$\,s$^{-1}$) the pressure of material is so high that the
shock can be buried in the WD atmosphere and that the heating of the
WD from below produces an intense blackbody-like emission.  This
component does not enter in the energy balance, thus solving the soft
X-ray puzzle.  In AM\,Her most of the reprocessed radiation was found
to emerge in the FUV/UV range \citep{Gaensicke95}.  In this context,
we note that the recent modelling of the accretion heated spot of the
WD using a temperature distribution across the region can well
reproduce the soft X-ray spectrum in AM\,Her \citep{Beuermann12}.

\begin{figure}[t]
\includegraphics[width=\columnwidth]{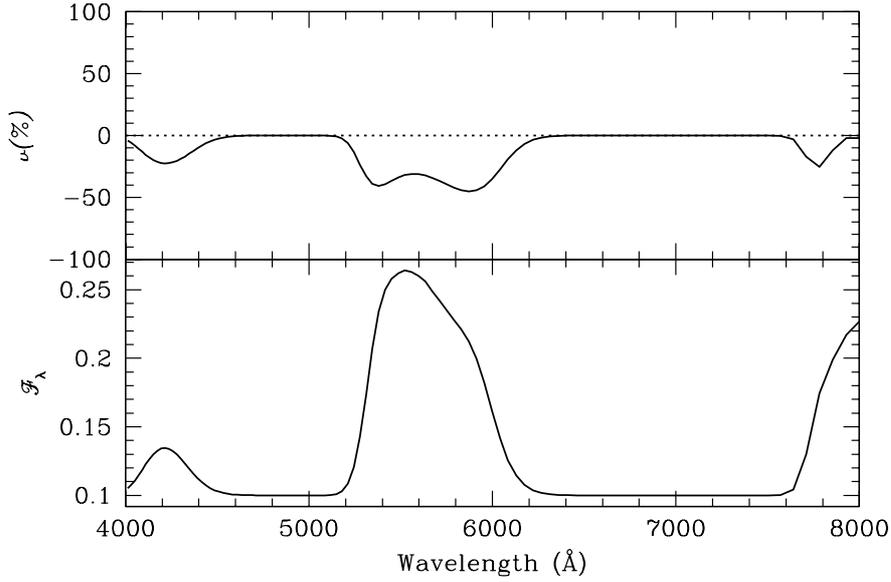}
\caption{Phase-averaged cyclotron model for the PREP PZ\,Vir
  (=SDSS\,J1324+0320) corresponding to a magnetic field
  strength $B=64$\,MG. Lower panel panel: flux. Upper panel:
  circular polarisation}
\label{fig:SDSS1324_mod}
\end{figure}

The structure of the post-shock region has been the subject of several
studies.  Detailed one-dimensional two-fluid hydrodynamic calculations
coupled with radiative transfer equations for cyclotron and
bremsstrahlung were performed for different regimes by
\citet{woelk_beuermann92,woelk_beuermann96,ThompsonCawthorne87,kuijpers_pringle82}
and \citet{fischer_beuermann01}. They also include the so-called
bombardment regime, which is associated with very low specific mass
accretion rates and high field strengths. These very low accretion
rates do not allow the formation of a hydrodynamic shock, and the
photosphere is heated directly by particle bombardment. This is the
typical accretion regime of PREPS, as indicated by the observations of
systems such as PZ\,Vir (=SDSS~J1324+0320,
Fig.\,\ref{fig:SDSS1324_obs}). The very low electron temperatures and
specific accretion rates derived by the modelling of these systems
\citep[see Fig.\,\ref{fig:SDSS1324_mod} and ][]{Ferrario2005} are
found to be consistent with the bombardment regime. The absence of
Zeeman features from the photosphere of the WD further strengthens the
hypothesis that these systems are not polars in a low state of
accretion but pre-MCV with very cool, old, WDs.

When fitting the X-ray spectra, early computations of the structure of
the post-shock flow using one-temperature calculations
\citep{cropper99} were later demonstrated to provide higher WD masses
than those obtained using hydrodynamical models that included 
two-temperature effects \citep{Saxton05} and dipolar field geometry
\citep{Saxton07}.

In the IPs, accretion generally occurs via magnetically confined curtains (see
Fig.\, \ref{fig:curtains}) forming arc-shaped shock regions around the
WD magnetic poles \citep{Ferrarioetal93, FerrarioWick93, hellier95}.
Since these systems are strong hard X-ray emitters, the post-shock
region mainly cools via thermal bremsstrahlung. Because of the large
arc-shaped footprints, the reprocessed radiation was initially
expected to emerge in the EUV range. However, the \emph{ROSAT} survey
revealed the existence of a few IPs with a soft X-ray component
similar to that observed in polars.  Observations with the
\emph{XMM-Newton} satellite have increased the number of IPs that
exhibit a soft blackbody component to $\sim18$, although these cover a
wider range of temperatures than those in polars \citep[][and
reference therein]{Anzolin08, Anzolin09, Bernardini12}. Since the soft
X-ray component is only a small fraction of the hard X-ray luminosity,
it is consistent with reprocessing.

Interestingly, recent X-ray observations with \emph{XMM-Newton} of
polars in high states of accretion have revealed an increasing number
of systems that do not exhibit a distinct soft X-ray component but
rather a more `IP-like' X-ray spectrum
\citep{ramsay04,Vogel08,Ramsay09,Bernardini14}. However, the
magnetic fields and orbital periods of these polars do not appear to
be very dissimilar from all other polars with a more classic type of
behaviour.  Hence, the distinction between the two subclasses now
appears less marked than ever before, requiring further
investigations.

\section{Conclusions}

To date there are about $\sim 250$ MWDs with well determined fields
(see Table \ref{tab:mwds}) and over $\sim 600$ if we also count
objects with no or uncertain field determination
\citep[see][]{Kepler2013,Kepler2015}. These MWDs have been discovered
following surveys such as the SDSS, HQS and the Cape Survey. The
enlarged sample has shown that the field distribution of MWDs is in
the range $10^3-10^9$\,G.  While the high field cut-off appears to
be real, the low field one is currently determined solely by the
sensitivity of current spectropolarimetric surveys. Observations also
indicate that MWDs may be divided into two groups: a high field group
($1-1\,000$\,MG), where most objects are found, and a low field group
($<0.1$\,MG), whose importance still needs to be determined by
much more sensitive spectropolarimetric surveys conducted on 8\,m
class telescopes. 

The high field group of MWDs differs from the low field group in terms
of average mass \citep[see also][]{Kepler2013}. That is, high field
MWDs exhibit a higher average mass ($\sim 0.85$\,M$_\odot$) than
weakly magnetic or non-magnetic WDs ($\sim 0.66$\,M$_\odot$). High
field MWDs also have a relatively strong tail that extends to the
Chandrasekhar limit.

The significant increase in the number of MWDs has also led to new
insights on the nature of magnetism.  However, we still need to
construct (i) more realistic model atmospheres that allow for the
presence of magnetic fields and (ii) stellar evolution tracks of
intermediate mass stars that take into consideration both fossil and
dynamo generated fields. Such calculations may be able to tell us
whether all WDs are magnetic at some level.

The origin of fields in highly magnetic WDs is currently being
debated. Although the newly proposed scenario that all high field MWDs
(single and in binaries) are the result of close binary evolution and
mergers is gaining momentum, the fossil field hypothesis cannot be
totally dismissed. The attractiveness of the merger hypothesis lies
mostly in its ability to explain why there are no wide binaries consisting
of a MWD with a non-degenerate companion star and why MWDs are on
average more massive than their non-magnetic or weakly magnetic
counterparts.

Observations of magnetic WDs in interacting binaries obtained in the
last decade have also opened interesting questions on their evolution,
accretion and emission properties.  Forthcoming surveys such as
SDSS-IV and VPHAS+ \citep{Drew2014} in the optical and in the X-rays,
{such as the one expected to be conducted by \emph{eROSITA}
\citep[e.g.,][]{Schwope2012} will allow the discovery of new
systems providing new exciting challenges.}

The study of isolated and binary MWDs is likely to remain at the
forefront of research for many years to come.

\section*{Acknowledgements}
The authors gratefully acknowledge the organisers of this workshop for
the timely opportunity to review our understanding of the strongest
magnetic fields in the Universe, and ISSI for their warm hospitality. 
The research leading to these results has received funding from the
European Research Council under the European Union's Seventh Framework
Programme (FP/2007-2013) / ERC Grant Agreement n. 320964 (WDTracer).
DDM acknowledges financial support by ASI/INAF under contract I/037/12/0.

\bibliographystyle{spr-mp-nameyear-cnd}  
\bibliography{mwd,MWD_list,bibliotables_mcvs_new}                
\newpage
.
\vskip 6truecm
\centering{\bf \huge APPENDIX: Tables of Magnetic White Dwarfs and Magnetic
  Cataclysmic Variables}
\begin{flushleft}
\footnotesize
\begin{landscape}

\pagestyle{empty}
\voffset4cm
\begin{longtable}{llllllll}
\caption{\label{tab:mwds} Magnetic White Dwarfs}\\
\hline\hline
\noalign{\smallskip}
WD & Other names&Comp& $B_p$&$T_{\rm eff}$&$M$&$P_{\rm rot}$& References\\
      &                     &          & (MG)  &  (K)            & (M$_\odot$)&      & Notes\\
\noalign{\smallskip}
\hline
\endfirsthead
\endfoot
\caption{continued.}\\
\hline
\noalign{\smallskip}
WD & Other names&Comp& $B_p$&$T_{\rm eff}$&$M$&$P_{\rm rot}$& References\\
      &                     &          & (MG)  &  (K)            & (M$_\odot$)&      & Notes\\
\noalign{\smallskip}
\hline
\noalign{\smallskip}
\endhead
\endfoot
\noalign{\smallskip}
0003-103 & SDSS J000555.91-100213.4&DQ&1.47$^*$&19420$\pm$920&$\log\,g=(8.0)$&2.110$\pm0.045$\,d&       {1,2,3} \\

0005-148&NLTT\,347, &DA& -0.0046$\pm$0.0019&6400$\pm$180&0.59:& $\cdots$&       {4}\\
  &SDSS\,J010319.70−032501.0  & & & & & & \\

0009+501&LHS\,1038, G\,217-37, NLTT\,574&DA&0.316$^*$& 6540$\pm$150&0.74$\pm$0.04 & hrs--2.5\,d&       {5,6,4} \\
  
0011-134&LHS\,1044, G\,158-045&DA&16.7$\pm$0.6&6010$\pm$120&0.71$\pm$0.07&30\,min-days&       {7,8,9}\\
  
0015+004 &SDSS\,J001742.44+004137.4  &DB &8.3  &  15000  &  $\cdots$  &  $\cdots$  &         {10}\\
  
0018+147 &SDSS\,J002129.00+150223.7  &DA& 530.69$\pm$63.56 &  7000  &  $\cdots$  &  $\cdots$  &         {11,12}\\

0038-084 & NLTT\,2219 & DA & 0.307$^*$ & 6000$\pm$180 & 0.59:  & $\cdots$ &        {4} \\
  
0040+000 &SDSS\,J004248.19+001955.3  &DA &2 &11000&$\cdots$  &  $\cdots$  & DD,        {10,12} \\
  
0041-102 &Feige\,7,L\,795-7&DBA &  35  & 20000  & $\log\,g=(8.0)$  &  131.606\,min &        {13,14,15}\\

0051+115 & HS\,0051+1145, PHL\,886& DA & 0.240$\pm$0.010 & $\cdots$  & $\cdots$ & $\cdots$ &        {16} \\

0058-044 &GD\,9, GR\,407, PHL\,940 & DA & 0.325$\pm$0.035 & 16700 & $\log g=8.07$ & $\cdots$  &         {17,16}\\

0104+149&SDSS\,J010647.92+151327.8& DQ& not confirmed& 23430$\pm$1680  & $\log\,g=(8.0)$  &  $\cdots$ &       {2} \\
  
0140+130  &SDSS\,J014245.37+131546.4  &DB &     4  &  15000  &  $\cdots$  &  $\cdots$  &        {10} \\
  
0155+003 &SDSS\,J015748.15+003315.1  &DZ & 3.49$\pm$0.05$^*$ &5700&$\cdots$&$\cdots$&       {10,18}\\
  
0159-032  &1H\,0201-029 &DA &  6  &  26000  &  $\log\,g=(8.0)$  &   $\cdots$  &         {19}\\
   
0208+002 &SDSS\,J021116.34+003128.5  &DA  &341.31$\pm$54.34  &  9000  &  $\cdots$  &  $\cdots$  &        {10,12} \\
  
0209+210 &SDSS\,J021148.22+211548.2  &DA  & 166.16$\pm$7.41  &  12000  &  $\cdots$  &$\cdots$&       {11,12} \\
          
0231+263 &SDSS\,J023420.63+264801.7 &DA & 32.82$\pm$6.26  & 13500 &
   $\cdots$&  $\cdots$     &           {12} \\ 
       
0233-083 &SDSS\,J023609.40-080823.9   &DQA&      5  &  10000  &  $\cdots$  &  $\cdots$  &         {11} \\
  
0236-269 &HE\,0236-2656 &DB &  $\cdots$ &  6000-7000  &  $\cdots$  &  $\cdots$  &         {20}\\

0239+109 &G\,4-34, LTT\,10886 & DA  & 0.725$\pm$0.025  & 10060 & $\log g=8.73$  & $\cdots$&DD,        {17,16,21}\\
  
0253+508  &KPD\,0253+508    &DA    &     17  &  15000  &  $\log\,g=(8.0)$  &   3.79$\pm$0.05\,hr &         {22,23,24}\\
  
0257+080&LHS\,5064, G\,76-48&DA&0.1$^*$&6680$\pm$150 & 0.57$\pm$0.09&6\,d:&       {25,16,4,9}\\
   
0301-006  &SDSS\,J030407.40-002541.7  &DA  & 10.95$\pm$0.98 &  15000  &  $\cdots$  &  $\cdots$  &         {26,12}\\
  
0307-428 &1H0307-426&DA& 10& 25000  &  $\log\,g=(8.0)$ &$\cdots$ &        {19} \\

0315-293&NLTT\,10480, LHS\,5070, LP\,887-66 & DAZ&0.5$^*$&5340$\pm$190 & 0.58: &$\cdots$ &       {27,4,28}\\

0315+422&SDSS\,J031824.19+422651.0 &DA & 10.12$\pm$0.10 &10500 & $\cdots$ & $\cdots$ &         {12}\\ 

0321-026 & KUV\,03217-0240   & DA & $<1:$  &27370 &  $\log g=8.45$ & $\cdots$ &        {21} \\

0322-019&G\,77-50, NLTT\,10871, LHS\,1547&DAZ&0.120$^*$&5310$\pm$100&0.60$\pm$0.01&28-33\,d:&       {29,4}\\
  
0323+051&SDSS\,J032628.17+052136.3 &DA & 16.87$\pm$2.41 & 25000& $\cdots$ & $\cdots$ &         {12}\\ 
   
0325-857&RE\,J0317-853, EUVE\,J0317-855&DA&185-450&33000&1.34$\pm$0.03&725\,s& DD,  {30,31,32,33}\\

0329+005&KUV\,0329+0035, &DA &13.13$\pm$1.00  & 15500  & $\cdots$ &$\cdots$ &        {26,10,12}\\ 
  &SDSS\,J033145.69+004517.0 & & & & & & \\

0330-000 &HE\,0330-0002, &DB&$\cdots$  &6000-7000& $\cdots$&$\cdots$&       {20,26,10} \\

$\cdots$  &SDSS\,J033320.36+000720.6 &DA &849.30$\pm$51.75  &  7000:  &  $\cdots$ &  $\cdots$     &         {12}  \\ 
  
0340-068&SDSS\,J034308.18-064127.3&DA & 9.96$\pm$2.06 & 13000:  &  $\cdots$  &  $\cdots$  &         {10,12}\\
   
0342+004 &SDSS\,J034511.11+003444.3 &DA  &  1.96$\pm$0.42    & 8000  &  $\cdots$  &  $\cdots$  &       {26,10,12}\\

0350+098 & 1RXS\,J035315.5+095700 & $\cdots$& $\cdots$ & $\cdots$ & $\cdots$ & $\cdots$ &         {21}\\

0410-114&G\,160-51,  NLTT\,12758&DA&1.7$\pm$0.2&7440$\pm$150&0.83&$\cdots$& DD,        {34,4}\\
      
0413-077& 40\,Eri\,B &DA & 0.0073$^*$&16490$\pm$84&0.497$\pm$0.005&$\cdots$&       {35,4} \\
 
0416-096 & NLTT\,13015, LP\,714-52   & DA & 6-7.5 & 5745$\pm$405 & 0.59: & Variable: &   {4}  \\
          
0446-789   &BPM\,3523  &DA & 0.0135$^*$ & 23450$\pm$20 & 0.49$\pm$0.01  &  $\cdots$  &         {36,16}\\
   
0503-174   &LHS\,1734, LP\,777-001&DA&7.3$\pm$0.2  & 5300$\pm$120 & 0.37$\pm$0.07  &  $\cdots$  &        {37,7}\\
  
0548-001&G\,99-37&DQB&7.3$\pm$0.3&6200$\pm$200&0.69$\pm$0.02&4.117\,hr&  {38,13,39,40,41,42}\\
  
0553+053&G\,99-47,LTT\,17891&DA & 20$\pm$3& 5790$\pm$110&0.71$\pm$0.03  & $26.8\pm0.7$\,min&       {13,43,44,39,37,9}\\
   
0616-649 &EUVE\,J0616-649 &DA &  14.8      &  50000  &  $\log\,g=(8.0)$  &   $\cdots$  &         {45}\\
  
0637+477   &GD\,77 &DA &1.2$\pm$0.2 & 14870$\pm$120 & 0.69  &  $\cdots$  &        {1,46}\\
  
0728+642   &G\,234-4 &DA  & 0.125$^*$ &  4500$\pm$500  &  $\cdots$  &  $\cdots$  &         {8}\\

$\cdots$      &SDSS\,J073549.19+205720.9 & DZ & 6.12$\pm$0.06 &6000&$\cdots$&$\cdots$&       {18} \\

0745+302 &SDSS\,J074850.48+301944.8 &DA  & 6.75$\pm$0.41  &  22000:  &  $\cdots$  &  $\cdots$  &         {11,12}\\

0745+303 &SDSS\,J074853.07+302543.5 & DA&11.4 &21000$\pm$2000&0.81$\pm$0.09&$\cdots$ &       {47} \\
 
0746+172 &SDSS\,J074924.91+171355.4 &DA & 13.99$\pm$1.30 & 20000 &  $\cdots$  &  $\cdots$ &     {12}\\ 

0749+173 &SDSS\,J075234.96+172525.0 &DA & 10.30$\pm$1.23 & 9000&  $\cdots$ & $\cdots$ &     {12}  \\ 
  
0755+358 &SDSS\,J075819.57+354443.7  &DA & 26.40$\pm$3.94&  22000  &  $\cdots$  &  $\cdots$  &         {10,12}\\
  
0756+437 &G\,111-49  &DA &  180-220 & 8500$\pm$500&$\cdots$& $6.68$\,hr &       {44,8,9}\\

0801+124 &SDSS\,J080359.93+122944.0 &DA & 40.70$\pm$2.13  & 9000 &  $\cdots$ & $\cdots$      &     {12} \\ 
   
0801+186 &SDSS\,J080440.35+182731.0  &DA & 48.47$\pm$2.93 &  11000  &  $\cdots$  &  $\cdots$  &        {11,12} \\
  
0802+220 &SDSS\,J080502.29+215320.5  &DA &  6.11$\pm$1.29  &  28000:  &  $\cdots$  &  $\cdots$  &        {11,12}  \\
  
0804+397 &SDSS\,J080743.33+393829.2  &DA & 65.75$\pm$18.52  &  13000  &  $\cdots$  &  $\cdots$  &         {10,12}  \\ 
    
0806+376&SDSS\,J080938.10+373053.8  &DA &  39.74$\pm$5.41 &  14000  &  $\cdots$  &  $\cdots$  &         {11,12}  \\
  
0814+043&SDSS\,J081648.71+041223.5  &DA  & 10.13$\pm$8.03 & 11500  &  $\cdots$  &  $\cdots$ &         {11,12} \\ 

0814+201&SDSS\,J081716.39+200834.8 &DA &  3.37$\pm$0.44  &  7000 &  $\cdots$ & $\cdots$ &           {12}  \\ 
     
0816+376&GD\,90  &DA &   9  &  14000 & $\log\,g=(8.0)$  &   $\cdots$  &         {48,13,43,49,8}\\

0821-252 &EUVE\,J0823-254 &DA & 2.8-3.5  &  43200$\pm$1000  &  1.20$\pm$0.04  &  $\cdots$  &         {50}\\
  
0825+297&SDSS\,J082835.82+293448.7 &DA  & 33.40$\pm$10.53  &  19500  &  $\cdots$  &  $\cdots$  &          {11,12}\\

 $\cdots$    & SDSS\,J083200.38+410937.9 & DZ &2.35$\pm$0.11$^*$ & 5900 &  $\cdots$  &  $\cdots$  &         {18}\\
   
0825+822 &SDSS\,J083448.63+821059.1 &DA &14.44$\pm$4.57 & 27000 &$\cdots$ & $\cdots$ &        {12}  \\ 

0836+201 & EG\,59 (Mislabeled as EG\,61)  &DA  & 2.83$\pm$0.19  & 17000$\pm$500  & 0.82$\pm$0.05&$\cdots$ &  {51,52}\\

0837+273 &SDSS\,J084008.50+271242.7 &DA &10 & 12250&$\cdots$&$\cdots$&        {11} \\

 $\cdots$   &SDSS\,J083945.56+200015.7 &DA &  3.38$\pm$0.49 & 15000:& $\cdots$& $\cdots$ &           {12} \\ 

0839+026 &SDSS\,J084155.74+022350.6 &DA & 5.00$\pm$0.99 &  7000  &  $\cdots$  &  $\cdots$  &         {10,12}\\

0843+488 &SDSS\,J084716.21+484220.4 &DA & $\sim$\,3  &  19000  &  $\cdots$  &  $\cdots$  &        {10}\\

0848+121 &SDSS\,J085106.12+120157.8 &DA &  2.03$\pm$0.10  &  11000  &  & $\cdots$ &      {12}\\ 
   
0853+163&PG\,0853+164,LB\,8915 &DBA&0.75-1.0&21200-27700&$\log\,g=(8.0)$& 2-24\,hr&       {8,53,9,4}\\

0853+169 & SDSS\,J085523.87+164059.0  &DA &  12.6$\pm$1.0  &  20000$\pm$500&1.12$\pm$0.11 & $\cdots$ &       {12,52} \\ 

 $\cdots$    & SDSS\,J085550.67+824905.3&DA & 10.82$\pm$2.99  & 25000 & & $\cdots$ &       {12} \\ 
  
0855+416   &SDSS\,J085830.85+412635.1  &DA &  3.38$\pm$0.19  &  7000  &  $\cdots$  &  $\cdots$  &          {10,12}\\

 $\cdots$   & SDSS\,J090222.98+362539.6 & DZ & 1.92$\pm$0.05$^*$ & 6300 & $\cdots$ &  $\cdots$ &        {18} \\
 
0903+083   &SDSS\,J090632.66+080716.0  &DA & 5.98$\pm$3.02 & 17000  &  $\cdots$  &  $\cdots$  &        {11,12}\\
   
0904+358   &SDSS\,J090746.84+353821.5  &DA   &   22.40$\pm$8.80  &  16500  &  $\cdots$  &  $\cdots$  &        {11,12} \\

0907+083  &SDSS\,J091005.44+081512.2 &DA &1.01&  25000 &$\cdots$ &$\cdots$ &       {12} \\ 
  
0908+422  &SDSS\,J091124.68+420255.9  &DA  & 35.20$\pm$5.83 & 10250  &  $\cdots$  &  $\cdots$  &        {11,12}\\
  
0911+059  &SDSS\,J091437.40+054453.3  &DA  &   9.16$\pm$0.77 &  17000  &  $\cdots$  &  $\cdots$  &          {11,12}\\

0912+536&G195-19&DB&    $\sim$\,100  &  7160$\pm$190 & 0.75$\pm$0.02&1.3301\,d &       {37,13,54}  \\
   
0915+211  &SDSS\,J091833.32+205536.9 &DA & 2.04$\pm$0.10 &  14000  & $\cdots$  &  $\cdots$  &        {12} \\ 
   
0922+014  &SDSS\,J092527.47+011328.7  &DA    &  2.04 &  10000  &  $\cdots$  &  $\cdots$  &         {10,12}\\

$\cdots$ &SDSS\,J092646.88+132134.5&DA&210$\pm$25.1&9500$\pm$500&0.62$\pm$0.10 & $\cdots$& DD,        {55} \\
            
0930+010&SDSS\,J093313.14+005135.4&DQB: &$\cdots$ & $\cdots$ &$\cdots$ &$\cdots$ &Like LHS\,2229,        {10}\\
   
0931+105 &SDSS\,J093356.40+102215.7  &DA  & 2.11$\pm$0.49 & 8500  &  $\cdots$  &  $\cdots$  &        {11,12}  \\

0931+394 &SDSS\,J093409.90+392759.3 &DA &  1.01  &  10000  & $\cdots$  & $\cdots$ &       {12} \\ 
  
0931+507 &SDSS\,J093447.90+503312.2  &DA    &  7.35$\pm$2.21  &  8900  &  $\cdots$  &  $\cdots$  &        {11,12} \\

0939+211 &SDSS\,J094235.02+205208.3 &DA & 39.21$\pm$4.55 & 20000  & $\cdots$ & $\cdots$ &        {12} \\ 
  
0941+458   &SDSS\,J094458.92+453901.2  &DA    &  15.91$\pm$9.10 &  15500:  &  $\cdots$  &  $\cdots$  &         {11,12} \\

0945+246   &LB\,11146a  &DA   & 670  &  16000$\pm$2000  &  0.90$^+_{-0.14}$  &  $\cdots$  & DD,      {56,57}\\
  
0952+094   &SDSS\,J095442.91+091354.4  & DQ   & $\cdots$  &  $\cdots$&$\cdots$  &  $\cdots$  &        {11}\\  
  
0957+022   &SDSS\,J100005.67+015859.2  &DA   & 19.74$\pm$10.26&  9000  &  $\cdots$  &  $\cdots$  &         {10,12}\\

$\cdots$    & SDSS\,J100346.66−003123.1 & DZ & 4.37$\pm$0.05& 6300 & $\cdots$  &  $\cdots$  &          {18}\\

1001+058  &SDSS\,J100356.32+053825.6  &DA &672.07$\pm$118.63 &  23000  &  $\cdots$  &  $\cdots$ &       {11,12}\\ 

1004+304  &SDSS\,J100657.51+303338.1 &DA &  1.0$\pm$0.1  &  10000&$\cdots$ & $\cdots$ &        {12} \\ 

1004+128 &SDSS\,J100715.55+123709.5&DA & 5.41: &18000& $\cdots$&$\cdots$&Complex $B$,  {11,12}\\

1005+163 &SDSS\,J100759.80+162349.6 &DA &   19.18$\pm$3.36  & 11000 & $\cdots$ &$\cdots$  &        {12} \\ 

1011+371   &SDSS\,J101428.09+365724.3 &DA &   11.09$\pm$1.50  & 10500  &  $\cdots$& $\cdots$ &         {12} \\ 
  
1008+290 &LHS\,2229 &DQB &  $\sim$\,100 & 4600  &  $\cdots$  &  $\cdots$  &        {58,59}   \\
   
1012+093 &SDSS\,J101529.62+090703.8 &DA &  4.09$\pm$0.86 &  7200  &  $\cdots$  &  $\cdots$  &        {11,12}\\ 
  
1013+044  &SDSS\,J101618.37+040920.6 &DA&  2.01  & 10000  &  $\cdots$  &  $\cdots$  &        {10,12}\\ 

1015+014&PG\,1015+014,&DA&120$\pm$10&14000&$\log\,g=(8.0)$ & $105^{+12}_{-8}$\,min &    {60,23,61,9}\\
   &SDSS\,J101805.04+011123.5 & & & & & & \\

1019+200 &SDSS\,J102239.06+194904.3 &DA&  2.94$\pm$0.71 & 9000 &  $\cdots$ & $\cdots$  &         {12} \\ 
     
1019+274  &SDSS\,J102220.69+272539.8 &DA& 4.91$\pm$0.31  & 11000 &  $\cdots$ & $\cdots$  &        {12} \\ 

1017+367  &GD\,116, Ton\,1206 &DA   &    65$\pm$5  &  16000  &  $\cdots$  &  $\cdots$  &        {62}\\

1018-103  & EC\,10188-1019&  DA & 3:  & 17720 & $\log g=8.52$ & $\cdots$  &        {21} \\

1026+117&LHS\,2273&DA& 18  &  7160$\pm$170  &  0.59: &  35-45\,min &         {25,9} \\  

$\cdots$  & HS\,1031+0343 & DA &6.1$\pm$0.3 &$\cdots$ &$\cdots$ & $\cdots$&      {16}\\
  
1031+234  &PG\,1031+234, Ton\,527 &DA & $\sim$\,200-1000 &  $\sim$\,15000  &  $\cdots$  & $5.53\pm0.05$\,hr &       {63,9}\\

1032+214 &LP\,372-41,NLTT\,24770&DA &  2.96$\pm$0.33  &  $7000-8000$: & $\cdots$& $\cdots$&       {34,12} \\ 
&  SDSS\,J103532.53+212603.5 & & & & & & \\
 
1033+656  &SDSS\,J103655.38+652252.0 & DQ   & 4:$^*$&$\cdots$  & $\cdots$& $\cdots$ &        {64,10} \\
  
1036-204 &LP\,790-29&DQB? &50 & 7800&$\log\,g=(8.0)$ & 24-28\,yrs &       {58,65,66,59}\\
  
1043-050 &HE\,1043-0502      &DB& $\sim$\,820 & $\sim$\,15000& $\cdots$&$\cdots$&       {20,67}\\
   
1045-091 &HE\,1045-0908      &DA&  16 & 10000$\pm$1000  &  $\log\,g=(8.0)$  &  2.7 hr &       {68} \\
  
1050+598 &SDSS\,J105404.38+593333.3  &DA& 17.41$\pm$7.90  &  9500  &  $\cdots$  &  $\cdots$  &         {10,12}\\
  
1053+656 &SDSS\,J105628.49+652313.5 &DA & 29.27$\pm$5.78 &  16500  &  $\cdots$  &  $\cdots$  &         {10,12}\\

1054+042&SDSS\,J105709.81+041130.3 &DA & 2.03 & 8000 & $\cdots$ & $\cdots$ &        {12} \\ 
   
1105-048  &LTT\,4099   &DA &  0.0123$^*$ &  15280$\pm$20 & 0.52$\pm$0.01&$>$3 &       {36,69,16,4}\\

1107+602  &SDSS\,J111010.50+600141.4  &DA  & 6.5  &  30000  &  $\cdots$  &  $\cdots$  &         {10,12}\\
  
1111+020&LSPM\,J1113+0146, &DQB?&$\cdots$&$\cdots$&$\cdots$&$\cdots$&Like LP\,790-29,       {10,59}\\
  & SDSS\,J111341.33+014641.7 & & & & & & \\
   
1115+101 &SDSS\,J111812.67+095241.4 &DA &  3.38$\pm$0.72 &  10500  &  $\cdots$  &  $\cdots$  &         {11,12}\\

1117-113 &SDSS\,J112030.34-115051.1 &DA & 8.90$\pm$1.02 & 20000 & $\cdots$ & $\cdots$ &        {12} \\

1120+324 &SDSS\,J112257.10+322327.8 &DA & 11.38$\pm$3.42  &12500  & $\cdots$ & $\cdots$ &        {12} \\ 

$\cdots$    & SDSS\,J232538.93+044813.1& DZ & 6.56$\pm$0.09$^*$ & 7200 &  $\cdots$  &  $\cdots$  &         {18}\\

1120+101   &SDSS\,J112328.49+095619.3 &DA &1.21&9500  &  $\cdots$ &  $\cdots$ &        {12} \\ 
   
1126-008   &SDSS\,J112852.88-010540.8  &DA &  2  &  11000  &  $\cdots$  &  $\cdots$  &         {10,12}\\
  
1126+499   &SDSS\,J112924.74+493931.9  &DA & 5.31$\pm$0.64  &  10000  &  $\cdots$  &  $\cdots$  &        {11,12} \\

1129+284     &SDSS\,J113215.38+280934.3 &DA & 3.01$\pm$0.82 & 7000: &  $\cdots$ & $\cdots$ &        {12} \\ 
   
1131+521   &SDSS\,J113357.66+515204.8 &DA  & 8.64$\pm$0.78& 22000 &  $\cdots$  &  $\cdots$  &         {10,12}\\
  
1135+579   &SDSS\,J113756.50+574022.4 &DA &5.00$\pm$0.34  &  7800  &  $\cdots$  &  $\cdots$  &        {11,12} \\

1136-015   &  LBQS\,1136-0132 &DA& 22.71$\pm$1.26 &  10500  &  $\cdots$  &  $\cdots$  &       {70,10,12}\\
      & SDSS\,J113839.51-014903.0 & & & & & & \\
  
1137+614   &SDSS\,J114006.37+611008.2 &DA & 50.19$\pm$17.78 &  13500  &  $\cdots$  &  $\cdots$  &        {10,12} \\
  
1145+487   &SDSS\,J114829.00+482731.2 & DA    & 32.47$\pm$7.11  &  27500  &  $\cdots$  &  $\cdots$  &         {11,12}\\

 $\cdots$    & SDSS\,J115224.51+160546.7 & DZ & 2.72$\pm$0.04$^*$ & 6500 & $\cdots$  &  $\cdots$  &         {18}\\
 
1151+015   &SDSS\,J115418.14+011711.4 & DA  & 33.47$\pm$2.07 &  27000:  &  $\cdots$  &  $\cdots$  &         {10,12} \\
   
1156+619 &SDSS\,J115917.39+613914.3 & DA&  20.10$\pm$6.70&  23000  &  $\cdots$  &  $\cdots$  &         {10,12}  \\
  
1159+619 &SDSS\,J120150.10+614257.0 &DA  &  11.35$\pm$1.53 &  10500  &  $\cdots$  &  $\cdots$&       {11,12}  \\
  
1203+085   &SDSS\,J120609.80+081323.7 &DA& 760.63$\pm$281.66  &  13000  &  $\cdots$  &  $\cdots$  &       {11,12}  \\
   
1204+444   &SDSS\,J120728.96+440731.6 & DA &       2.03  &  16750  &  $\cdots$  &  $\cdots$  &       {11,12}  \\

1209+018   &SDSS\,J121209.31+013627.7 &DA & 10.12$\pm$0.93   &  10000  &  $\cdots$  &  90\,min  & L5-L8 comp.     {71,12}\\

1211-171&HE\,1211-1707 &DB & 50 & $\sim$\,12000 & $\cdots$  &    $\cdots$  &  {20,9}\\  
   
1212-022&LHS\,2534, &DZ&1.92$^*$& 5200&$\cdots$&$\cdots$&          {72,10,18} \\

  &SDSS\,121456.39−023402.7  & & & & & & \\

1214-001   &SDSS\,J121635.37-002656.2 &DA  & 59.70$\pm$10.23 & 15000 &  $\cdots$  &  $\cdots$  &         {26,10,12} \\ 
  
1219+005   &SDSS\,J122209.44+001534.0&DA &14.70$\pm$4.70  &  14000  &  $\cdots$  &  $\cdots$  &         {26,10,12}\\
  
1220+234   &PG\,1220+234&DA  &3:  & 26540 &  0.81  &  $\cdots$  &    {73,21} \\

1220+484   &SDSS\,J122249.14+481133.1  &DA & 8.05$\pm$2.24  & 9000  &  $\cdots$  &  $\cdots$  &         {11,12}\\
 
1221+422   &SDSS\,J122401.48+415551.9 &DA &  22.36$\pm$3.02&  9500  &  $\cdots$  &  $\cdots$  &         {11,12}\\
   
1231+130   &SDSS\,J123414.11+124829.6 &DA &    4.32$\pm$0.27    &  8200  &  $\cdots$  &  $\cdots$  &        {11,12} \\

1233-052   & HE\,1233-0519 & DA & 0.61$\pm$0.01 &$\cdots$ & $\cdots$&$\cdots$ &        {17,16}\\

1235+422  & LHS\,5222, NLTT\,31347& DQ & $\cdots$ &$\cdots$ & $\cdots$&$\cdots$ &  {112}\\
  
1245+413   &SDSS\,J124806.38+410427.2 &DA &  7.03$\pm$1.19  &  7000  &  $\cdots$  &  $\cdots$  &   {11,12}\\

1246+296   &SDSS\,J124836.31+294231.2 &DA  &  3.95$\pm$0.25  & 7000:  & $\cdots$   &  $\cdots$ &        {12} \\ 
  
1246-022   &SDSS\,J124851.31-022924.7 &DA  &   7.36$\pm$2.19 & 13500  &  $\cdots$  &  $\cdots$  &         {10,12}\\
   
1248+161  &SDSS\,J125044.42+154957.4 &DA  &20.71$\pm$3.66  &  10000  &  $\cdots$  &  $\cdots$  &        {11,12} \\
  
1252+564  &SDSS\,J125416.01+561204.7 & DA &  38.86$\pm$9.03&  13250  &  $\cdots$  &  $\cdots$  &         {11,12}\\

 $\cdots$   &SDSS\,J125434.65+371000.1 &DA &  4.10$\pm$0.35 & 10000  & $\cdots$ & $\cdots$  &        {12} \\ 

1254+345 &HS\,1254+3440, &DA & 11.45$\pm$0.71 & 8500  & $\cdots$& $\cdots$ &        {74,12} \\ 
                  & SDSS\,J125715.54+341439.3&& & & & & \\
1300+590&SDSS\,J130033.48+590407.0&DA &$\sim$6&6300$\pm$300&0.54$\pm$0.06  &  $\cdots$ &DD, {75}\\

1309+853  &G256-7&DA  & 4.9$\pm$0.5  &  $\sim$\,5600  &  0.5 &$\cdots$&       {44,8,73}\\
  
1312+098  &PG\,1312+098 &DA & 10  &  $\sim$\,20000  &  $\cdots$  &  5.42839\,hr  &        {23,8}\\
  
1317+135 &SDSS\,J132002.48+131901.6  &DA&2.02 &  14750  &  $\cdots$  &  $\cdots$  &         {11,12}\\
   
1327+594  &SDSS\,J132858.20+590851.0 &DA&  18  &  25000  &  $\cdots$  & $\cdots$&         {11} \\
 
1328+307&G165-7, &DZ& 0.65$^*$&  6440$\pm$210 & 0.57$\pm$0.17 & $\cdots$ &        {76,18} \\
  &SDSS J133059.26+302953.2  & & & & & & \\

1330+015 &G62-46 &DA & 7.36$\pm$0.11  & 6040  &  0.25 & $\cdots$  & DD,        {77}\\
   
1331+005&SDSS\,J133359.86+001654.8 &DQB? &$\cdots$&$\cdots$&$\cdots$&$\cdots$ &Like LHS\,2229        {10,59} \\
  
1332+643   &SDSS\,J133340.34+640627.4 & DA  &  10.71$\pm$1.03  &  13500  &  $\cdots$  &  $\cdots$  &         {10,12} \\  

1334+486 & GD\,359, & DA& 2.7 & $\cdots$  &  $\cdots$  & $\cdots$  &            {78}\\
   & SDSS J170751.91+353239.97  & & & & & & \\

1339+659   &SDSS\,J134043.10+654349.2 & DA  & 4.32$\pm$0.76 &  15000  &  $\cdots$  &  $\cdots$  &         {10,12}\\

1346+383  &SDSS\,J134820.79+381017.2 &DA  & 13.65$\pm$2.66 & 35000 &  $\cdots$ &  $\cdots$ &        {12} \\ 

1349+545&SBS\,1349+5434, &DA&761.00$\pm$56.42&12000&$\cdots$&$\cdots$&       {79,11,12}\\
  & SDSS\,J135141.13+541947.4 & & & & & & \\
  
1350-090&LP\,907-037&DA &0.268$^*$&9520$\pm$140& 0.83$\pm$0.03 & $\cdots$ &        {5,80}\\

1405+501&SDSS\,J140716.66+495613.7 &DA & 12.49$\pm$6.20 & 20000 & $\cdots$ & $\cdots$ &        {12} \\ 

1416+256&SDSS\,J141906.19+254356.5 &DA &  2.03$\pm$0.10  &9000 &$\cdots$&  $\cdots$ &       {12} \\ 
 
 $\cdots$   & SDSS\,J142625.71+575218.3&DQB& $\sim 1.2$ &19830$\pm$750 & $\log g=9.0$ & $\cdots$& Pulsating DQ,        {81}\\

1425+375 &SDSS\,J142703.40+372110.5  &DA    & 27.04$\pm$3.20&  19000  &  $\cdots$  &  $\cdots$  &         {11,12}\\

1428+282 &SDSS\,J143019.05+281100.8 &DA &  9.34$\pm$1.44  &9000 &  $\cdots$ & $\cdots$ &        {12} \\ 
   
1430+432   &SDSS\,J143218.26+430126.7 &DA &  1.01  &  24000  &  $\cdots$  &  $\cdots$  &        {11,12} \\
  
1430+460   &SDSS\,J143235.46+454852.5 &DA &  12.29$\pm$6.98 &  16750  &  $\cdots$  &  $\cdots$  &         {11,12}\\
  
1440+753   &  EUVE\,J1439+750    &DA & 14-16  &  20000-50000  &  0.88-1.19  &  $\cdots$  &  DD,        {45}\\
   
1444+592   &SDSS\,J144614.00+590216.7  &DA    &4.42$\pm$3.79  &  12500  &  $\cdots$  &  $\cdots$  &         {10,12}\\
  
1452+435   &SDSS\,J145415.01+432149.5  &DA &  2.35$\pm$0.88&  11500  &  $\cdots$  &  $\cdots$ &         {11,12}\\

 $\cdots$  &LSPM\,J1459+0851 & DA & $\sim$\,2 & 5535$\pm$45 & $\log g=(8.0)$ & $\cdots$&T4.5$\pm$0.5 companion  {82}\\
  
1503-070   &GD\,175       &DA    &      2.3 &  6990  &  0.70$\pm$0.13&  $\cdots$  &  DD,        {37}\\
   
$\cdots$& SDSS\,J150746.80+520958.0 & DA & 65.2$\pm$0.3& 18000$\pm$1000 &0.99$\pm$0.05&$\cdots$&DD, {55}\\
 
1506+399& CBS\,229, & DA & 18.9 & 18000$\pm$2000&0.81$\pm$0.09& $\cdots$ &DD,        {11,12,47}\\
  & SDSS\,J150813.24+394504.9 & & & & & &\\
1509+425 &SDSS\,J151130.20+422023.0 &DA & 22.40$\pm$9.41 &  9750  &  $\cdots$  &  $\cdots$  &         {11,12}\\

1511+076 &SDSS\,J151415.65+074446.5 &DA & 35.34$\pm$2.80  &10000 & $\cdots$ & $\cdots$ &        {12} \\ 
  
1516+612 &SDSS\,J151745.19+610543.6 &DA    &  13.98$\pm$7.36 &  9500  &  $\cdots$  &  $\cdots$  &         {10,12} \\

1521+191  &SDSS\,J152401.60+185659.2 &DA & 11.96$\pm$1.85 &13500&$\cdots$  &  $\cdots$  &        {12} \\ 
   
1531-022&GD\,185&DA& 0.035$\pm$0.016$^*$ &18620$\pm$285 & 0.88$\pm$0.03 & $\cdots$ &Uncertain,        {83,80}\\
  
1533+423 &SDSS\,J153532.25+421305.6  &DA& 5.27$\pm$4.05& 18500  &  $\cdots$  &  $\cdots$  &        {10,12} \\

 $\cdots$ & SDSS\,J153642.53+420519.2 & DZ &9.59$\pm$0.04$^*$&5500&  $\cdots$  &  $\cdots$ &       {18}\\
  
1533-057&PG\,1533-057&DA&31$\pm$3&20000$\pm$1040&0.94$\pm$0.18&$1.89\pm0.001$\,hr&       {84,80,22,9}\\
   
1537+532 &SDSS\,J153829.29+530604.6  &DA & 13.99$\pm$3.82& 13500  &  $\cdots$  &  $\cdots$  &        {10,12} \\

1536+085& SDSS\,J153843.10+084238.2  &DA & 13.20$\pm$4.34 & 9500& $\cdots$ & $\cdots$ &        {12} \\ 
  
1539+039   &SDSS\,J154213.48+034800.4  &DA    & 8.35$\pm$2.60 &  8500  &  $\cdots$  &  $\cdots$  &         {10,12} \\
         
1541+344  & SDSS\,J154305.67+343223.6 &DA &  4.09$\pm$2.67  & 25000 &  $\cdots$  &  $\cdots$  &        {12} \\ 
   
 $\cdots$   & SDSS\,J155708.04+041156.52 & DA & 41 & $\cdots$  &  $\cdots$  & $\cdots$  &       {78}\\

1603+492   &SDSS\,J160437.36+490809.2  &DA &  59.51$\pm$4.64  &  9000  &  $\cdots$  &  $\cdots$  &         {10,12}\\

1610+330 & CBS\,418 & $\cdots$ & $\cdots$ & $\cdots$ & $\cdots$ &$\cdots$  &         {21}\\
   
1639+537 &GD\,356 &DA &13&7510$\pm$210 & 0.67$\pm$0.07&  0.0803\,d  & H in emission,   {85,86,87,88}\\
  
1641+241   &SDSS\,J164357.02+240201.3  &DA    &  2 &  16500  &  $\cdots$  &  $\cdots$  &        {11,12} \\
  
1645+372   &SDSS\,J164703.24+370910.3  &DA    & 2.10$\pm$0.67 &  16250  &  $\cdots$  &  $\cdots$  &         {11,12} \\

1647+591   & G\,226-29, V$^*$\,DN\,Dra, NLTT\,43637&  DA & $<$0.005: &$\cdots$   & $\cdots$  &$\cdots$   & Pulsating DA,        {89,69}\\
 
1648+342   &SDSS\,J165029.91+341125.5  &DA    & 3.38$\pm$0.67  &  9750  &  $\cdots$  &  $\cdots$  &        {11,12} \\
  
1650+355   &SDSS\,J165203.68+352815.8  &DA    & 7.37$\pm$2.92  &  11500  &  $\cdots$  &  $\cdots$  &         {10,12} \\ 

1650+334  &  SDSS\,J165249.09+333444.9 &DA &  5.07$\pm$4.18& 9000& $\cdots$ & $\cdots$ &        {12} \\ 

1653+385  & NLTT\,43806,& DAZ & 0.07$^*$ & 5900 &$\cdots$ & $\cdots$ &        {90} \\
    & SDSS J165445.69+382936.5  & & & & & & \\

1658+440&PG\,1658+440&DA&2.3$\pm$0.2&30510$\pm$200&1.31$\pm$0.02&6\,hr–4\,d&       {91,50,9}\\
   
1702+322 &SDSS\,J170400.01+321328.7 &DA &   50.11$\pm$25.08&  23000  &  $\cdots$  &  $\cdots$  &         {11,12} \\
  
1713+393   &  NLTT\,44447 &DA &  1.3  & 7000$\pm$1000 & 0.59: &  $\cdots$  &         {92}\\
  
1715+601   &SDSS\,J171556.29+600643.9  &DA    & 2.03  &  13500  &  $\cdots$  &  $\cdots$  &        {11,12} \\
   
1719+562   &SDSS\,J172045.37+561214.9  &DA    & 19.79$\pm$5.42 &  22500  &  $\cdots$  &  $\cdots$  &         {26,10,12} \\ 
  
1722+541   &SDSS\,J172329.14+540755.8  &DA    & 32.85$\pm$3.56  &  10000  &  $\cdots$  &  $\cdots$  &        {26,10,12} \\
  
1728+565   &SDSS\,J172932.48+563204.1  &DA    &  27.26$\pm$7.04  &  10500  &  $\cdots$  &  $\cdots$  &        {10,12}   \\
        
1743-520   &BPM\,25114 &DA &      36 & $\sim$\,20000  &$\log\,g=(8.0)$  &  2.84\,d  &        {93,43,94}\\
  
1748+708&G\,240-72&DB&$\gtrsim$\,100&5590$\pm$90&0.81$\pm$0.01&$\gtrsim$\,100\,yr &       {95,96,97,37}\\ 
  
1814+248&G\,183-35&DA&$\sim$\,14 &6500$\pm$500&$\log\,g=(8.0)$&$\sim$\,50\,min-yr &       {44,8}\\
   
1818+126&G\,141-2 &DA &  $\sim$\,3 & 6340$\pm$130  &  0.26$\pm$0.12&$\cdots$& DD,        {98,25}\\

1820+609  &LP\,103-294, G\,227-28 &DBA& $\le$\,0.1& 4780$\pm$140& 0.48$\pm$0.05& months-–years &        {37,8,9}\\
  
1829+547   &G\,227-35  &DBA & 170-180 & 6280$\pm$140 & 0.90$\pm$0.07  &  $\gtrsim$\,100\,yr  &        {99,100,44,37}\\
  
1900+705&Grw\,+70$^\circ$8247&DA&320$\pm$20 & 16000  &  0.95$\pm$0.02&$\gtrsim$\,100\,yr&       {101,85,102,103,37} \\
      
1939+401 &NGC\,6819-8 &DA & 10.3$\pm$1.1 & 19000$\pm$1000 & 0.50$\pm0.05$&$\cdots$&       {104,52} \\

1953-011&NLTT\,48454, G\,92-40&DA&0.1-0.5&7920$\pm$200&0.74$\pm$0.03&1.44176\,d&Magnetic spot,        {37,105,106}\\
  
2010+310 &GD\,229 &DB&520&  18000&$\gtrsim$\,1&$\gtrsim$\,100\,yr &          {101,107,97,67} \\
        
2022+130  & SDSS\,J202501.10+131025.6 &DA &  10.10$\pm$1.76  & 17000& $\cdots$ &  $\cdots$ &        {12} \\ 
  
2039-682  & GJ\,2149, LTT\,8190 &DA  & 0.05$^*$ & 16050 & $\cdots$ & $\cdots$ &        {83,4}\\

2043-073   &SDSS\,J204626.15-071037.0  &DA &  2.03  &  8000  &  $\cdots$  &  $\cdots$  &         {10,12}\\
   
2049-004   &SDSS\,J205233.52-001610.7  &DA    &  13.42$\pm$3.73  &  19000  &  $\cdots$  &  $\cdots$  &        {10,12}  \\ 

2051-208 & HK\,22880−134& DA & 0.22-0.29 &$\cdots$ &$\cdots$ & Variable &        {16}\\
  
2105-820&L\,24-52, LTT\,8381, G\,J820.1 &DAZ &  0.043$^*$ &10800$\pm$290&0.75$\pm$0.03& $\cdots$&        {37,83,16,108}\\

2146+005   &SDSS\,J214900.87+004842.8 &DA & 10.09$\pm$4.71&  11000  &  $\cdots$  &  $\cdots$  &         {11,12}\\
   
2146-077   &SDSS\,J214930.74-072812.0  &DA &  44.71$\pm$1.92 &  22000  &  $\cdots$  &  $\cdots$  &        {10,12} \\
 
2149+002&SDSS\,J215135.00+003140.5&DA&$\sim$\,300&9000&$\cdots$&$\cdots$&       {10}\\

2149+126   &SDSS\,J215148.31+125525.5 &DA  & 20.76$\pm$1.39&14000  &  $\cdots$  &  $\cdots$  &        {10,12} \\

2153-512&GJ\,841B, BPM\,27606&DQ&1.3&6100$\pm$200 &$\cdots$ & $\cdots$&       {109,58,42} \\
     
 $\cdots$  &SDSS\,J220029.09-074121.5 & DQ & Very weak?  & 21240$\pm$180  &  $\log\,g=(8.0)$ &$\cdots$&       {2}  \\

2202-000 &SDSS\,J220435.05+001242.9 &DA&  1.02$\pm$0.10  & 22000 &$\cdots$ &  $\cdots$  &         {12} \\ 
   
2215-002 &SDSS\,J221828.59-000012.2 &DA &  257.54$\pm$48.71 & 15500  &  $\cdots$  &  $\cdots$  &         {10,12}\\
 
2225+176 & NLTT\,53908 & DAZ & 0.334$\pm$0.003 & 6250$\pm$70& $\log g=(7.87\pm0.12)$&$\cdots$&       {28} \\

2245+146 &SDSS\,J224741.46+145638.8 &DA &42.11$\pm$2.83&18000&$\cdots$&$\cdots$  &       {10,12}\\
       
2254+076   &SDSS\,J225726.05+075541.7 &DA &  16.17$\pm$2.81  & 40000 & $\cdots$ & $\cdots$ &        {12} \\ 
  
2329-291 &$\cdots$ & DA& 0.031 &24000&$\cdots$ &$\cdots$ &          {83,4}\\

2316+123 & KUV\,813-14, KUV\,23162+1220&DA&  45$\pm$5 &11000$\pm$1000 & $\log\,g=(8.0)$ & 17.856\,d  &        {23,44}\\
   
2317+008 &SDSS\,J231951.73+010909.3  &DA &9.35: &  8300  &  $\cdots$  &  $\cdots$  &         {11,12}\\
  
2320+003 &SDSS\,J232248.22+003900.9 &DA& 21.40$\pm$3.36&20000-39000  &  $\cdots$  &  $\cdots$  &        {26,10,12}\\
  
2321-010 &SDSS\,J232337.55-004628.2 &DB & 4.8  & 15000  &  $\cdots$  &  $\cdots$  &         {10} \\
   
2329+267&PG\,2329+267&DA & 2.31$\pm$0.59 & 9400$\pm$240 & 0.61$\pm$0.16&2.767\,hr&       {110,37,9}\\
  
2343+386 &SDSS\,J234605.44+385337.7  &DA & 798.1$\pm$163.6& 26000 &  $\cdots$  &  $\cdots$  &        {11,12} \\
  
2343-106 &SDSS\,J234623.69-102357.0 &DA & 9.17$\pm$1.58&8500 &$\cdots$& $\cdots$&       {11,12}\\
   
2359-434 & LP\,988-088, LTT\,9857, LHS\,1005 &DA & 0.0098$^*$ & 8570$\pm$50 &0.98$\pm$0.04&2.69479& Different B detected,      {83,36,111,16,108,4,112} \\ \\
\hline

\end{longtable}

\textit{References:}
\footnotesize

(1)\,\citet{Schmidt1992_subMG}; (2)\,\citet{DufourHotDQ2008}; (3)\,\citet{Lawrie2013}; 
(4)\,\citet{Kawka2012}; 
(5)\,\citet{SchmidtSmith1994}; (6)\,\citet{Valyavin2005}; 
(7)\,\citet{Bergeron1992}; (8)\,\citet{Putney1997}; (9)\,\citet{Brinkworth2013}; 
(10)\,\citet{Schmidt2003}; 
(11)\,\citet{Vanlandingham2005}; (12)\,\citet{Kulebi2009}; 
(13)\,\citet{Angel1977}; (14)\,\citet{Achilleos1992b}; (15)\,\citet{Liebert1977}; 
(16)\,\citet{Koester2009};  (17)\,\citet{Koester2001}; 
(18)\,\citet{Hollands2015};  (19)\,\citet{Achilleos1991}; 
(20)\,\citet{Schmidt2001};  (21)\,\citet{Gianninas2011}; 
(22)\,\citet{Achilleos1989}; (23)\,\citet{SchmidtNorsworty1991}; (24)\,\citet{Friedrich1997}; 
(25)\,\citet{Bergeron1997};  (26)\,\citet{Gaensicke2002}; 
(27)\,\citet{Kawka2011}; (28)\,\citet{Kawka2014}; 
(29)\,\citet{Farihi2011}; 
(30)\,\citet{Barstow1995}; (31)\,\citet{Ferrario1997}; (32)\,\citet{Vennes2003}; (33)\,\citet{Burleigh1999}; 
(34)\,\citet{Arazimova2009};  (35)\,\citet{Fabrika2000}; 
(36)\,\citet{Aznar2004};  (37)\,\citet{Bergeron2001}; 
(38)\,\citet{Angel1974a}; 
(39)\,\citet{Bues1989}; 
(40)\,\citet{Dufour2005}; 
(41)\,\citet{berdyuginaetal07-1}; 
(42)\,\citet{Vornanen2010}; 
(43)\,\citet{WickMartin1979}; 
(44)\,\citet{putney+jordan95-1}; 
(45)\,\citet{Vennes1999};  (46)\,\citet{Giovannini1998}; 
(47)\,\citet{Dobbie2013}; 
(48)\,\citet{Angel1974c}; (49)\,\citet{Martin1984}; 
(50)\,\citet{Ferrario1998}; 
(51)\,\citet{Claver2001}; (52)\,\citet{Kulebi2013}; 
(53)\,\citet{Wesemael2001};  (54)\,\citet{Angel1972}; 
(55)\,\citet{Dobbie2012}; 
(56)\,\citet{Liebert1993}; (57)\,\citet{Glenn1994}; 
(58)\,\citet{schmidtetal99-1}; (59)\,\citet{Hall2008}; 
(60)\,\citet{WickCrop1988}; (61)\,\citet{euchneretal05-1}; 
(62)\,\citet{Saffer1989};  (63)\,\citet{Schmidt1986}; 
(64)\,\citet{Liebert2003DQ}; 
(65)\,\citet{Bues1999}; (66)\,\citet{Jordan2002}; 
(67)\,\citet{Wickramasinghe2002};  (68)\,\citet{Euchner2005}; 
(69)\,\citet{Valyavin2006};  
(70)\,\citet{Foltz1989}; 
(71)\,\citet{Schmidt2005};
(72)\,\citet{Reid2001};  (73)\,\citet{Liebert2003}; 
(74)\,\citet{Hagen1987};  (75)\,\citet{Girven2010}; 
(76)\,\citet{Dufour2006} (77)\,\citet{Bergeron1993}; 
(78)\,\citet{Kepler2015};  (79)\,\citet{Liebert1994}; 
(80)\,\citet{Liebert2005b};  (81)\,\citet{Dufour2008}; 
(82)\,\citet{Day-Jones2011};  (83)\,\citet{Koester1998}; 
(84)\,\citet{Liebert1985}; 
(85)\,\citet{Greenstein1985}; (86)\,\citet{Ferrario1997_GD356}; (87)\,\citet{Brinkworth2004}; (88)\,\citet{Wickramasinghe2010}; 
(89)\,\citet{Schmidt1997};  (90)\,\citet{Zuckerman2011}; 
(91)\,\citet{Schmidt1992_ultramass};  (92)\,\citet{Kawka2006}; 
(93)\,\citet{MartinWick1978}; (94)\,\citet{Wegner1977}; 
(95)\,\citet{Angel1974b}; (96)\,\citet{Angel1978}; (97)\,\citet{Berdyugin1999}; 
(98)\,\citet{Greenstein1986}; 
(99)\,\citet{Angel1975}; (100)\,\citet{Cohen1993}; 
(101)\,\citet{Angel1985}; (102)\,\citet{Wickramasinghe1988}; (103)\,\citet{Jordan1992}; 
(104)\,\citet{Kalirai2008}; 
(105)\,\citet{Maxted2000}; (106)\,\citet{Brinkworth2005}; 
(108)\,\citet{Landstreet2012};  (109)\,\citet{WickBessell1979}; 
(110)\,\citet{Moran1998}; 
(111)\,\citet{Kawka2007}; (112)\,\citet{Gary2014}; 
(112)\,\citet{Vornanen2013}
\end{landscape}

\newpage
\footnotesize
\begin{landscape}

\pagestyle{empty}
\voffset4cm

\begin{longtable}{l c c c c r c c c}

\caption{\label{tab:mcvs} Magnetic white dwarfs in synchronous cataclysmic 
variables}\\
\hline\hline
\noalign{\smallskip}
Name  & B$_{\rm cyc,1}$ & B$_{\rm cyc,2}$ & B$_{\rm Zeem,phot}$ & B$_{\rm Zeem,halo}$ & T$_{\rm eff}$ & P$_{\rm rot}$ & Mass & References\\
      &  (MG)    &  (MG)    &  (MG)   & (MG)    &    (K)    &  (min)    &  (M$_{\odot}$) &  Notes \\                   
\noalign{\smallskip}
\hline
\endfirsthead
\endfoot
\caption{continued.}\\
\hline
\noalign{\smallskip}
Name  & B$_{\rm cyc,1}$ & B$_{\rm cyc,2}$ & B$_{\rm Zeem,phot}$ & B$_{\rm Zeem,halo}$ & T$_{\rm eff}$ & P$_{\rm rot}$ & Mass & References\\
      &  (MG)    &  (MG)    &  (MG)   & (MG)    &    (K)    &  (min)    &  (M$_{\odot}$) &  Notes \\
\noalign{\smallskip}
\hline
\noalign{\smallskip}
\endhead
\endfoot
\noalign{\smallskip}
 & & & & & & & & \\
{\bf Polars} & & & & & & & &  \\
EQ Cet(= RX\,J0128.8-2339)  & 34  & 45  & $\cdots$ & $\cdots$ & $\cdots$   & 92.8  & $\cdots$  & 1,2\\ 
CV Hyi(= RX\,J0132.7-6554)  & 68  & $\cdots$ & $\cdots$ & $\cdots$ & $\cdots$   & 77.8  & $\cdots$  & 3 \\ 
BL Hyi(= H\,0139-68)        & 23  & $\cdots$ & 21  & 12  & 13300 & 113.6 & 0.71 & 4,5,6\\ 
RX\,J0154.0-5947            & $\cdots$ & $\cdots$ & $\cdots$ & $\cdots$ & $\cdots$   & 80:   & $\cdots$  & 7\\
FL Cet(=SDSS\,J0155+0028)   & 29  & $\cdots$ & $\cdots$ & $\cdots$ & $\cdots$   & 87.1  & $\cdots$ & 8\\
AI Tri(=RX\,J0203.8+2959)   & 38  & $\cdots$ & $\cdots$ & $\cdots$ & $\cdots$   & 275.5 & $\cdots$ & 9\\
BS Tri(=RX\,J0209.4+2832)   & $\cdots$ & $\cdots$ & $\cdots$ & $\cdots$ & $\cdots$   & 96.3  & $\cdots$ & 10\\
CW Hyi(=RBS0324)            & 13  & $\cdots$ & $\cdots$ & $\cdots$ & $\cdots$   & 181.8 & $\cdots$ & 11\\
WW Hor(=EXO\,023432-5232)   & 25  & $\cdots$ & $\cdots$ & $\cdots$ & $\cdots$   & 114.6 & 1.1 & 12\\
EF Eri(=2A0311-227)         & 21  & 17  & $\cdots$ & 15  & 9500  &  81   & 0.9 & 13,14,15\\
CSS091109:\,J035759+102943  & $\cdots$ &  $\cdots$ & $\cdots$  & $\cdots$ & 114.0 & $\cdots$ &  16\\

SDSS\,J032855.00+052254.2   & 33  & $\cdots$ & $\cdots$ & $\cdots$ & $\cdots$   & 122.0 & $\cdots$ & 17\\
VY For(=EXO\,0329.9-2606)   & $\cdots$ & $\cdots$ & $\cdots$ & $\cdots$ & $\cdots$   & 228.0 & $\cdots$ & 18\\
UZ For(=EXO\,0333.3-2554)   & 56  & 75: & $\cdots$ & $\cdots$ & $\cdots$   & 126.5 & 0.7 & 19,20\\
RX\,J0425.6-5714          & $>$50 & $\cdots$ & $\cdots$ & $\cdots$ & $\cdots$   & 85.8  & $\cdots$ & 21\\
IW Eri(=RBS0541)            & $\cdots$ & $\cdots$ & $\cdots$ & $\cdots$ & $\cdots$   & 87.1  & $\cdots$ & 11\\ 
RS Cae(=RX\,J0453.4-4213)   & 36  & $\cdots$ & $\cdots$ & $\cdots$ & $\cdots$   & 102.1 & $\cdots$ & 22,23\\
HY Eri(=RX\,J0501.7-0359)   & 25: & $\cdots$ & $\cdots$ & $\cdots$ & $\cdots$   & 171.3 & 0.4 & 24\\ 
V1309 Ori(=RX\,J0515.6+0105) & 61: & $\cdots$& $\cdots$ & $\cdots$ & $<$20000 & 479.0 & 0.6-0.7 & 25,26,27\\
IPHAS\,J052832.69+283837.6  & $\cdots$ & $\cdots$ & $\cdots$ & $\cdots$ &  $\cdots$  & $\cdots$   & $\cdots$ & 28\\
UW Pic(=RX\,J0531.5-4624)   & 19  & $\cdots$ & $\cdots$ & $\cdots$ &  $\cdots$  & 133.4 & $\cdots$  &  29,30\\ 
BY Cam(=H\,0538+608)        & 41  & $\cdots$ & $\cdots$ & $\cdots$ &  $\cdots$  & 199.3 & 1.04 & 31,32,33\\   
                            &     &     &     &     &       &       &      & $\rm P_{orb}$=201.3 \\ 
RX\,J0649.8-0737            & $\cdots$ & $\cdots$ & $\cdots$ & $\cdots$ &  $\cdots$  & 265:  & $\cdots$  & 34\\
LW Cam(=RX\,J0704.2+6203)   & 20  & $\cdots$ & $\cdots$ & $\cdots$ &  $\cdots$  & 97.3  & $\cdots$  & 35\\
CSS081231:\,J071126+440405  & $\cdots$ & $\cdots$ & $\cdots$ & $\cdots$ & $\cdots$   & 117.2 & $\cdots$  & 36\\

HS Cam(=RX\,J0719.2+6557)   & $\cdots$ & $\cdots$ & $\cdots$ & $\cdots$ &  $\cdots$  & 98.2  & 0.75 & 37\\
V654 Aur(=SDSS\,J072910.2+365838) & $\cdots$ & $\cdots$ & $\cdots$ & $\cdots$  & $\cdots$ &  150   & $\cdots$  & 38\\
RX\,J0749.1-0549            & $\cdots$ & $\cdots$ & $\cdots$ & $\cdots$ & $\cdots$   & 215:  & $\cdots$  & 34\\
EU Lyn(=SDSS\,J075240.45+362823.2) & $\cdots$ & $\cdots$ & $\cdots$ & $\cdots$  & $\cdots$  & 164  & $\cdots$ & 38\\
V516 Pup(=RX\,J0803.4-4748) & 39  & $\cdots$ & $\cdots$ & $\cdots$ & $\cdots$   & 136.8  & $\cdots$ & 39\\ 
CSS100108:\,J081031+002429  & $\cdots$ & $\cdots$ & $\cdots$ & $\cdots$ & $\cdots$   & 116.2  & $\cdots$ & 40\\
VV Pup(=2E\,0812.8-1853)    &  31 & 56  & $\cdots$ & $\cdots$ & 12100 & 100.4 & 0.73 & 6,41\\
EG Lyn(=RBS0696)            & $\cdots$ & $\cdots$ & $\cdots$ & $\cdots$ &  $\cdots$  &  99.4 & $\cdots$  & 11\\
EU Cnc(=G\,186)             & 41  & $\cdots$ & $\cdots$ & $\cdots$ &  $\cdots$  &  125.4 & $\geq$0.68 &  42,43\\
FR Lyn(=SDSS\,J085414.02+390537.2) & 44  & $\cdots$ & $\cdots$ & $\cdots$  &  $\cdots$ & 113.3 & $\cdots$ & 44,45\\ 
SDSS\,J085909.18+053654.5   & $\cdots$ & $\cdots$ & $\cdots$ & $\cdots$ & $\cdots$   & 143.8 & $\cdots$ & 45\\
SDSS\,J092122.84+203857.1   & 32  & $\cdots$ & $\cdots$ & $\cdots$ & $\cdots$   &  84.2 & $\cdots$ & 46,47\\
HU Leo(=SDSSJ\,092444.48+080150.9) & $\cdots$ & $\cdots$ & $\cdots$ & $\cdots$ & $\cdots$ & 131.24 & $\cdots$ & 48\\
MN Hya(=RX\,J0929.1-2404)   & 42  & $\cdots$ & $\cdots$ & $\cdots$ & $\cdots$   &  203.4 & $\cdots$ & 49,50\\
SDSSJ\,093839.25+534403.8    & $\cdots$ &  $\cdots$ &  $\cdots$ &  $\cdots$ &  $\cdots$ &  $\lesssim 120$ & 51\\
RX\,J0953.1+1458            & $\cdots$ & $\cdots$ & $\cdots$ & $\cdots$ & $\cdots$   &  102   & $\cdots$ & 7\\
1RXS\,J100211.4-192534      & $\cdots$ & $\cdots$ & $\cdots$ & $\cdots$ & $\cdots$   & 107    & $\sim$0.5 & 52\\
1RXS\,J100734.4-201731      & 94  & $\cdots$ & $\cdots$ & $\cdots$ & $\cdots$   & 208.6  & 0.8 & 52,53\\
GG Leo(=RBS0842)            & 23  & $\cdots$ & $\cdots$ & $\cdots$ & $\cdots$   & 79.9   & $\sim$1.0 & 54,55\\
V381 Vel(=RX\,J1016.9-4103) & 52  & $\cdots$ & $\cdots$ & $\cdots$ & $\cdots$   & 134    &  $\cdots$ & 56\\
FH UMa(=RBS0904)           & $>$20  & $\cdots$ & $\cdots$ & $\cdots$ & $\cdots$ &  80    & $\cdots$  & 57\\
EK UMa(=1E\,1048.5+5421)    & 47  & $\cdots$ & $\cdots$ & $\cdots$ & $\cdots$   & 114.5  & $\cdots$  & 58,59\\
AN UMa(=PG\,1101+453)       & 36  & $\cdots$ & $\cdots$ & $\cdots$ & $\cdots$   & 114.8  & $\cdots$  & 60,61\\
ST LMi(=CW\,1103+254)       & 12  &  21 & 19  & $\cdots$ & 10800 & 113.9  & 0.45 & 6,62,63,64\\
AR UMa(=1ES\,1113+432)  & $\gtrsim$160 & 240  & 230 & $\cdots$  & $\sim$20000 & 115.9 & $\cdots$ & 65,66\\
DP Leo(=1E\,1114+18.2)      & 31  & 76 & $\cdots$  & $\cdots$ &  13500 &  89.8 & $\cdots$ & 11,67,68\\
V1033 Cen(=RX\,J1141.3-6410) & 20 & $\cdots$ & $\cdots$ & $\cdots$ & $\cdots$   & 189.4 & $\cdots$  & 69,70\\
EU UMa(=RX\,J1149.9+2844)   & 43  & $\cdots$ & $\cdots$ & $\cdots$ & $\cdots$   & 90.1  & $\cdots$  & 71,72\\
SDSSJ\,120724.69+223529.8   & $\cdots$ & $\cdots$ & $\cdots$ & $\cdots$ & $\cdots$ & $\cdots$ & $\cdots$ & 73\\
V379 Vir(=SDSS\,J121209.31+013627.7) & 7 & $\cdots$ & $\cdots$ & $\cdots$ & $\sim$10000 & 88.4 & $\cdots$ & 74,75,76\\
                                     &   &     &     &     &             &      &      & LARP\\
WD\,1248+161(=SDSS\,J125044.42+154957.4) & 20 & $\cdots$ & $\cdots$ & $\cdots$ & 10000 & 86.3 & $\cdots$ & 77\\
                                         &    &     &     &     &       &      &      &   LARP\\     
EV UMa(=RX\,J1307.8+5351)    & 30: & $\cdots$ & $\cdots$ & $\cdots$  & $\cdots$ & 79.7 & $\cdots$ & 78\\
2XMM\,J131223.4+173659      & $\leq$10 & $\cdots$ & $\cdots$ & $\cdots$   & $\cdots$ & 91.9 & $\cdots$ & 79\\
V1043 Cen(=RX\,J1313.2-3259) & 56 & $\cdots$ & $\cdots$ & $\cdots$ & 15000  & 251.4 & $\cdots$ & 80,81\\
SDSS\,J133309.19+143706.9   & $\cdots$ & $\cdots$ & $\cdots$ & $\cdots$ & $\cdots$  &  132 & $\cdots$  & 46,47\\
SDSS\,J134441.83+204408.4   & 65: & $\cdots$ & $\cdots$ & $\cdots$ & $\cdots$  &  110: & $\cdots$ & 82\\
V834 Cen(=1E\,1405-45.1)    & 23  & $\cdots$ & 22  & 23  & 14300 & 101.5 & 0.66 & 6,83,84,85,86\\
SDSS\,J142256.31-022108.0   & $\cdots$ & $\cdots$ & $\cdots$ & $\cdots$ & $\cdots$   & 202:  & $\cdots$ &  87\\
V895 Cen(=EUVE\,J1429-38.0) & $\cdots$ & $\cdots$ & $\cdots$ & $\cdots$ & 13800 & 285.9 & $\cdots$ & 6,88\\
IGR\,J14536-5522            & 20 & $\cdots$ & $\cdots$ & $\cdots$  & $\cdots$   & 189.4 & $\cdots$ & 89\\
CSS100216:\,J150354-220711  & $\cdots$ & $\cdots$ & $\cdots$ & $\cdots$ &  $\cdots$  & 133.4 & $\cdots$ & 40\\
SDSS\,J151415.65+074446.5   & 36  & $\cdots$ & $\cdots$ & $\cdots$ & 10000 &  88.7 & $\cdots$ & 77\\
                              &     &     &     &     &       &       &     & LARP\\
SDSS\,J153023.64+220646.4   & $\cdots$ & $\cdots$ & $\cdots$ & $\cdots$ & $\cdots$ & $\cdots$ & 90\\
BM CrB(=SDSS\,J154104.66+360252.9) & 33 & $\cdots$ & $\cdots$ & $\cdots$   & $\cdots$ & 84    & $\cdots$ & 44,45\\
2XMM\,J154305.5-522709      & $\cdots$ & $\cdots$ & $\cdots$ & $\cdots$ & $\cdots$   & 143:  & $\cdots$ & 91\\
MR Ser(=PG\,1550+191)       & 25  & $\cdots$ & 28  & 25  & 14000 & 113.5 & 0.5 & 6,64,92\\ 
AP CrB(=RX\,J1554.2+2721)   & 110 & $\cdots$ & 144 & $\cdots$ & $\sim$20000 & 151.9 & $\cdots$ & 93,94,95\\
V519 Ser(=1RXS\,J161008.0+035222) & 15: & $\cdots$ & $\cdots$ & $\cdots$ & $\cdots$   & 190.5 & $\cdots$ & 96,97\\
V1189 Her(=SDSS\,J162936.53+263519.5)   & $\cdots$ & $\cdots$ & $\cdots$ & $\cdots$ & $\cdots$ & 134 & $\cdots$ & 44\\
1RXS\,J170053.7+400354      & $\cdots$ & $\cdots$ & $\cdots$ & $\cdots$ & $\cdots$ & 116.4 & $\cdots$ & 38\\
V1007 Her(=RBS1646)         & 50  & $\cdots$ & $\cdots$ & $\cdots$ & $\cdots$ & 119.9 & $\cdots$ & 98\\
V2301 Oph(=1H\,1752+081)    & $\cdots$ & $\cdots$ & $\cdots$ &  7  & 10000 & 113.0 & 0.8 & 99,100,101\\
V884 Her(=RX\,J1802.1+1804) & 150 & $\cdots$ & $\cdots$ & 150 & $\cdots$ & 113.3 & $\cdots$ & 102,103\\
AM Her(=3U1809+50)          & 14  & $\cdots$ & 13  & $\cdots$ & 19800 & 185.6 & 0.78 & 104,105,106\\
XGPS-I\,J183251-100106      & $\cdots$ & $\cdots$ & $\cdots$ & $\cdots$ & $\cdots$ & 89.0 & $\cdots$ & 107\\
V347 Pav(=RX\,J1844.7-7418) & 10: & 20  & $\cdots$ & $\cdots$ & 12300 & 90.1 & $\cdots$ & 6,108,109\\
1RXS\,J184542.4+483134      & $\cdots$ & $\cdots$ & $\cdots$ & $\cdots$ & $\cdots$ & 79.1 & $\cdots$ & 110\\
MT Dra(=RX\,J1846.9+5538)   & 15: & $\cdots$ & $\cdots$ & $\cdots$ & $\cdots$   & 128.7  & $\cdots$ & 111\\
EP Dra(=1H\,1903+689)       & $\cdots$ & $\cdots$ & $\cdots$ & 16  & $\cdots$ & 104.6 & 0.43 & 112\\
CTCV\,J1928-5001            & 20: & $\cdots$ & $\cdots$ & $\cdots$ & $\cdots$ & 101.0  & $\cdots$ & 113,114\\
QS Tel(=RX\,J1938.6-4613)   & 47  & 75: & 60  & $\cdots$ & 17500 & 139.9 & $\cdots$ & 115\\

V1432 Aql(=RX J1940.2-1025) & 30: & $\cdots$ & $\cdots$ & $\cdots$& $\sim$35000 &  202.5 & 1.2 & 116,117,118\\
                            &      &     &     &    &             &        &     &  $\rm P_{orb}$=201.9\\ 
CSS100805:J194428-420209    & $\cdots$ & $\cdots$ & $\cdots$ & $\cdots$ & $\cdots$         & 91.9   & $\cdots$ & 119\\
V393 Pav(=RX\,J1957.1-5738) & $\cdots$ & $\cdots$ & $\cdots$ & 16  & $\cdots$         & 98.8   & $\cdots$ & 120\\
QQ Vul(=1E\,2003+22.5)      & 30: & $\cdots$ & $\cdots$ & $\cdots$ & $\cdots$ & 222.5 & 0.58 & 64, 121, 122\\
V349 Pav(=Drissen V211b)    & $\cdots$ & $\cdots$ & $\cdots$ & $\cdots$ & $\cdots$ & 159.7 & $\cdots$ & 123\\
V4738 Sgr(=RX\,J2022.6-3954)& 67  & $\cdots$ & $\cdots$ & $\cdots$ & $\cdots$ & 78.0  & $\cdots$ & 3\\
SDSS\,J205017.83-053626.7   & $\cdots$ & $\cdots$ & $\cdots$ & $\cdots$ & $\cdots$ & 94.2  & $\cdots$ & 124\\
HU Aqr(=RX\,J2107.9-0517)   & 37  & $\cdots$ & 20: & $\cdots$ & 14000 & 125.0 & 0.9 & 6,125,126,127\\
V1500 Cyg(=Nova Cyg 1975)   & $\gtrsim$25 & $\cdots$ & $\cdots$ & $\cdots$  & 70000 & 197.5 & 0.9 & 128,129\\
                            &      &     &     &    &             &        &     & $\rm P_{orb}$=201.0 \\ 
CD Ind(=RX\,J2115.7-5840)   & 11  & $\cdots$ & $\cdots$ & $\cdots$ &   $\cdots$ & 109.6 & 0.79 & 130,131,132\\
                            &      &     &     &    &             &        &     & $\rm P_{orb}$=110.9\\
CE Gru(=Grus V1)            & 32  & $\cdots$ & $\cdots$ & $\cdots$ & $\cdots$   & 108.5 &  $\sim$1.0 & 133,134,135\\
SDSS\,J215427.19+155713.0   & $\cdots$ & $\cdots$ & $\cdots$ & $\cdots$ & $\cdots$   & 96.9  & $\cdots$ & 82\\
V388 Peg(=RX\,J2157.5+0855) & 20: & $\cdots$ & $\cdots$ & $\cdots$ & $\cdots$   & 202.5 & $\cdots$ & 136\\
SWIFT\,J2218.4+1925         & $\cdots$ & $\cdots$ & $\cdots$ & $\cdots$ & $\cdots$ & 129.5 & 0.97 & 137,138\\
2XMM\,J225036.9+573154      & $\cdots$ & $\cdots$ & $\cdots$ & $\cdots$ & $\cdots$ & 174.2 & $\cdots$  & 139\\
CP Tuc(=AX\,J2315-592)      & $\cdots$ & $\cdots$ & 10  & $\cdots$ & $\cdots$ & 89.0  & 0.68 & 33, 140, 141\\
1RXS\,J231603.9-052713      & 25  & $\cdots$ & $\cdots$ & $\cdots$ & $\cdots$ & 209.5 & 1.0  &  96\\
SWIFT\,J2319.4+2619         & $\cdots$ & $\cdots$ & $\cdots$ & $\cdots$ & $\cdots$ & 180.6 & $\cdots$ & 142\\

\noalign{\smallskip}
{\bf Pre-polars} & & & & & & & & \\
SDSS\,J030308.35+005444.1  & $\cdots$  & $\cdots$ &  8 & $\cdots$ & 9150 & 193.6 & 0.84 & 143\\
SDSS\,J083751.0+383012.5   & $\cdots$  & $\cdots$ & $\cdots$ & $\cdots$ &  $\cdots$  &  178.8  & $\cdots$ & 87\\
HS 0922+1333                & 66  & 81  & $\cdots$ & $\cdots$ & $\leq$8000 & 242.4 & $\cdots$ & 144,145,146\\
WX LMi(=HS\,1023+3900 )     & 61  & 70  & $\cdots$ & $\cdots$ & $\cdots$        & 166.9 & $\cdots$ & 147,148,149\\
IL Leo(=SDSS\,J103100.55+202832.2) & 42 & $\cdots$ & $\cdots$ & $\cdots$ & 9500   &  83.2 & $\cdots$ & 150\\ 
SDSS\,J105905.06+272755.4          & 57 & $\cdots$ & $\cdots$ & $\cdots$ & $\leq$ 8500 & 150: & $\cdots$ & 150\\
SDSS\,J120615.73+510047.0        & 108  & $\cdots$ & $\cdots$ & $\cdots$ & 9000 & 197: & $\cdots$ & 151\\
PZ Vir(=J132411.57+032050.4)     & 63:  & $\cdots$ & $\cdots$ & $\cdots$ & $\lesssim$ 7500 & 158.7 & $\cdots$ & 152,153,154\\
MQ Dra(=SDSS\,J155331.11+551614.4) & 58 & $\cdots$ & $\cdots$ & $\cdots$ & $\lesssim$10000 &  263.5 & $\cdots$ & 152,154\\ 
SDSS\,J204827.91+005008.9        & 62:  & $\cdots$ & $\cdots$ & $\cdots$ & 7500 & 252 & $\cdots$ & 154\\
\noalign{\smallskip}
\hline

\end{longtable}

\textit{References:}
\footnotesize
(1)~\citet{Schwope99}; 
(2)~\citet{Campbell2008a}; 
(3)~\citet{Burwitz97}; 
(4)~\citet{Ferrario96}; 
(5)~\citet{Schwope95a}; 
(6)~\citet{Araujo05}; 
(7)~\citet{beuermann99}; 
(8)~\citet{Szkody02}; 
(9)~\citet{Schwarz98}; 
 (10) ~\citet{Denisenko06}; 
(11)~\citet{Schwope02a}; 
(12)~\citet{Bailey88}; 
(13)~\citet{Ferrario96}; 
(14)~\citet{Achilleos1992a}; 
(15)~\citet{Schwope10}; 
(16)~\citet{Schwope_Thinius12} 
(17)~\citet{Szkody07}; 
(18)~\citet{Beuermann89}; 
(19)~\citet{Ferrario89}; 
(20)~\citet{Schwope90}; 
(21)~\citet{Halpern98}; 
(22)~\citet{Burwitz96}; 
(23)~\citet{Traulsen14}; 
(24)~\citet{Burwitz99}; 
(25)~\citet{Staude01}; 
(26)~\citet{Shafter95}; 
(27)~\citet{deMartino98}; 
(28)~\citet{Skinner14}; 
(29)~\citet{Reinsch94}; 
(30)~\citet{Romero-Colmenero03}; 
(31)~\citet{Cropper89}; 
(32)~\citet{Mason98}; 
(33)~\citet{ramsay00a}; 
(34)~\citet{Motch98}; 
(35)~\citet{Tovmassian01}; 
(36)~\citet{Schwope15}; 
(37)~\citet{Tovmassian97}; 
(38)~\citet{Homer05}; 
(39)~\citet{Schwarz99}; 
(40)~\citet{Woudt12a};  
(41)~\citet{Wickramasinghe89}; 
(42)~\citet{Pasquini94}; 
(43)~\citet{Williams13}; 
(44)~\citet{Gaensicke09}; 
(45)~\citet{Szkody05}; 
(46)~\citet{Schmidt08}; 
(47)~\citet{Southworth15}; 
(48)~\citet{Southworth10}; 
(49)~\citet{Ramsay98}; 
(50)~\citet{Buckley98}; 
(51)~\citet{Szkody09}; 
(52)~\citet{Ramsay03}; 
(53)~\citet{Thomas12}; 
(54)~\citet{Burwitz98}; 
(55)~\citet{ramsay04}; 
(56)~\citet{Greiner98}; 
(57)~\citet{Singh95}; 
(58)~\citet{Beuermann09}; 
(59)~\citet{Cropper90b}; 
(60)~\citet{Schneider80}; 
(61)~\citet{Cropper89}; 
(62)~\citet{Ferrario93}; 
(63)~\citet{Schmidt83}; 
(64)~\citet{Mukai87}; 
(65)~\citet{Schmidt96}; 
(66)~\citet{Gaensicke01}; 
(67)~\citet{Cropper93}; 
(68)~\citet{Schwope02b}; 
(69)~\citet{Cieslinski97}; 
(70)~\citet{Buckley00}; 
(71)~\citet{Howell95}; 
(72)~\citet{Schwope95b}; 
(73))~\citet{Szkody11}; 
(74)~\citet{Schmidt05}; 
(75)~\citet{Burleigh06}; 
(76)~\citet{Farihi08}; 
(77)~\citet{Breedt12}; 
(78)~\citet{Osborne94}; 
(79)~\citet{Vogel08}; 
(80)~\citet{Thomas00}; 
(81)~\citet{Gaensicke00}; 
(82)~\citet{Szkody14}; 
(83)~\citet{Wickramasinghe87}; 
(84)~\citet{Ferrario92}; 
(85)~\citet{Schwope90a}; 
(86)~\citet{Cropper86}; 
(87)~\citet{Hilton09}; 
(88)~\citet{Howell97}; 
(89)~\citet{Potter10}; 
(90)~\citet{Breedt14}; 
(91)~\citet{Servillat12}; 
(92)~\citet{Schwope93}; 
(93)~\citet{Thorstensen02}; 
(94)~\citet{Gaensicke04}; 
(95)~\citet{Schwope06}; 
(96)~\citet{Rodrigues06}; 
(97)~\citet{Thorstensen15}; 
(98)~\citet{Greiner98b}; 
(99)~\citet{Ferrario95}; 
(100)~\citet{Silber94}; 
(101)~\citet{Barwig94}; 
(102)~\citet{Szkody95}; 
(103)~\citet{Schmidt01}; 
(104)~\citet{Wickramasinghe_Martin85}; 
(105)~\citet{Bailey91}; 
(106)~\citet{Gaensicke06}; 
(107)~\citet{Hui12}; 
(108)~\citet{Ramsay96}; 
(109)~\citet{Potter00}; 
(110)~\citet{Shchurova13}; 
(111)~\citet{Schwarz02}; 
(112)~\citet{Schwope_Mengel97}; 
(113)~\citet{Tappert04}; 
(114)~\citet{Potter05}; 
(115)~\citet{Schwope95c}; 
(116)~\citet{Friedrich96}; 
(117)~\citet{Rana05}; 
(118)~\citet{Schmidt01b}; 
(119)~\citet{Coppejans14}; 
(120)~\citet{Thomas96}; 
(121)~\citet{Schwope00}; 
(122)~\citet{Cropper98}; 
(123)~\citet{Wickramasinghe93}; 
(124)~\citet{Potter06}; 
(125)~\citet{Schwope93b}; 
(126)~\citet{Schwope97}; 
(127)~\citet{Glenn94}; 
(128)~\citet{Schmidt_Stockman91}; 
(129)~\citet{Schmidt95}; 
(130)~\citet{Ramsay99}; 
(131)~\citet{Schwope97b}; 
(132)~\citet{Ramsay00b}; 
(133)~\citet{Tuohy88}; 
(134)~\citet{Wickramasinghe91}; 
(135)~\citet{Ramsay02}; 
(136)~\citet{Tovmassian00}; 
(137)~\citet{Thorstensen13}; 
(138)~\citet{Bernardini14}; 
(139)~\citet{Ramsay09}; 
(140)~\citet{Beuermann07}; 
(141)~\citet{Ramsay99b}; 
(142)~\citet{Shafter08}; 
(143)~\citet{Parsons13}; 
(144)~\citet{Reimers00}; 
(145)~\citet{Tovmassian07}; 
(146)~\citet{Vogel11}; 
(147)~\citet{Reimers99}; 
(148)~\citet{Schwarz01}; 
(149)~\citet{Vogel07}; 
(150)~\citet{Schmidt07}; 
(151)~\citet{Schwope09}; 
(152)~\citet{Szkody03}; 
(153)~\citet{Southworth15}; 
(154)~\citet{Schmidt05}; 

\normalsize

\newpage

\footnotesize
\begin{longtable}{lccccc}
\caption{\label{tab:ips} Magnetic white dwarfs in asynchronous cataclysmic variables}\\
\hline\hline
\noalign{\smallskip}
Name  & B        & P$_{rot}$ & P$_{orb}$ & Mass           & References\\
      &  (MG)    &  (min)    &  (min)    &  (M$_{\odot}$) &  Notes \\ 
\noalign{\smallskip}
\hline
\endfirsthead
\endfoot
\caption{continued.}\\
\hline
\noalign{\smallskip}
Name  & B$_{\rm cyc}$        & P$_{\rm rot}$ & P$_{\rm orb}$ & Mass           & References\\
      &  (MG)    &  (s)    &  (min)    &  (M$_{\odot}$) &  Notes \\ 
\noalign{\smallskip}
\hline
\noalign{\smallskip}
\endhead
\endfoot
\noalign{\smallskip}
 & & & & & \\
V1033 Cas(=IGR\,J00234+6141) & $\cdots$ & 563.5 &  242.0 & 0.9 & 1,2\\
V709 Cas(=RX\,J0028.8+5917)  & $\cdots$ & 312.8 &  320.0 & 0.96 & 3,4,2\\  
V515 And(=XSS\,J00564+4548)  & $\cdots$ & 465.5 &  163.9 & 0.79 & 5,6,7\\
XY Ari(=H\,0253+193)         & $\cdots$ & 206.3 &  363.9 & 1.0 & 8,9,2\\
GK Per(=Nova Persei 1901)    & $\cdots$ & 351.3 & 2875.4 & 0.9 & 10,11,2\\
V1062 Tau(=H\,0459+246)      & $\cdots$ & 3704  &  598.9 & 0.7 & 2,4,12\\
UU Col(=RX\,J0512.2-3241)    & 10-30: & 863.5 &  207.0 & 0.6  & 13,14,15\\
Paloma(=1RXS\,J052430.2+424449) &  $\cdots$  & 8175.4:  & 157.2 & $\cdots$ & 16\\ 
TV Col(=2A\,0526-328)        & $\cdots$ & 1909.7 & 329.2 &  0.78 & 17,18,2\\
TX Col(=1\,H0542-407)        & $\cdots$ & 1911  & 343.2 & 0.7  & 19,2\\
V405 Aur(=RX\,J0558.0+5353)  & 32  & 545.4 & 249.6 & 0.89 &  20,21,22,2\\ 
MU Cam(=1RXS\,J062518.2+733433)  & $\cdots$ & 1187.2 & 283.1 & 0.74 & 23,2\\
V902 Mon(=IPHAS\,J062746.41+014811.3) & $\cdots$ &  2210 & 489.6 & $\cdots$ & 23b\\
V647 Aur(=1RXS\,J063631.9+353537) & $\cdots$ &  932.9 & 207.9 & 0.74 & 24,7\\
V418 Gem(=1RXS\,J070407.9+262501) & $\cdots$ &  480.7 & 262.8 & $\lesssim$0.5 & 25,26\\
BG CMi(=3\,A0729+103)             & $\sim$4   & 913.5 & 194.1 & 0.7 & 27,28,2\\
V667 Pup(=Swift\,J0732.5-1331)    & $\cdots$ & 512.4 & 336.2 & $\cdots$ & 4\\
PQ Gem(=RX\,J0751.2+1444)         & 9-21:  & 833.4 & 311.6  & 0.65 & 29,30,31,2\\
HT Cam(=RX\,J0757.0+6306)         & $\cdots$ & 515.1 & 86.0   & 0.6 & 32,33\\
DW Cnc(=HS\,0756+1624)            & $\cdots$ & 2314.7 & 86.1  & $\cdots$ & 34\\
WX Pyx(=1E\,0830.9-2238)          & $\cdots$ & 1559.2 & 318:  & $\cdots$ & 35\\
EI UMa(=1H0832+488)               & $\cdots$ & 741.6  & 386.1 & $\cdots$ & 36,37\\
IGR\,J08390-4833                  & $\cdots$ & 1480.8 & 480:  & 0.95 &  7\\
VZ Sex(=1RXS\,J094432.1+035738)   & $\cdots$ & 2450   & 214.1 & $\cdots$  & 38,4\\
YY Dra(=DO Dra)                   & $\cdots$ & 529.3  & 238.1 & 0.8  & 39\\
V1025 Cen(=RX\,J1238.2-3842)      & $\cdots$ & 2146.6 &  84.6 & 0.5  & 40,2\\
EX Hya(=4U\,1249-28)              & $\cdots$ & 4021.6 &  98.3 & 0.79 & 41,42\\
IGR\,J15094-6649         & $\gtrsim$ 10 & 809.4  & 353.4 & 0.89 & 43,44,7\\ 
NY Lup(=1RXS\,J154814.5-452845)   & $>$4 & 693.0 & 591.8 & 1.09 & 44,45,2\\
IGR\,J16500-3307                  & $\cdots$  & 571.9  & 217.0 & 0.92 & 7\\
1RXS\,J165443.5-191620            & $\cdots$  & 546.7  & 222.9 & $\cdots$ & 46\\
V2400\,Oph(=RX\,J1712.6-2414)     & 9-27 & 927.7  & 205.8 & 0.8 & 47,48,2\\
IGR\,J17195-4100                  & $\cdots$  & 1053.7 & 240.3 & 0.86 & 49,50,7\\
V2731 Oph(=1RXS\,J173021.5-055933) & 5: & 128.0  & 925.3 & 0.96 & 51,52,53 \\ 
AX\,J1740.1-2847                  & $\cdots$ & 730    & 125:  & $\cdots$ & 54,55\\
AX\,J1740.2-2903                  & $\cdots$ & 628.6  & 343.3 & $\cdots$ &  56\\
V1323 Her(=1RXS\,J180340.0+401214) & $\cdots$ & 1520.5 & 264.1 & 0.69 & 51,26\\
1RXS\,J180431.1-273932            & $\cdots$ & 494    & $\cdots$  & 0.8 & 57\\
DQ Her(=Nova Her 1934)            & $\cdots$ & 70.8   & 278.8 & 0.60 & 58,59,\\
IGR\,J18173-2509                  & $\cdots$ & 1663.4  &  91.9 & 0.96 & 7,56\\
IGR\,J18308-1232                  & $\cdots$ & 1820  & 322.4 &  0.85 & 7,56\\ 
AX\,J1832.3-0840                  & $\cdots$ & 1552.3 & $\cdots$ & $\cdots$ & 54\\
AX\,J1853.3-0128                  & $\cdots$ & 477.6 & 87.2 & $\cdots$ & 56\\
V1223 Sgr(=4\,U1851-31)           & $\cdots$ &  745.5  & 201.9 & 0.65 & 60,61,2\\
IGR\,J19267+1325                  & $\cdots$ &  935.1 & 206.9 & $\cdots$ & 56\\
V2306 Cyg(=WGA\,J1958.2+3232)     & 8:  &  1466.7 & 261.0 & 0.8 & 62,63,64,2\\
IGR\,J19552+0044                  & $\cdots$ & 4960 & 101.7 & 0.77 & 56,65\\
AE Aqr(=1E\,2037.5-0102)          & $\cdots$ & 33.1  &  592.8 & 0.63 & 66,67,68\\ 
V2069 Cyg(=RX\,J2123.7+4217)      & $\cdots$ & 743.1 & 448.8 & 0.82 & 7,69\\
1RXS\,J213344.1+510725            & $\gtrsim$20  & 570.8 & 431.6 & 0.93 & 70,71,72\\
FO Aqr(=H\,2215-086)              &  $\cdots$ & 1254.5 & 290.9 & 0.61 & 73,2\\
AO Psc(=H\,2252-035)              & $\cdots$ & 805.2 & 215.5 & 0.55 & 74,75,2\\
CC Scl(=1RXS J231532.3-304855)    & $\cdots$ & 389.5 & 840.9 & $\cdots$ & 76\\
V598 Peg(=SDSS\,J233325.92+152222.1) & $\cdots$ & 2500 & 83.12 & $\cdots$ & 77,78\\
V455 And(=HS\,2331+390)           & $\cdots$    & 67.6 & 81.1  & $\cdots$ & 79\\
\noalign{\smallskip}

\hline

\end{longtable}
 
\textit{References:}
\footnotesize
(1)~\citet{Bonnet-Bidaud07}; 
(2)~\citet{Brunschweiger09}; 
(3)~\citet{demartino01};  
(4)~\citet{Thorstensen10}; 
(5)~\citet{Kozhevnikov12}; 
(6)~\citet{Bonnet-Bidaud09}; 
(7)~\citet{Bernardini12};  
(8)~\citet{Allan96}; 
(9)~\citet{Hellier97}; 
(10)~\citet{Crampton86}; 
(11)~\citet{Mauche04}; 
(12)~\citet{Hellier02};  
(13)~\citet{Burwitz96b}; 
(14)~\citet{demartino06}; 
(15)~\citet{Katajainen10};  
(16)~\citet{Schwarz07};  
(17)~\citet{Augusteijn94};  
(18)~\citet{Rana04}; 
(19)~\citet{Buckley89}; 
(20)~\citet{Harlaftis99}; 
(21)~\citet{Skillman96}; 
(22)~\citet{piirolaetal08}; 
(23)~\citet{Staude03}; 
(24)~\citet{Kozhevnikov14}; 
(25)~\citet{Patterson11}; 
(26)~\citet{Anzolin08}; 
(27)~\citet{Kim05}; 
(28)~\citet{Chanmugam90}; 
(29)~\citet{Hellier97b}; 
(30)~\citet{evans06}; 
(31)~\citet{potter97}; 
(32)~\citet{Kemp02}; 
(33)~\citet{deMartino05}; 
(34)~\citet{Patterson04}; 
(35)~\citet{Joshi11}; 
(36)~\citet{Baskill05}; 
(37)~\citet{Thorstensen86};  
(38)~\citet{demartino07}; 
(39)~\citet{Haswell97}; 
(40)~\citet{Buckley98b}; 
(41)~\citet{Mauche09};  
(42)~\citet{Beuermann_Reinsch08};  
(43)~\citet{Pretorius09};  
(44)~\citet{potter12};  
(45)~\citet{demartino06a},  
(46)~\citet{Scaringi11}; 
(47)~\citet{buckleyetal95}; 
(48)~\citet{Vaeth97}; 
(49)~\citet{Girish12}; 
(50)~\citet{Pretorius09}; 
(51)~\citet{Gaensicke05}; 
(52)~\citet{demartino08}; 
(53)~\citet{butters09}; 
(54)~\citet{Kaur10}; 
(55)~\citet{Britt13}; 
(56)~\citet{Thorstensen13}; 
(57)~\citet{Masetti12}; 
(58)~\citet{Zhang95}; 
(59)~\citet{Bloemen10}; 
(60)~\citet{Osborne85}; 
(61)~\citet{Jablonski87}; 
(62)~\citet{Norton02}; 
(63)~\citet{Zharikov02}; 
(64)~\citet{Uslenghi01}; 
(65)~\citet{Bernardini13}; 
(66)~\citet{Welsh93}; 
(67)~\citet{deJager94}; 
(68)~\citet{Echevarria08}; 
(69)~\citet{Thorstensen01}; 
(70)~\citet{katajainen07}; 
(71)~\citet{bonnetbidaud06}; 
(72)~\citet{Anzolin09}; 
(73)~\citet{Patterson98}; 
(74)~\citet{Kaluzny88}; 
(75)~\citet{Hellier91};  
(76)~\citet{Woudt12};  
(77)~\citet{Southworth07}; 
(78)~\citet{Hilton09};  
(79)~\citet{araujo-betancoretal05-1} 

\end{landscape}

\end{flushleft}
\end{document}